\documentclass[cmp]{svjour}  
\usepackage{amsmath}
\usepackage{amsfonts,amssymb}
\usepackage{graphicx}

\usepackage[paperwidth=595pt,paperheight=842pt,margin=5pt]{geometry}
\journalname{Communications in Mathematical Physics}

\usepackage{xcolor}\definecolor{RRED}{rgb}{1,0.2,0.1}

\def\SSECSPA{\vspace{-15.015pt}}

\def\WSemi{Semi}
\def\Wsemi{semi}
\def\anWsemi{a semi}

\def\TSeDi{\WSemi-Differential }
\def\TSedi{\WSemi-differential }
\def\Tsedi{\Wsemi-differential }
\def\TsediPRODterm{\Wsemi-differential product }
\def\TSediPRODterm{\WSemi-differential product }
\def\anTsedi{\anWsemi-differential }
\newcommand{\Tsemi}[1]{{#1}--\Wsemi-differential}
\def\PRODterm{--product }
\def\DMoD{{\rm D}}

\newcommand{\ADDCONT}[1]{\addcontentsline{toc}{section}{\hspace{15pt}{\small \arabic{section}.\arabic{subsection} #1}}${}$ \medskip}

\def\ProdOper{\mu}
\def\FldEla{\phi}
\def\FldElb{\psi}
\def\AUGM{\varepsilon}
\def\AUGm{\varepsilon}

\DeclareMathAlphabet{\mathbbm}{U}{bbm}{m}{n}

\DeclareSymbolFont{ltrs}     {OT1}{pzc}{m}{it}
\DeclareSymbolFont{ltrsa}     {OMS}{cmsy}{m}{n}
\DeclareSymbolFont{ltrsA}{U}{txmia}{m}{it}
\DeclareSymbolFont{symbolsC}{U}{txsyc}{m}{n}
\DeclareSymbolFont{ltrsB}{U}{rsfs}{m}{n}

\DeclareSymbolFontAlphabet{\mfrak}{ltrsA}
\DeclareMathAlphabet{\mathpzc}{OT1}{pzc}{m}{it}
\DeclareMathAlphabet{\mathrsfs}{U}{rsfs}{m}{n}

\def\beq{\begin{equation}}
\def\eeq{\end{equation}}
\def\beqn{\begin{equation*}}
\def\eeqn{\end{equation*}}
\def\beqa{\begin{eqnarray}}
\def\eeqa{\end{eqnarray}}
\def\beqs{\begin{eqnarray*}}
\def\eeqs{\end{eqnarray*}}
\def\podr{&& \hspace{-2pt}}
\def\nnb{\nonumber \\}
\def\bnn{\\ \nonumber}
\def\nnbnn{\nonumber \\ \nonumber}
\def\nn{\nonumber}

\def\N{\mathbb{N}}
\def\Z{\mathbb{Z}}
\def\R{\mathbb{R}}

\def\C{\mathbb{C}}

\def\GRF{\mathbb{K}}

\def\x{\mathrm x}
\def\Ttx{\widetilde{\x}}
\def\y{\mathrm y}
\def\z{\mathrm z}

\def\r{\mathrm r}

\def\DX{\mathrm{x}}
\def\Dx{x}
\newcommand{\DXn}[1]{\DX_{#1,n}}
\newcommand{\Dxn}[1]{\Dx_{#1,n}}
\newcommand{\DXsn}[1]{\DX_{#1,n+1}}

\newcommand{\DNu}[2]{{#1}_{(#2)}}

\def\Hom{\mathit{Hom}}

\def\di{\partial}

\newcommand{\ALGcal}[1]{\mathcal{#1}}
\def\Aalg{\ALGcal{A}}
\def\Balg{\ALGcal{B}}
\def\Calg{\ALGcal{C}}
\def\Salg{\ALGcal{C}}
\def\CSalg{\widehat{\Salg}}
\def\Oalg{\ALGcal{O}}
\def\COalg{\widehat{\Oalg}}

\newcommand{\Salgn}[1]{\Salg_{#1}}

\newcommand{\COalgn}[1]{\COalg_{#1}}
\def\Jalg{\mathcal{J}}
\newcommand{\TISalgn}[1]{\Salg_{#1}^{\text{\rm \,t.i.}}}
\newcommand{\TIOalgn}[1]{\Oalg_{#1}^{\text{\rm 
\,t.i.}}}
\def\Jalg{\mathcal{J}}
\newcommand{\Jalgn}[1]{\Jalg^{\hspace{1pt}\raisebox{1pt}{\tiny $\bullet$}\hspace{1pt}#1}}

\newcommand{\MODcal}[1]{\mathcal{#1}}
\def\Mmod{\MODcal{M}}
\def\Nmod{\MODcal{N}}
\def\Pmod{\MODcal{P}}

\def\GRF{\mathbb{K}}
\def\Hom{\mathrm{Hom}}

\def\Drv{\di}
\def\DMN{D}

\newcommand{\DfrOp}[2]{\text{\rm D}^{\hspace{1.1pt}#1}_{\hspace{-1.1pt}#2}\hspace{1pt}}
\newcommand{\ADfrOp}[1]{\widetilde{\,\text{\rm D}\,}{\hspace{-2pt}}_{\hspace{0.1pt}#1}\hspace{1pt}}

\def\Tra{\mathit{T}}
\def\TRa{\mathrm{T}}
\def\vac{\widehat{1}}
\def\VA{V}
\def\vA{\mathcal{V}}
\def\TAvA{\vA^{\hspace{1pt}\textit{t.i.}}}
\def\YB{Y}
\def\YM{\mathcal{Y}}

\def\TYM{\widetilde{\YM}}
\def\FM{\mathcal{F}}
\newcommand{\VaRep}[1]{\text{\rm OPE}_{#1}}

\newcounter{DEFINi}

\newenvironment{DEFIN}[1]{\refstepcounter{DEFINi}\label{#1}\noindent
{\bf Definition \arabic{section}.\arabic{DEFINi}}\,}{}

\newcounter{STATEM}
\renewcommand{\theSTATEM}{\arabic{section}.\arabic{STATEM}}

\newenvironment{THEOR}[1]{\refstepcounter{STATEM}\label{#1}\noindent
{\bf Theorem \theSTATEM}\,}{}

\newenvironment{PROPO}[1]{\refstepcounter{STATEM}\label{#1}\noindent
{\bf Proposition \arabic{section}.\arabic{STATEM}}\,}{}

\newenvironment{LMA}[1]{\refstepcounter{STATEM}\label{#1}\noindent
{\bf Lemma \arabic{section}.\arabic{STATEM}}\,}{}

\newenvironment{COROL}[1]{\refstepcounter{STATEM}\label{#1}\noindent
{\bf Corollary \arabic{section}.\arabic{STATEM}}\,}{}

\def\PRF{\noindent{\it Proof. }}

\newcounter{EXAMp}
\renewcommand{\theEXAMp}{\arabic{section}.\arabic{EXAMp}}

\newenvironment{EXMP}[1]{\refstepcounter{EXAMp}\label{#1}\noindent
{\bf Example  \theEXAMp}\,}{}

\newcounter{Remk}
\renewcommand{\theRemk}{\arabic{section}.\arabic{Remk}}

\newenvironment{REMAR}[1]{\refstepcounter{Remk}\label{#1}\noindent
{\it Remark \theRemk}\,}{}

\def\SETCNTR{%
\setcounter{DEFINi}{0}%
\setcounter{STATEM}{0}%
\setcounter{EXAMp}{0}%
\setcounter{Remk}{0}%
\setcounter{equation}{0}%
}

\renewcommand{\theequation}{\arabic{section}.\arabic{equation}}

\newcommand{\DEFemph}[1]{\textbf{#1}}
\newcommand{\REMemph}[1]{\textit{#1}}
\newcommand{\THEemph}[1]{{\rm #1}}

\def\DOpu{u}
\def\DOpv{v}
\newcommand{\DMo}[2]{{#1}_{\{#2\}}}
\newcommand{\DAl}[2]{{#1}^{\{#2\}}}

\newcommand{\Jet}[2]{\text{\cal{J}}_{\hspace{-0pt}#1\hspace{1pt}}#2}
\def\Blt{\text{\tiny $\bullet$}}

\def\FrE{r}

\newcommand{\TXfrac}[2]{\frac{\textstyle #1\raisebox{-4pt}{}}{\textstyle #2\raisebox{9pt}{}}}
\newcommand{\TXUfrac}[2]{\frac{\textstyle #1\raisebox{-4pt}{}\raisebox{6pt}{}}{\textstyle #2\raisebox{9pt}{}}}
\newcommand{\TXDfrac}[2]{\frac{\textstyle #1\raisebox{-4pt}{}}{\textstyle #2\raisebox{9pt}{}\raisebox{-2pt}{}}}
\newcommand{\TXXfrac}[2]{\frac{\textstyle #1\raisebox{-4pt}{}\raisebox{6pt}{}}{\textstyle #2\raisebox{9pt}{}\raisebox{-2pt}{}}}

\newcommand{\EXTOP}[1]{{#1}^{ext}}

\def\RES{\mathit{Res}}
\def\Ker{\mathit{kernel}}
\newcommand{\KER}[1]{\Ker\left(#1\right)}
\def\DOLP{\lambda}
\def\LorPol{g}

\def\OpDiff{\mathfrak{D}}
\def\OpADiff{\mathfrak{V}}

\def\OpEnd{\mathfrak{End}}
\def\OpGen{\mathfrak{P}}
\def\OpCom{\mathfrak{Com}}
\def\OpComDer{\mathfrak{ComDer}}
\def\TIOpDiff{\OpDiff^{\textrm{\hspace{1pt}t.i.}}}
\def\TIOpADiff{\OpADiff^{\textrm{\hspace{1pt}t.i.}}}

\def\OperCp{\gamma}
\newcommand{\OperCmp}[1]{\OperCp_{#1}}
\newcommand{\OperComp}[3]{\OperCp_{#1}\left(#2;#3\right)}

\def\PDOa{\Phi}
\def\APDOa{\Gamma}

\def\BQu{q}
\def\TQu{Q}
\def\HQu{\widehat{q}}
\def\PermGr{\mathrsfs{S}}

\newcommand{\COMPLET}[1]{\widehat{#1}}
\newcommand{\COMPLEt}[1]{{#1}\raisebox{0pt}{\hspace{-2pt}}^{\widehat{\hspace{10pt}}}\,}
\newcommand{\COMPLEtt}[1]{{#1}\raisebox{8pt}{\hspace{-2pt}}^{\widehat{\hspace{10pt}}}\,}
\newcommand{\TRANS}[1]{\widetilde{#1}}
\newcommand{\TRANs}[1]{{\bigl(#1\bigr)}\raisebox{0pt}{\hspace{-2pt}}^{\widetilde{\hspace{10pt}}}\,}
\newcommand{\OPEPROD}[1]{\mathop{*}\limits_{#1}}
\def\AXi{\Lambda}

\def\UPin{\rotatebox{90}{$\in$}}

\def\repimorf{\longrightarrow\hspace{-8pt}\to}

\begin{document}

\setcounter{footnote}{3}

\title{{\TSeDi Operators and the Algebra of Operator Product Expansion of Quantum Fields}\footnote{%
to my parents}}
\titlerunning{\TSeDi Operators and the Algebra of Operator Product Expansion}

\author{Nikolay M. Nikolov}
\institute{Institute for Nuclear Research and Nuclear Energy of Bulgarian Academy of Sciences, Tzarigradsko chaussee 72, BG-1784, Sofia, Bulgaria.
\\ \email{mitov@inrne.bas.bg}}
\authorrunning{N.M. Nikolov}

\date{November 5, 2019} 

\maketitle

\setcounter{footnote}{0}

\begin{abstract}
We introduce a symmetric operad whose algebras are the Operator Product Expansion (OPE) Algebras of quantum fields.
There is a natural classical limit for 
the algebras over
this operad and 
they are
commutative associative algebras with derivations.
The latter are the algebras of classical fields.
In this paper we 
completely 
develop our approach to models of quantum fields, which come from vertex algebras in higher dimensions.
However, our approach to OPE algebras can be extended to general quantum fields even over curved space--time.
We introduce a notion of OPE operations based on the new notion of \Tsedi operators. 
The latter are linear operators $\APDOa:\Mmod\to\Nmod$ between two modules of a commutative associative algebra $\Aalg$, such that for every $m \in \Mmod$ the assignment $a\mapsto\APDOa(a \cdot m)$ is a differential operator $\Aalg\to\Nmod$ in the usual sense. 
The residue of a meromorphic function at its pole is an example of \anTsedi operator.
\end{abstract}


\tableofcontents

\section{Introduction}

This paper is addressed to both, physicists working in the area of mathematical models of Quantum Field Theory (QFT), and mathematicians, working in the areas of vertex algebras, differential algebras and algebraic operads.
We do not assume in our paper any prior knowledge of QFT.
However, since the main applications of this work are intended to this area, we start this introduction with subsections that contain our physical motivation. 
This part of the paper is relatively independent on the rest of the paper.
However, they can be also useful for mathematicians as a brief survey to the physical problems related to the present work.

\SSECSPA

\subsection{General motivation}
\ADDCONT{General motivation}

\noindent
The quantum fields are singular functions $\phi(\x)$ over the space-time.
A precise mathematical formulation of this feature is the description of these fields as \REMemph{operator-valued distributions}, which is done in the Wightman axiomatic approach.
A direct consequence from the singular nature of quantum fields is the impossibility to multiply them at 
the same point 
of the space--time, i.e., the non--existence of $\phi(\x)\psi(\x)$.
This multiplication operation plays a fundamental 
role
in classical field theory when we construct important physical fields like the Lagrangian density, the stress--energy tensor, the electromagnetic current etc.
Another important operations on classical fields, the de\-ri\-va\-ti\-ons, remain defined for quantum fields in the sense of Schwartz theory of distributions.

From mathematical point of view, we can say that the classical fields form a \REMemph{differential algebra}, or, more precisely, a commutative, associative algebra with derivations.
However, one of the basic operations in this algebraic structure, the products of fields at a point, fails to exist when we pass from classical to quantum fields.
One of the main problems, which we solve in this paper is what algebraic structure replaces the structure of differential algebra in the passage to QFT.
We intend to use this structure in the future as a basic axiomatic structure in QFT, which is an alternative to the so far used axiomatic schemes of Wightman and Haag--Kastler.

The main approach to the problem what replaces the missing pro\-ducts at a point of quantum fields is based on using the so called \REMemph{Operator Product Expansions} (OPE) of quantum fields, proposed by K. Wilson (\cite{W69}).
This concept has far reaching applications in QFT, related to a purely theoretical, as well as a phenomenological, analysis of behaviours at short distances or at high energies.
We shall address one of the main aspects of OPE's -- the construction of composite fields that are generalized products of quantum fields at a point.

From a heuristic point of view the OPE consists of the assumption that the product of two quantum fields at different points (which always exists in the sense of tensor product of distributions) has the following asymptotic expansion
\beq\label{OPE-1oo}
\phi(\x) \, \psi(\y) \, = \, 
\mathop{\sum}\limits_{\AXi \, = \, 0}^{\infty}
C_{\AXi}(\x,\y) \, \vartheta_{\AXi}(\y)
\quad \text{as} \quad
\x \, \to \, \y
\,,
\eeq
where $\vartheta_{\AXi}(\x)$ are again quantum fields that are subject, like $\phi$ and $\psi$, to the principle of locality (i.e., they are \REMemph{local quantum fields}) and $C_{\AXi}(\x,\y)$ are number-valued distributions.
Thus, the distributions $C_{\AXi}(\x,\y)$ keep the singular character of the product $\phi(\x)\psi(\y)$ when $\x\to\y$.
We will accept the following point of view to the usage of (\ref{OPE-1oo}).
Let us assume that we can choose a general, universal, basic system of two point OPE functions $\{C_{\AXi}(\x,\y)\}_{\AXi}$, by which we can write any OPE (\ref{OPE-1oo}) for arbitrary local quantum fields in the theory.
For example, within the frame of perturbative QFT it follows that such a system can be
\beqa\label{C-Xi}
&
C_{\AXi} (\x,\y) =
\bigl((\x-\y)^2\bigr)^n
\,
\bigl(\log \, (\x-\y)^2\bigr)^{\ell}
\,
h_{m,\sigma} (\x-\y)
\,,
\quad
\AXi = (n,\ell,m.\sigma)\,,
& \nnb &
n \in \Z \,,\qquad
\ell,m \,=\, 0,1,\dots\,,\qquad
\sigma \,=\, 1,\dots,H_m
\,,
&
\eeqa
where $(\x-\y)^2$ is the pseudo--Euclidean interval\footnote{%
To be more precise, we need to consider it in the so called forward tube domain (in the sense of boundary value of analytic functions).}
between two points $\x$ and $\y$ of Minkowski space and $\{h_{m,\sigma}\}_{\sigma}$ is a basis of harmonic homogeneous polynomials of degree $m$.
Then, for every symbol $\AXi$ of a basic OPE function $C_{\AXi}$ we obtain a binary operation
\beq\label{OPE-op1}
( \phi \,,\, \psi ) \,\mapsto\, \vartheta_{\AXi} \,=:\, \phi
\,\OPEPROD{\AXi}\,
\psi
\,,
\eeq
called $\AXi$-th \REMemph{OPE product}.
These we call also \REMemph{OPE operations}.
Note that in classical field theory the expansion (\ref{OPE-1oo}) is replaced by the Taylor expansion, 
where the contribution is restricted to only
these basic OPE functions that have indices $\AXi=(n,\ell,m,\sigma)$ for $n \geqslant 0$ and $\ell = 0$.
One can easily derive that in this case the $\AXi$-th product is obtained from the product of $\psi(\x)$ by a particular linear partial differential operator on $\phi$ at the point $\x$.
For quantum fields the new products for $\AXi=(n,\ell,m,\sigma)$ with $n < 0$ or $\ell > 0$ are called singular products.

There are many details, which we have skipped above.
For example, under what axiomatic conditions the expansion  (\ref{OPE-1oo}) exists?
Can one expect some kind of convergence for this expansions?
In this paper we shall focus on what kind of algebraic structure one can expect that the OPE operations (\ref{OPE-op1}) will form.
A test for our axiomatic model will be the so called Globally Conformal Invariant (GCI) QFT that was first time introduced in \cite{NT01}.
This is the only class of quantum fields' models, which admit complectly equivalent formulation in terms of OPE \cite{N05}.
These are the \REMemph{vertex algebras} that originate from two dimensional conformal field theories (cf. \cite{B86,FLM88,K98} and others).
Beyond  GCI QFT there are almost no known self-contained axiomatic formulations for vertex algebras, and even less, such that correspond to the already existing axiomatic field approach of Wightman.
Algebraic structures based on axiomatized OPE are also generally called ``\REMemph{OPE algebras}''.

The new aspects in the present work, comparing to the above cited results from the theory of vertex algebras, is that we develop here an \REMemph{operadic approach}
(see the next subsection for the notion of operad).
The suggested operadic view on vertex algebras is of interest in itself as the axioms for vertex algebras go beyond the frame of Universal Algebra and this makes difficult to formulate them as algebras over an operad.
In the previous years there have been already suggested 
other
operadic approaches to vertex algebras \cite{HL93,BD04,BDHK18}.
Important features of our approach are:
\begin{itemize}
\item[(i)] 
it has a natural generalization beyond the limitations of conformal field theories with ``rational singularities'' (like GCI QFT).
\item[(ii)] 
The constructed operad appears also in the description of the renormalization group (see Sect.~\ref{Ssec1.4-20190922} below).
In this way we obtain a very interesting connection between OPE and renormalization theory.
\end{itemize}
In addition to the latter, this connection was a generating idea for our investigations and it was developed in a series of works \cite{N09,N14,N16}.
The upgrading element in this paper, compared to the previous works, is the notion of  \REMemph{\Tsedi operators}, which is the exact and very concise model for OPE operations.

\SSECSPA

\subsection{Operations, algebras and operads}\label{ssec1.2nn0913n1}
\ADDCONT{Operations, algebras and operads}

\noindent
Different types of algebras, like associative algebras and Lie algebras, are usually defined by specifying some basic operations and identities among these operations.
The identities are formed by using compositions of operations and possibly permutation of arguments.
This formation of types of algebras is formulated in the operad theory in such a way that to any type of algebras there is a corresponding operad.
An operad\footnote{%
In this paper ``operad'' will always stand for ``\REMemph{symmetric} operad'' unless otherwise stated.}
contains a sequences of sets labeled by positive integers and the elements of the $n$-th operadic space are also called the $n$-ary operations of the corresponding type of algebras.
The $n$-ary operation of a given type of algebras include all possible $n$-ary operations that are obtained from the basic operations under compositions and permutations of the arguments.
In this way the operadic spaces encode the identities among the basic operations.
The structures of compositions of operations and permutation group actions on operations are axiomatized in the operad theory in a similar way as the category theory axiomatizes classes of structures and morphisms between them.
Another analogy is provided by the group theory, where every group encodes some symmetry and when a group is mapped homomorphically to the group of all maps of a set we obtain a realization of this symmetry as actual transformations of this set.
Continuing the latter analogy, an algebra over an operad is a representation of this operad in a specific operad built from the underlying set of this algebra.
For a further introduction to the operad theory we refer the reader to \cite[Sect. B]{N14} where we made a brief review of this subject, which is specially prepared for the need of our applications.
Our main reference for the operad theory is the book~\cite{LV12}.

We shall make the following convention in this paper.
When we speak about an ``operation'' we shall have in mind an element of an operad.
Although the operads in this work will be built from operators (e.g., differential operators), when these operators are considered as elements of an operad we shall call them ``operations''.
In particular, we shall call ``differential\PRODterm operations''\footnote{%
In other words, these are compositions of the multiplication operation at the same point and differential operators.
The differential\PRODterm operations can be called also \REMemph{poly--differential operations}, as the corresponding differential operators contain an evaluation at the total diagonal (presenting the multiplication operation) and for this reason are called sometimes  \REMemph{poly--differential operators}.}
the elements of the operad corresponding to commutative associative algebras with derivations (cf. Sect.~\ref{Se8nn}).

\SSECSPA

\subsection{Brief presentation of OPE operations and \Tsedi operators}
\ADDCONT{Brief presentation of OPE operations and \Tsedi operators}

\noindent
We construct in this paper an operad that will govern the OPE algebras and its elements we call ``OPE operations''.
As this operad will be mathematically constructed from ``\Tsedi operators'' that generalize the notion of differential operators we shall call the OPE operations also ``\TsediPRODterm operations''.

Before the formal mathematical presentation let us give a more detailed, but still intuitive description of our concept for the OPE operations (\ref{OPE-op1}).
To this end, 
let us view a QFT model as specified via all the correlation functions of its local fields
(according to the Wightman's reconstruction theorem).
Then,
another point of view on the OPE (\ref{OPE-op1}) is provided by pluging it inside correlation functions:
\beqa\label{INtr-1}
\podr
\bigl\langle \Phi_1 \bigr|
\varphi_1(\z_1) \cdots \varphi_k(\z_k)\,
\phi(\x) \, \psi(\y) \,
\varphi_{k+1}(\z_{k+1})\cdots \varphi_n(\z_{n+1})
\, \Phi_2 \bigr\rangle
\nnb
\podr
\mapsto \,
\bigl\langle \Phi_1 \bigr|
\varphi_1(\z_1) \cdots \varphi_k(\z_k)\,
\vartheta_{\AXi}(\y) \, 
\varphi_{k+1}(\z_{k+1})\cdots \varphi_n(\z_{n+1})
\, \Phi_2 \bigr\rangle
\,.\qquad
\eeqa
In this way, the binary OPE operations induce maps from $(n+1)$--point functions to $n$--point functions.
In classical field theory these operations are the linear partial differential operators in $\x$ and $\y$ evaluated at the diagonal $\x=\y$.
In QFT the correlation functions in (\ref{INtr-1}) are in general singular for $\x=\y$ and we cannot evaluate them at $\x=\y$.
However some ``trace'' of differential operators on the diagonal still remains.
Namely, it turns out that for every fixed correlation function in the left hand side of (\ref{INtr-1}) when we multiply it by an arbitrary smooth function 
$F(\z_1,$ $\dots,$ $\z_k,$ $\x,$ $\y,$ $\z_{k+1},$ $\dots,$ $\z_{n+1})$
then the resulting function is a value of a differential operator on the diagonal $\x=\y$ acting on $F$.

The abstract algebraic situation that we meet here is the following. 
We have two modules $\Mmod$ and $\Nmod$ of a commutative associative algebra $\Aalg$ and we consider a linear map $\APDOa : \Mmod \to \Nmod$ such that for every element $m \in \Mmod$ the assignment
$$
\Aalg \to \Nmod \,:\, a \mapsto \APDOa(a \cdot m)
$$
is a differential operator.\footnote{%
Namely, $\Aalg$ is the commutative associative algebra of smooth functions in $(\z_1,$ $\dots,$ $\z_k,$ $\x,$ $\y,$ $\z_{k+1},$ $\dots,$ $\z_{n+1})$ and the modules $\Mmod$ and $\Nmod$ are the spaces of distributions where the left and the right hand sides of assignment (\ref{INtr-1}) belong to, respectively.}
This is exactly what we call in this paper 
\REMemph{\anTsedi operator}.
Let us point out that one has a direct definition for differential operators $\Mmod \to \Nmod$ (cf. Sect.~\ref{SectDiffOp}) and they will be \Tsedi operators,
however, the converse is not true in general (cf. Sect.~\ref{S5}).

Our general notion of an OPE operation is provided by a sequence of modules $\Oalg_n$ of the commutative associative algebras $\Calg_n$ $=$ $\Calg_1^{\otimes n}$ of smooth $n$--point functions and a $n$--ary OPE operation $\APDOa$ will be modelled by \anTsedi operator 
\beq\label{Intr-eq2}
\APDOa : \Oalg_n \to \Calg_1
\,,
\eeq
where we consider $\Calg_1$ as a $\Calg_n$--module under the diagonal action.

\SSECSPA

\subsection{Relation to previous works}\label{Ssec1.4-20190922}
\ADDCONT{Relation to previous works}

\noindent
The previous works of the author on the problems of the present paper include \cite{N09,N14,N16}.
Since we have tried to make this paper self contained, as much as possible, we do not assume a detailed knowledge of the reader on these previous works.
However, we will make now a brief review on the way how the ideas of our work have developed over the last 10 years.

In \cite[Sect.~2.6]{N09} we have first introduced linear maps\footnote{%
we denoted there $\gamma$ by the letter $Q$, cf. \cite[Theorem~2.9]{N09}}
\beqa\label{Intr-eq2-1}
\podr
\gamma :
\Oalg_n \to 
\{
\text{distributions supported at the total diagonal}
\}
\,, \quad \text{where:}
\\ \label{Intr-eq2-2-20190922}
\podr
\Oalg_n
\, = \,
\biggl\{
\begin{array}{l}
\text{space of $n$--point rational functions on Minkowski space $\R^D$}
\\
\text{with light--cone singularities}
\end{array}
\biggl\}
\,\qquad
\eeqa
for $n=2,3,\dots$ in order to describe 
the \REMemph{finite changes of the renormalization prescriptions}.\footnote{%
The renormalization prescriptions are described by the so called \REMemph{renormalization maps} of \cite{N09,NST14}).}
These maps $\gamma$ obey the property that they are $\Salg_n$--module maps (cf. Eq.~(\ref{eq1.10-20190922})), where
\beq\label{Intr-eq2-3-20190922}
\Salg_n
\, = \,
\bigl\{
\begin{array}{l}
\text{space of $n$--point polynomial functions on Minkowski space $\R^D$}
\end{array}
\bigl\}
\,.\qquad
\eeq
Thus, one can consider the values $\gamma(G)$ ($G \in \Oalg_n$) of the maps $\gamma$ (\ref{Intr-eq2-1}) as \REMemph{kernels} of linear partial differential operators on the total diagonal.
Applying $\gamma(G)$, as a differential operator, on a polynomial $F \in \Salg_n$ would then mean to integrate $\gamma(G)$ as a distribution against $F$, i.e.,
\beq\label{eq1.9-20190922}
\gamma(G) [F] \,= \mathop{\int}\limits_{\R^{Dn}} 
\hspace{-4pt}
\gamma(G)(\x_1,\dots,\x_n;\x)
\,.\,F(\x_1,\dots,\x_n) \, d^Dx_1 \cdots d^Dx_n
\,,\quad
\gamma(G)[F] \in \Salg_1
\eeq
($\text{supp} \, \gamma(G)$ $\subseteq$ $\{\x_1=\cdots=\x_n=\x\}$).
The $\Salg_n$--module property of $\gamma$ means that
\beq\label{eq1.10-20190922}
\gamma(F_1.G)[F] \,=\, \gamma(G) [F_1.F]
\eeq
for every $F,F_1 \in \Salg_n$ and $G \in \Oalg_n$
and it is a consequence of the physical requirement that renormalization prescriptions commute with the operators of multiplications by polynomials (or smooth functions, cf. \cite[Sect.~2.4]{N09} and \cite[Sect.~5]{NST14}).
It follows from (\ref{eq1.10-20190922}) that the map
\beq\label{eq1.11-20190922}
\Gamma : \Oalg_n \to \Salg_1 : G \,\mapsto\, \gamma(G)[1]
\eeq
completely characterizes $\gamma$ and it is \anTsedi operator, as introduced in the previous subsection.
In other words, the finite changes of renormalization introduced in \cite[Sect.~2.6]{N09} are nothing but \Tsedi operators.
In the above presentation we make one step further comparing to \cite{N09}: we consider not only translation invariant distributions for the values of $\gamma(G)$ but arbitrary distributions supported at $\x_1$ $=$ $\cdots$ $=$ $\x_n$ with polynomial coefficients in $\x$.
This is convenient in order to make manifest the equivariance of our constructions under the permutations' action on the points.

Then, in \cite[Sect.~2.6]{N09} it was obtained a composition law for the sequences of finite changes in renormalization (cf. \cite[Eq.~(2.48)]{N09}).
In this way, in the framework of \cite{N09,NST14} one introduces the notion of \REMemph{renormalization group}.
Later, in \cite[Theorem~1]{LN12} (cf. also \cite[Proposition~5.4.3.]{LV12}) it was found a general functorial construction that can produce the invented in \cite{N09} renormalization group structure as coming from a symmetric operad.
In \cite{N14} this new operad structure was first described on the level of non--symmetric operads.
Before that this operad structure was described in \cite{LN12} on a combinatorial level.
A first attempt to extend the operad structure to a symmetric operad structure was done in \cite{N16}.
In the present paper we completely solve the problem concerning the permutation equivariance.
The final construction of this operad is contained in Sect.~\ref{s6.2} of the present paper.
Having found an operad coming from renormalization group a natural question appears: what are the algebras over this operad.
In \cite{N14} we also made a connection between these algebras and vertex algebras.

Thus, the idea for \Tsedi operators and the operads that they can form came from renormalization theory.
We are leaving this important relation to the renormalization theory to a separate work.

\SSECSPA

\subsection{Overview of the paper}
\ADDCONT{Overview of the paper}

\noindent
This paper contains two main parts.
The first part, which includes Sects.~\ref{SectDiffOp} -- \ref{S5}, is devoted to the general theory of \Tsedi operators.
The exposition in this part 
will be at a greater level of mathematical generality and abstraction.
In Sect. \ref{SectDiffOp} we remind the basic notions related to differential operators on modules of commutative associative algebras.
Two main types of extensions of such modules are of primary interest for us:
the modules of fractions (localized modules) and formal completions of modules.
To these we have devoted Sects.~\ref{LocMod} and \ref{FormalCompl}, respectively, where we have established that the differential operators posses unique prolongations on these extensions (Theorems~\ref{th3.1} and \ref{ThFS1}, respectively).
In Sect.~\ref{LocMod} we consider also the technique of pull-backs of modules via algebra homomorphism.
The pull-backs play a very important role in our work as the evaluation of a differential operator on the diagonal, which we mentioned in the previous subsection, is modeled exactly by a pull-back (this is contained in our notion of a $\vartheta$--differential operator in Definition~\ref{df2.2}~(b) for an algebra homomorphism $\vartheta$).
The results of Sects.~\ref{SectDiffOp} -- \ref{FormalCompl} are essentially known.
The new theory about \Tsedi operators starts with Sect.~\ref{SectADiffOp}.
There we have established analogous properties for uniqueness of prolongations for the above 
two
types of extensions.
Another central result is that the class of all \Tsedi operators stay closed under the composition.
The latter statement is established for modules of free commutative associative algebras (i.e., polynomial algebras), which are of main interest for the applications of this work.
It seem that a larger generality is possible, but this requires a separate and profound investigation on the theory of \Tsedi operators.
The last Sect.~\ref{S5} of the first part is devoted to the main example of \anTsedi operator: the residue functional.

Part 2 is devoted to the ``\TsediPRODterm operations''.
Mathematically speaking, an ``operation'' is an element of an operad (an $n$-ary operation belongs to the $n$-th operadic space), and so we work within the frame of operad theory.
In Sect.~\ref{s6} we have shown how to construct an operad from spaces of \Tsedi operators.
This construction directly generalizes the operad of ``differential\PRODterm operations'', whose algebras are the differential algebras considered in Sect.~\ref{Se8nn}.
In fact, the operad of differential\PRODterm operations is a quotient of the operad of \TsediPRODterm operations.
Next, Sects.~\ref{Se9nn} and \ref{Se10nn} contain the main application of the present work: considering the vertex algebras as algebras over the operad of \TsediPRODterm operations.
Since the vertex algebras formalize not only the notion of OPE of quantum fields but also they combine this structure with the translation symmetry of the flat Minkowski space, we needed to separate the suboperad of translation invariant \TsediPRODterm operations.
This construction is contained in Sect.~\ref{Se9nn}.
In Sect.~\ref{Se10nn} we establish that every vertex algebra induces an algebra over the operad of (translation invariant) \TsediPRODterm operations.
We leave open the converse problem, i.e., whether all algebras over the operad of \TsediPRODterm operations correspond to vertex algebras and if not under what additional conditions this is true.
Section
\ref{Se13nn} 
contains 
further directions for investigations related to the present work and conclusions.

\SSECSPA

\subsection{Some general notations and conventions}
\ADDCONT{Notations and conventions\vspace{4.4444pt}}

\noindent
$\Z$, $\R$ and $\C$ stands for the sets of integer, real and complex numbers, respectively.
In the first part of the work we also worked over an arbitrary field $\GRF$ of characteristics $0$.
More specific notations we have also introduced at the beginning of some sections.

All the algebras that we will consider from Sect.~\ref{SectDiffOp} to Sect.~\ref{s6} will be, in particular, associative algebras and so, we will omit sometimes ``associative''.

\addcontentsline{toc}{section}{\it Part I. Theory of \Tsedi operators\vspace{4.4444pt}}

\section{Differential operators}\label{SectDiffOp}
\SETCNTR

All differential operators in this work will be \REMemph{linear} differential operators. They will include the ordinary notion of linear partial differential operators from analysis.
We follow \cite[Sect. 2]{HS69}, \cite{Sm86} and we have included here some of their results in the way how they will be used in the paper.

One of the simplest examples of differential operators are the derivations acting on smooth functions on a manifold.
These are first order differential operators and the zeroth order differential operators in this case are the operators of multiplication by smooth functions.
A widely used fact in quantum mechanics is that the commutator of a derivation with the operator of multiplication by a smooth function is again an operator of multiplication by a smooth function, i.e., it is a zeroth order differential operator.
In general, when we commute the operator of multiplication by a smooth function and an arbitrary differential operator the degree of the latter differential operator decreases by one.
This is the key to the general inductive definition of a differential operator, which we give in this section.

\medskip

Let $\Aalg$ be a commutative associative algebra over a ground field $\GRF$ of characteristics $0$.\footnote{%
We shall not assume everywhere in this paper that $\Aalg$ has a unit but in all our applications there will be a unit in $\Aalg$ (i.e., $\Aalg$ will be unital).}
Let $\Mmod$ and $\Nmod$  be $\Aalg$--modules and $\Hom_{\Aalg} (\Mmod,\Nmod)$ stands for the $\Aalg$--module of all $\Aalg$--linear maps, while $\Hom_{\GRF} (\Mmod,\Nmod)$ is just the vector space of all $\GRF$--linear maps.
Thus, $\Hom_{\GRF} (\Mmod,\Nmod)$ is an $(\Aalg$--$\Aalg)$--bimodule:
\beqa\label{Bi-act}
(a \cdot \DOpu )(m) \podr := \, a \cdot \DOpu (m)
\quad \text{(left action)}
\,,\quad
\nnb
(\DOpu \cdot a) (m) \podr := \, \DOpu(a \cdot m)
\quad \text{(right action)}
\eeqa
for $a \in \Aalg$, $m \in \Mmod$, $\DOpu \in \Hom_{\GRF} (\Mmod,\Nmod)$,
where\footnote{%
we shall write almost everywhere a dot to indicate our module actions}
$$
a \cdot m \, \equiv \, am
$$
stands for the module action of $\Aalg$ on $\Mmod$.
We define inductively,
\beqa
\nn \podr
\DfrOp{0}{\Aalg} (\Mmod,\Nmod)
\, := \, \Hom_{\Aalg} (\Mmod,\Nmod)
\quad \text{and for} \quad k \, = \, 1,2,\dots:
\bnn
\podr
\DfrOp{k}{\Aalg} (\Mmod,\Nmod)
\, := \, \bigl\{\DOpu \in \Hom_{\GRF} (\Mmod,\Nmod) \, \bigl| \,
\DAl{\DOpu}{a} \in \DfrOp{k-1}{\Aalg} (\Mmod,\Nmod)
\
\bnn
\podr
\phantom{\DfrOp{k}{\Aalg} (\Mmod,\Nmod)
\, := \, \bigl\{\DOpu \in \Hom_{\GRF} (\Mmod,\Nmod) \, \bigl| \,}
\text{for all } a \in \Aalg\bigr\}
\,,
\bnn
\podr
\text{where} \quad
\DAl{\DOpu}{a} (m)
\, := \,
[a,\DOpu] (m) \, := \,  (a \cdot \DOpu - \DOpu \cdot a) (m)
\bnn
\podr
\phantom{\text{where} \quad
\DAl{\DOpu}{a} (m):}
\equiv \,
a \cdot \DOpu(m) - \DOpu (a \cdot m)
\,.
\eeqa
Note, that one can start the above induction from $k=-1$ by setting
$\DfrOp{-1}{\Aalg} (\Mmod,\Nmod)$ $=$ $0$.
Note also that
\beqa\label{MissEq}
&
\DAl{(\DAl{\DOpu}{a})}{b}
\,=\,
\DAl{(\DAl{\DOpu}{b})}{a}
\,, &
\\ \label{MissEq2}
&
a \cdot (\DAl{\DOpu}{b}) \,=\, \DAl{(a \cdot \DOpu)}{b}
\,,\quad
(\DAl{\DOpu}{a}) \cdot b \,=\, \DAl{(\DOpu \cdot b)}{a}
\,, &
\eeqa
for every $a,b \in \Aalg$ as well as that we have the property:
\beq\label{MissEq1}
\DOpu \in \DfrOp{k}{\Aalg}(\Mmod,\Nmod)
\quad \text{iff} \quad
\DAl{\bigl(\cdots \DAl{(\DAl{\DOpu}{a_1})}{a_2} \cdots \bigr)}{a_{k+1}} \,=\, 0
\,,
\eeq
for all $a_1,\dots,a_{k+1} \in \Aalg$.
It follows that
$$
\DfrOp{0}{\Aalg} (\Mmod,\Nmod) \, \subseteq \,
\DfrOp{1}{\Aalg} (\Mmod,\Nmod) \, \subseteq \, \cdots \,
\, \subseteq \, \DfrOp{k}{\Aalg} (\Mmod,\Nmod) \, \subseteq \, \cdots \,
$$
and each $\DfrOp{k}{\Aalg} (\Mmod,\Nmod)$ is an $(\Aalg$--$\Aalg)$--sub-bimodule of
$\Hom_{\GRF} (\Mmod,\Nmod)$.
We set
$$
\DfrOp{}{\Aalg} (\Mmod,\Nmod) \, = \, \mathop{\bigcup}\limits_{k \, = \, 0}^{\infty}
\DfrOp{k}{\Aalg} (\Mmod,\Nmod)
\,
$$
and the elements of $\DfrOp{}{\Aalg} (\Mmod,\Nmod)$ are generally called \DEFemph{differential operators}.
If $u$ $\in$ $\DfrOp{k}{\Aalg} (\Mmod,\Nmod)$ we say that $u$ is a \DEFemph{differential operator of order not exceeding $k$} and if additionally $u$ $\notin$ $\DfrOp{k-1}{\Aalg} (\Mmod,\Nmod)$ we say that $u$ is a \DEFemph{differential operator of order $k$}.
As in \cite{HS69}, when we speak about $\DfrOp{k}{\Aalg} (\Mmod,\Nmod)$ just as $\Aalg$--module we shall have in mind the left $\Aalg$--module action $a \cdot \DOpu$ (defined above).

\medskip

The spaces of differential operators stay closed with respect to the compositions:

\medskip

\begin{PROPO}{pr2.1}
\textrm{\cite[Lemma 2.1.1 a and c]{HS69}}
$(a)$
{\it Let $\Mmod$, $\Nmod$ and $\Pmod$ be $\Aalg$--modules.
Then the composition map
$$
\Hom_{\GRF} (\Mmod,\Nmod) \times \Hom_{\GRF}(\Nmod, \Pmod) \,\to\, \Hom_{\GRF}(\Mmod, \Pmod) \,:\,
(\DOpu,\DOpv) \,\mapsto\, \DOpv \,\circ\, \DOpu
$$
maps
$\DfrOp{k}{\Aalg} (\Mmod,\Nmod) \times \DfrOp{\ell}{\Aalg} (\Nmod,\Pmod)$
to
$\DfrOp{k+\ell}{\Aalg} (\Mmod,\Pmod)$.\raisebox{-8pt}{}

$(b)$
Let $\Mmod_j$ and $\Nmod_j$ be modules of the commutative algebra $\Aalg_j$ for $j=1,2$.
Then $\Mmod_1 \otimes \Mmod_2$ and $\Nmod_1 \otimes \Nmod_2$ are modules of $\Aalg_1 \otimes \Aalg_2$ and the natural inclusion
\beq\label{isom1a}
\hspace{0pt}
\Hom_{\GRF}(\Mmod_1,\Nmod_1) \otimes \Hom_{\GRF}(\Mmod_2,\Nmod_2) \,\hookrightarrow\,
\Hom_{\GRF}(\Mmod_1\otimes\Mmod_2,\Nmod_1\otimes\Nmod_2)
\eeq
restricts to an inclusion
\beq\label{isom1b}
\hspace{0pt}
\DfrOp{k_1}{\Aalg_1}(\Mmod_1,\Nmod_1) \otimes \DfrOp{k_2}{\Aalg_2}(\Mmod_2,\Nmod_2) \,\hookrightarrow\,
\DfrOp{k_1+k_2}{\Aalg_1\otimes\Aalg_2}(\Mmod_1\otimes\Mmod_2,\Nmod_1\otimes\Nmod_2)\,.
\eeq%
}%
\end{PROPO}%

A main example is the case when the commutative algebra $\Aalg$ is a \REMemph{coordinate ring}.
We shall consider this case within the context of algebraic geometry when the coordinate rings are polynomial algebras.

\medskip

\begin{EXMP}{ex2.1}
Let $\Aalg$ be the \REMemph{polynomial algebra} $\Aalg$ $:=$ $\GRF [t_1,$ $\dots,$ $t_N]$.
Let $\Mmod$ $:=$ $\Aalg$, with respect to the multiplication as an $\Aalg$--module action, and
let $\Nmod$ be an arbitrary $\Aalg$--module.
Then every differential operator $\DOpu$ $\in$ $\DfrOp{k} {\Aalg}(\Aalg,\Nmod)$  of order $k$ has the form
\beq\label{repU}
u(f) \,=\,
f \cdot U^{(0)} +
\mathop{\sum}\limits_{n \,=\, 1}^{k} \
\mathop{\sum}\limits_{\mu_1,\dots,\mu_n \,=\, 1}^N
\frac{\di^n f}{\di t_{\mu_1}\cdots \,\di t_{\mu_n}} \cdot U^{(n)}_{\mu_1,\dots,\mu_n}
\eeq
($f \in \Aalg$)
for unique coefficients $U^{(0)}$ and $U^{(n)}_{\mu_1,\dots,\mu_n}$ $\in$ $\Nmod$.
Note also, that Eq.~(\ref{repU}) always defines a differential operator of order not exceeding $k$.
These statements are proven by induction in $k$.
\end{EXMP}

\medskip

\begin{REMAR}{rm2.1}
In a more general situation of a commutative algebra $\Aalg$ than in Example \ref{ex2.1} one can construct a \REMemph{filtered} $\Aalg$--module $\Jet{\Blt}{\Aalg}$ $=$ $\mathop{\cup}\limits_{k \,=\, 0}^k$
$\Jet{k}{\Aalg}$ together with an element
$j_k$ $\in$ $\DfrOp{k}{\Aalg}(\Aalg,\Jet{k}{\Aalg})$ for every $k=0,1,\dots$
such that every $\DOpu \in \DfrOp{k}{
\Aalg}(\Aalg,\Nmod)$ is represented as a composition $\DOpu$ $=$ $\widetilde{\DOpu} \circ j_k$ for a unique $\Aalg$--linear map $\widetilde{\DOpu} \in \Hom_{\Aalg} (\Jet{k}{\Aalg},\Nmod)$ (cf. \cite[Theorem 2.2.6]{HS69}).
In the case of Example~\ref{ex2.1},
$\Jet{k}{\Aalg}$ is a free $\Aalg$--module with a basis $P^{(0)}$, $P^{(n)}_{\mu_1,\dots,\mu_n}$ such that
\beq\label{repJ}
j_k(f) \,=\,
f \cdot P^{(0)} +
\mathop{\sum}\limits_{n \,=\, 1}^{k} \
\mathop{\sum}\limits_{\mu_1,\dots,\mu_n \,=\, 1}^N
\frac{\di^n f}{\di t_{\mu_1}\cdots \,\di t_{\mu_n}} \cdot P^{(n)}_{\mu_1,\dots,\mu_n}
\eeq
and then we obtain the coefficients in Eq.~(\ref{repU}) just as
$U^{(0)}$ $=$ $\widetilde{\DOpu} (P^{(0)})$,
$U^{(n)}_{\mu_1,\dots,\mu_n}$ $=$ $\widetilde{\DOpu} (P^{(n)}_{\mu_1,\dots,\mu_n})$.
\end{REMAR}

\medskip

\begin{REMAR}{rm2.2}
A useful property of the assignment $\DOpu$ $\mapsto$ $\DAl{\DOpu}{a}$
for $\DOpu$ $\in$ $\DfrOp{}{\Aalg}(\Mmod,\Nmod)$ and $a \in \Aalg$
is the ``Leibniz property'':
\beq\label{Leib1}
\DAl{\DOpu}{ab} \,=\,
a \cdot (\DAl{\DOpu}{b})
+ (\DAl{\DOpu}{a}) \cdot b
\eeq
for $a,b \in \Aalg$.
\end{REMAR}

\section{Localizations and pullbacks}\label{LocMod}
\SETCNTR

When we focus our considerations to a domain outside the zeros of an element $\FrE \in \Aalg$ of the coordinate ring $\Aalg$ of our manifold we extend our modules by allowing  a division by $\FrE$.
In other words, we \REMemph{consider modules of fractions}.

In more details, let $\Aalg$ be  a commutative algebra and $\FrE \in \Aalg$. Let $\Mmod$ be an $\Aalg$--module.
We shall consider only the special case when $\FrE$ is \REMemph{not a zero divisor on $\Mmod$} (i.e., when $\FrE$ acts on $\Mmod$).
The latter means that
\beq\label{Mregel}
\FrE \cdot m \, = \, 0 \quad \Longrightarrow\quad m \, = \, 0
\eeq
for every $m \in \Mmod$
and in the field of commutative algebra one says in the situation of Eq.~(\ref{Mregel}) that $\FrE$ is an \DEFemph{$\Mmod$--regular} element of $\Aalg$.
Then, one sets
\beqs
\Aalg [1/\FrE]
\podr := \,
\Bigl\{\frac{a}{\FrE^n}\,\Bigr|\, a \,\in\, \Aalg,\, n \,=\, 1,2,\dots\Bigr\}
\,,
\nnb
\Mmod [1/\FrE]
\podr := \,
\Bigl\{\frac{m}{\FrE^n}\,\Bigr|\, m \,\in\, \Mmod,\, n \,=\, 1,2,\dots\Bigr\}
\,
\eeqs
with the identifications
$\TXfrac{a_1}{\FrE^{n_1}}$ $=$ $\TXfrac{a_2}{\FrE^{n_2}}$
and
$\TXfrac{m_1}{\FrE^{n_1}}$ $=$ $\TXfrac{m_2}{\FrE^{n_2}}$
if $\FrE^{n_2} a_1$ $=$ $\FrE^{n_1} a_2$
and
$\FrE^{n_2} m_1$ $=$ $\FrE^{n_1} m_2$,
respectively.
Under the operations
$\TXUfrac{a_1}{\FrE^{n_1}}$ $+$ $\TXUfrac{a_2}{\FrE^{n_2}}$
$=$
$\TXUfrac{\FrE^{n_2} a_1+\FrE^{n_1} a_2}{\FrE^{n_1+n_2}}$,
$\TXUfrac{m_1}{\FrE^{n_1}}$ $+$ $\TXUfrac{m_2}{\FrE^{n_2}}$
$=$
$\TXUfrac{\FrE^{n_2} m_1+\FrE^{n_1} m_2}{\FrE^{n_1+n_2}}$,
$\TXUfrac{a_1}{\FrE^{n_1}}$ $\TXUfrac{a_2}{\FrE^{n_2}}$
$=$
$\TXUfrac{a_1a_2}{\FrE^{n_1+n_2}}$
and
$\TXUfrac{a_1}{\FrE^{n_1}}$ $\cdot$ $\TXUfrac{m_2}{\FrE^{n_2}}$
$=$
$\TXDfrac{a_1 \cdot m_2}{\FrE^{n_1+n_2}}$
for $a_1,a_2 \in \Aalg$, $m_1,m_2 \in \Mmod$ and $n_1,n_2=1,2,\dots$
the sets
$\Aalg [1/\FrE]$ and $\Mmod [1/\FrE]$ become a commutative (associative) algebra and its module, respectively.
As $\Aalg$ is a subalgebra of $\Aalg [1/\FrE]$ under the natural inclusion
$a$ $\mapsto$ $\TXXfrac{ra}{r}$
we shall consider $\Mmod[1/r]$ as an $\Aalg$--module, which has in turn, $\Mmod$ as an $\Aalg$--submodule under a similar inclusion.

\medskip

\begin{THEOR}{th3.1}
{\it
$(a)$
Let $\Aalg$ be a commutative algebra, $\Mmod$ and $\Nmod$ be $\Aalg$--modules and $\FrE \in \Aalg$ be 
an $\Mmod$-- and $\Nmod$--regular
$($i.e., $\FrE$ is not a zero divisor for both $\Aalg$--modules, $\Mmod$ and $\Nmod$$)$.
Then every $\DOpu \in \DfrOp{}{\Aalg}(\Mmod,\Nmod)$ posses a unique extension
$\EXTOP{\DOpu} \in \DfrOp{}{\Aalg}(\Mmod[1/\FrE],\Nmod[1/\FrE])$
in the sense that
$\EXTOP{\DOpu}(m)$ $=$ $\DOpu(m)$ for every $m \in \Mmod$.
The differential operator $\EXTOP{\DOpu}$ has the same order as $\DOpu$.
One has also $\DAl{(\EXTOP{\DOpu})}{a}$ $=$ $\EXTOP{(\DAl{\DOpu}{a})}$ for every $a \in \Aalg$.

$(b)$
The map $\DOpu \mapsto \EXTOP{\DOpu}$ is an 
injection of
$(\Aalg$--$\Aalg)$--modules$:$
$\DfrOp{}{\Aalg}(\Mmod,$ $\Nmod)$ $\hookrightarrow$ $\DfrOp{}{\Aalg}(\Mmod[1/\FrE],$ $\Nmod[1/\FrE])$.
In other words, one has the identities
\beq\label{AAmod1}
\EXTOP{(a \cdot \DOpu)} \,=\, a \cdot \EXTOP{u}
\,,\quad
\EXTOP{(\DOpu \cdot a)} \,=\, \EXTOP{\DOpu} \cdot a
\,,
\eeq
for every $a \in \Aalg$.}
\end{THEOR}

\medskip

We \textit{prove} both statements of
Theorem~\ref{th3.1} in a common induction of the order $k$ of the differential operator $\DOpu$.
The base of the induction is at $k=-1$ when $\DOpu = 0$ and the theorem is trivially satisfied.

Assume by induction that both statements $(a)$ and $(b)$ are proven for all differential operators of order not exceeding $k-1=-1,0,1,\dots$.
Let $\DOpu$ be a differential operator of order $k$.
Noting that if we construct $\EXTOP{\DOpu}$ then it should satisfy the identities
$$
\EXTOP{(\DAl{\DOpu}{\FrE^n})}
\Bigl(\frac{m}{\FrE^n}\Bigr)
\,=\,
\DAl{(\EXTOP{\DOpu})}{\FrE^n}
\Bigl(\frac{m}{\FrE^n}\Bigr)
\,=\,
\FrE^n \cdot \EXTOP{\DOpu}\Bigl(\frac{m}{\FrE^n}\Bigr)
-
\EXTOP{\DOpu}\Bigl(\FrE^n \cdot \frac{m}{\FrE^n}\Bigr)
$$
we set
\beq\label{Ext-def}
\EXTOP{\DOpu}\Bigl(\frac{m}{\FrE^n}\Bigr)
\,:=\,
\frac{1}{\FrE^n} \cdot
\Bigl(
\EXTOP{(\DAl{\DOpu}{\FrE^n})}
\Bigl(\frac{m}{\FrE^n}\Bigr)
+
\DOpu(m)
\Bigr)
\,.
\eeq
The latter definition is correct:
if $n_1>n_2$ are positive integers and
$\TXDfrac{m_1}{\FrE^{n_1}}$ $=$ $\TXDfrac{m_2}{\FrE^{n_2}}$ for $m_1,$ $m_2$ $\in$ $\Mmod$
then $m_1=\FrE^{n_{12}} \cdot m_2$ for $n_{12} := n_1-n_2$ and hence
\def\SPECVSPACE{\vspace{-17.7pt}}
\beqs
\podr
\EXTOP{\DOpu}\Bigl(\frac{m_1}{\FrE^{n_1}}\Bigr)
\,=\,
\frac{1}{\FrE^{n_1}} \cdot
\Bigl(
\EXTOP{(\DAl{\DOpu}{\FrE^{n_{12}}\FrE^{n_2}})}
\Bigl(\frac{m_2}{\FrE^{n_2}}\Bigr)
+
\DOpu(\FrE^{n_{12}} \cdot m_2)
\Bigr)
\hspace{55pt}
\eeqs\SPECVSPACE\beqs
\podr=\,
\frac{1}{\FrE^{n_1}} \cdot
\Bigl(
\EXTOP{\bigl(
\FrE^{n_{12}} \cdot (\DAl{\DOpu}{\FrE^{n_2}})
+
(\DAl{\DOpu}{\FrE^{n_{12}}}) \cdot \FrE^{n_2}
\bigr)}
\Bigl(\frac{m_2}{\FrE^{n_2}}\Bigr)
+
\DOpu(\FrE^{n_{12}} \cdot m_2)
\Bigr)
\hspace{17pt}
\eeqs\SPECVSPACE\beqs
\podr=\,
\frac{1}{\FrE^{n_1}} \cdot
\Bigl(
\Bigl(\FrE^{n_{12}} \cdot
\EXTOP{
(\DAl{\DOpu}{\FrE^{n_2}})}
+
\EXTOP{(\DAl{\DOpu}{\FrE^{n_{12}}})} \cdot \FrE^{n_2}
\Bigr)
\Bigl(\frac{m_2}{\FrE^{n_2}}\Bigr)
+
\DOpu(\FrE^{n_{12}} \cdot m_2)
\Bigr)
\eeqs\SPECVSPACE\beqs
\podr=\,
\frac{1}{\FrE^{n_1}} \cdot
\Bigl(
\FrE^{n_{12}} \cdot
\EXTOP{
(\DAl{\DOpu}{\FrE^{n_2}})}
\Bigl(\frac{m_2}{\FrE^{n_2}}\Bigr)
+
\EXTOP{(\DAl{\DOpu}{\FrE^{n_{12}}})}
 (m_2)
+
\DOpu(\FrE^{n_{12}} \cdot m_2)
\Bigr)
\hspace{16pt}
\eeqs\SPECVSPACE\beqs
\podr=\,
\frac{1}{\FrE^{n_1}} \cdot
\Bigl(
\FrE^{n_{12}} \cdot
\EXTOP{
(\DAl{\DOpu}{\FrE^{n_2}})}
\Bigl(\frac{m_2}{\FrE^{n_2}}\Bigr)
+
\DAl{\DOpu}{\FrE^{n_{12}}} (m_2)
+
\DOpu(\FrE^{n_{12}} \cdot m_2)
\Bigr)
\hspace{36pt}
\eeqs\SPECVSPACE\beqs
\podr=\,
\EXTOP{\DOpu}\Bigl(\frac{m_2}{\FrE^{n_2}}\Bigr)
\,,
\hspace{249pt}
\eeqs
where
in the second equality we used Eq.~(\ref{Leib1}), and in the third and fifth equalities we used the inductive assumption applied to $\DAl{u}{\FrE^{n_{12}}}$ and $\DAl{u}{\FrE^{n_{2}}}$.
The linearity of the constructed map $\EXTOP{\DOpu}: \Mmod \to \Nmod$ follows by the linearity in $m$ of the right hand side in Eq.~(\ref{Ext-def}) (as we can pass to a common denominator).

Next, we check that Eq.~(\ref{Ext-def}) is indeed an extension of $\DOpu$:
\beqs
\EXTOP{\DOpu}\Bigl(\frac{\FrE m}{\FrE}\Bigr)
\podr =\,
\frac{1}{\FrE} \cdot
\Bigl(
\EXTOP{(\DAl{\DOpu}{r})}
\Bigl(\frac{\FrE m}{\FrE}\Bigr)
+
\DOpu(\FrE m)
\Bigr)
\nnb
\podr =\,
\frac{1}{\FrE} \cdot
\Bigl(
\DAl{\DOpu}{r}(m)
+
\DOpu(\FrE m)
\Bigr)
\,=\,
\frac{1}{\FrE} \cdot (r \cdot \DOpu(m))
\,=\, \DOpu(m)
\,.
\eeqs

Then, we check the property
$\DAl{(\EXTOP{\DOpu})}{a}$ $=$ $\EXTOP{(\DAl{\DOpu}{a})}$:
\beqs
\DAl{(\EXTOP{\DOpu})}{a}\Bigl(\frac{m}{\FrE^n}\Bigr)
\podr =\,
\frac{1}{\FrE^n} \cdot
\Bigl(
\DAl{\Bigl(\EXTOP{(\DAl{\DOpu}{\FrE^n})}\Bigr)}{a}
\Bigl(\frac{m}{\FrE^n}\Bigr)
+
\DAl{\DOpu}{a}(m)
\Bigr)
\eeqs\SPECVSPACE\beqs
\phantom{\DAl{(\EXTOP{\DOpu})}{a}\Bigl(\frac{m}{\FrE^n}\Bigr)}
\podr =\,
\frac{1}{\FrE^n} \cdot
\Bigl(
\EXTOP{\Bigl(\DAl{(\DAl{\DOpu}{\FrE^n})}{a}\Bigr)}
\Bigl(\frac{m}{\FrE^n}\Bigr)
+
\DAl{\DOpu}{a}(m)
\Bigr)
\eeqs\SPECVSPACE\beqs
\phantom{\DAl{(\EXTOP{\DOpu})}{a}\Bigl(\frac{m}{\FrE^n}\Bigr)}
\podr =\,
\frac{1}{\FrE^n} \cdot
\Bigl(
\EXTOP{\Bigl(\DAl{(\DAl{\DOpu}{a})}{\FrE^n}\Bigr)}
\Bigl(\frac{m}{\FrE^n}\Bigr)
+
\DAl{\DOpu}{a}(m)
\Bigr)
\eeqs\SPECVSPACE\beqs
\phantom{\DAl{(\EXTOP{\DOpu})}{a}\Bigl(\frac{m}{\FrE^n}\Bigr)}
\podr=\,
\EXTOP{(\DAl{\DOpu}{a})}\Bigl(\frac{m}{\FrE^n}\Bigr)
\,,
\hspace{105pt}
\eeqs
where the second equality is due to the induction and we also used Eq.~(\ref{MissEq}).

Now, the  established identity
$\DAl{(\EXTOP{\DOpu})}{a}$ $=$ $\EXTOP{(\DAl{\DOpu}{a})}$
(for every $a \in \Aalg$)
implies (by the induction) that the constructed $\EXTOP{\DOpu}$
is a differential operator of the same order as $\DOpu$.

For the uniqueness of $\EXTOP{\DOpu}$ it is enough to prove that
\\
{\it
if
$\DOpv$ $\in$ $\DfrOp{k_{\DOpv}}{\Aalg}(\Mmod[1/\FrE],\Nmod[1/\FrE])$ is such that
$\DOpv (m)$ $=$ $0$ for every $m \in \Mmod$ then $\DOpv = 0$.}
To this end, one establishes by induction in $k_{\DOpv}=1,2,\dots$ that
$\DOpv \Bigl(\TXXfrac{m}{\FrE^n}\Bigr)$ $=$ $0$.
Indeed,
$$
\FrE^n \cdot \DOpv \Bigl(\frac{m}{\FrE^n}\Bigr)
\,=\,
\DAl{\DOpv}{\FrE^n} \Bigl(\frac{m}{\FrE^n}\Bigr)
+
\DOpv (m)
\,=\, 0.
$$

It remains to prove Eq.~(\ref{AAmod1}).
It follows by the uniqueness of the extensions as it is trivially satisfied on $\Mmod \subset \Mmod[1/\FrE]$.

This completes the proof of the theorem.$\quad\Box$

\medskip

In what follows we shall use also another version of the argument for the uniqueness in the above proof.

\medskip

\begin{LMA}{Lm4.XX}
{\it
Let $\Aalg$ be a commutative algebra, $\Nmod$ be an $\Aalg$--module and $\FrE \in \Aalg$ be not a zero divisor on $\Nmod$ $($i.e., $\FrE$ is $\Nmod$--regular$)$.
Then if $\DOpv$ $\in$ $\DfrOp{}{\Aalg}(\Mmod,\Nmod)$ is such that there exists $n \in \{1,2,\dots\}$ for which $\DOpv(r^n \cdot m)$ $=$ $0$ for all $m \in \Mmod$ then $\DOpv =0$.}
\end{LMA}

\medskip

The {\it proof} is by induction in the order $k_{\DOpv}$ of $\DOpv$ as:
$$
\FrE^n \cdot \DOpv (m)
\,=\,
\DAl{\DOpv}{\FrE^n} (m)
+
\DOpv (\FrE^n \cdot m)
\,=\, 0. \quad\Box
$$

\medskip

\begin{REMAR}{uheid}
Under the conditions of Theorem~\ref{th3.1}
it follows that
$$
\EXTOP{\DOpu} \,\in\, \DfrOp{}{\Aalg[1/\FrE]}(\Mmod[1/\FrE],\Nmod[1/\FrE])
\,.
$$
However, we shall not need localization for the algebras in our applications but only for their modules and for this reason we do not extend the algebras in our extensions of differential operators.
\end{REMAR}

\medskip

\begin{DEFIN}{df2.2}
$(a)$
Let $\Aalg$ and $\Balg$ be commutative algebras and $\vartheta$ $:$ $\Aalg$ $\to$ $\Balg$ be an algebra homomorphism.
Let $\Nmod$ be a $\Balg$--module.
Then $\Nmod$ becomes also an $\Aalg$--module under the action
\beq\label{amodactpb}
(a,m) \,\mapsto \, \vartheta(a) \cdot m
\eeq
for $a \in \Aalg$ and $m \in \Nmod$.
This $\Aalg$--module is denoted by $\vartheta^*\Nmod$ and it is called a \DEFemph{pullback} of the $\Balg$--module $\Nmod$ under $\vartheta$.
(Note that as vector spaces $\Nmod$ and $\vartheta^*\Nmod$ are identical.)

$(b)$
For an $\Aalg$--module $\Mmod$, $\Balg$--module $\Nmod$ and an algebra homomorphism $\vartheta : \Aalg \to \Balg$
let us set
\beq\label{theta-diff}
\hspace{3pt}
\DfrOp{k}{\,\vartheta\,} (\Mmod,\Nmod)
\,:=\,
\DfrOp{k}{\Aalg} (\Mmod, \vartheta^* \, \Nmod)
\,,\ \
\DfrOp{}{\,\vartheta\,} (\Mmod,\Nmod)
\,:=\,
\DfrOp{}{\Aalg} (\Mmod, \vartheta^* \, \Nmod)
\,,
\eeq
as \REMemph{right} $\Aalg$--modules (cf.~(\ref{Bi-act})) and
$\DfrOp{k}{\,\vartheta\,} (\Mmod,\Nmod)$
(as well as
$\DfrOp{}{\,\vartheta\,} (\Mmod,\Nmod)$)
is additionally a left $\Balg$--module
(i.e., it is a $(\Balg$--$\Aalg)$--bimodule), where the left $\Balg$--module of 
$\DfrOp{k}{\,\vartheta\,} (\Mmod,\Nmod)$
is pulled back via $\vartheta$
to a left $\Aalg$--module in
$\DfrOp{k}{\Aalg} (\Mmod, \vartheta^* \, \Nmod)$.
We shall call the elements of 
$\DfrOp{}{\,\vartheta\,} (\Mmod,\Nmod)$
also
\DEFemph{$\vartheta$--differential operators}.

$(c)$ An \DEFemph{augmentation} of a unital associative algebra $\Aalg$ is a homomorphism of unital algebras $\AUGM : \Aalg \to \GRF$ (these are also called \DEFemph{characters} of $\Aalg$).
Then every vector space $V$ becomes an $\Aalg$--module via the pull-back $\AUGM^* V$.
These $\Aalg$--modules are called \DEFemph{trivial}.
Hence, every time in what follows when we speak about a trivial action of an algebra on its module we assume that this is with respect to some augmentation of the algebra.
\end{DEFIN}

\medskip

Let us mention some other immediate properties of the pullbacks:
\beq\label{pullbkq1}
\vartheta^*\,\psi^*\,\Mmod
\,=\,
(\psi \circ \vartheta)^*\,\Mmod
\eeq
for algebra homomorphisms 
$\Aalg$ $\mathop{\to}\limits^{\vartheta}$ $\Balg$ $\mathop{\to}\limits^{\psi}$ $\Calg$
of commutative algebras and a $\Calg$--module~$\Mmod$.
Also,
\beq\label{pullbkq2}
(\vartheta_1 \otimes \cdots \otimes \vartheta_n)^* 
(\Mmod_1 \otimes \cdots \otimes \Mmod_n)
\, = \,
\vartheta_1^*\,\Mmod_1 \otimes \cdots \otimes \vartheta_1^*\,\Mmod_n
\,,
\eeq
where $\vartheta_j : \Aalg_j \to \Balg_j$ is a homomorphism of commutative algebras and $\Mmod_j$, and $\Nmod_j$ are $\Aalg_j$, and $\Balg_j$--modules, respectively, for $j=1,\dots,n$.

One has also an immediate generalization of Proposition~\ref{pr2.1} and Theorem~\ref{th3.1}.

\medskip

\begin{PROPO}{1-pr2.1}
$(a)$
{\it Let $\Mmod$, $\Nmod$ and $\Pmod$ be modules over the commutative algebras 
$\Aalg$, $\Balg$ and $\Calg$, respectively, and we have a sequence of algebra homomorphisms
$\Aalg$ $\mathop{\to}\limits^{\vartheta}$ $\Balg$ $\mathop{\to}\limits^{\psi}$ $\Calg$.
Then the composition map
$$
\Hom_{\GRF} (\Mmod,\Nmod) \times \Hom_{\GRF}(\Nmod, \Pmod) \,\to\, \Hom_{\GRF}(\Mmod, \Pmod) \,:\,
(\DOpu,\DOpv) \,\mapsto\, \DOpv \,\circ\, \DOpu
$$
maps
$\DfrOp{k}{\,\vartheta\,} (\Mmod,\Nmod) \times \DfrOp{\ell}{\,\psi\,} (\Nmod,\Pmod)$
to
$\DfrOp{k+\ell}{\,\psi \,\circ\, \vartheta\,} (\Mmod,\Pmod)$.

$(b)$
Let $\Mmod_j$ and $\Nmod_j$ be modules of the commutative algebras $\Aalg_j$ and $\Balg_j$, respectively, for $j=1,2$.
Let us have algebra homomorphisms
$\vartheta_j : \Aalg_j \to \Balg_j$ $(j=1,2)$.
Then $\Mmod_1 \otimes \Mmod_2$ and $\Nmod_1 \otimes \Nmod_2$ are modules of $\Aalg_1 \otimes \Aalg_2$ and $\Balg_1 \otimes \Balg_2$, respectively, and the natural inclusion
$(\ref{isom1a})$
restricts to an inclusion
\beq\label{isom1b-2-1103}
\hspace{0pt}
\DfrOp{k_1}{\,\vartheta_1}(\Mmod_1,\Nmod_1) \,\otimes\, \DfrOp{k_2}{\,\vartheta_2}(\Mmod_2,\Nmod_2) \hookrightarrow
\DfrOp{k_1+k_2}{\,\vartheta_1\otimes\,\vartheta_2}(\Mmod_1\otimes\Mmod_2,\Nmod_1\otimes\Nmod_2)\,.
\eeq%

$(c)$
Let $\Mmod$ and $\Nmod$ be modules of the commutative algebras $\Aalg$ and $\Balg$, respectively, 
let
$\vartheta : \Aalg \to \Balg$ be an algebra homomorphism,
and let $\FrE \in \Aalg$, and $\vartheta(\FrE) \in \Balg$ be 
$\Mmod$-- and $\Nmod$--\THEemph{regular},\footnote{%
in partucular, $\vartheta(\FrE) \neq 0$}
respectively.
Then every $\DOpu \in \DfrOp{}{\,\vartheta}(\Mmod,\Nmod)$ posses a unique extension
$\EXTOP{\DOpu} \in \DfrOp{}{\,\vartheta}(\Mmod[1/\FrE],\Nmod[1/\vartheta(\FrE)])$
in the sense that
$\EXTOP{\DOpu}(m)$ $=$ $\DOpu(m)$ for every $m \in \Mmod$.}%
\end{PROPO}%

\section{Formal completions}\label{FormalCompl}
\SETCNTR

In this section we shall consider \REMemph{linear topologies} on commutative algebras and their modules, and associated (formal) \REMemph{completions}.
A typical example for such formal completions are the spaces of \REMemph{formal powers series}.
For more details we refer to
\cite[Chapt.~10]{AM69},
\cite{M80}[Chapt. 9],
\cite{SZ60}[Chapt. VIII].

Let $\Aalg$ be a commutative algebra and $\Jalg$ be a an ideal in $\Aalg$.
Let us consider the sequence
\beqa\label{Jseq}
&
\Jalg \,=\, \Jalgn{1} 
\,\supseteq\, 
\Jalgn{2} 
\,\supseteq\,
\,\cdots\,
\,\supseteq\,
\Jalgn{n} 
\,\supseteq\,
\,\cdots\,
\,,
&
\bnn
&
\text{where} \quad
\Jalgn{n} \, := \, \underbrace{\Jalg .\ \Jalg .\ \cdots .\ \Jalg}_{\text{$n$ times}}
\,,
&
\eeqa
as sequence of \REMemph{neighbourhoods} of zero in $\Aalg$.
This makes $\Aalg$ a topological vector space over the field $\GRF$ considered with the \REMemph{discrete} topology.
The resulting topology on $\Aalg$ is known as the \DEFemph{Krull topology}, or also as \DEFemph{$\Jalg$-adic topology}, or \DEFemph{linear topology}.
The algebra $\Aalg$ is called a 
\DEFemph{linearly topologized algebra}.
The \DEFemph{completion} with respect to the Krull topology is the \REMemph{inverse} limit
$$
\COMPLET{\Aalg}
\,=\,
\mathop{\lim}\limits_{\longleftarrow} \,
\Aalg
\hspace{1pt}\bigl/\hspace{1pt}
\Jalgn{n}
\,.
$$
It is again a commutative algebra.
Under this completion we add to $\Aalg$ all \DEFemph{Cauchy's sequences}, i.e., sequences $(a_j)_{j \, = \, 1}^{\infty}$ $\subseteq$ $\Aalg$ such that for every $n = 1,2,,\dots$ there exists $n'=1,2,\dots$ so that
$a_j-a_{j'}$ $\in$ $\Jalgn{n}$ for $j,j'$ $>$ $n'$.
When
$$
\mathop{\bigcap}\limits_{n \,=\, 1}^{\infty} \Jalgn{n} \,=\, 0
$$
the linear topology on $\Aalg$ is \REMemph{Hausdorff} and then the natural map
$$
\Aalg \,\to\, \COMPLET{\Aalg}
$$
is an inclusion.

Sometimes, the above topological notions are also called \DEFemph{formal} (i.e., we speak about \DEFemph{formal topology}, \DEFemph{formal completion} etc.) in order to distinguish these notions from the similar ones in Functional analysis (where the ground field $\GRF$ $=$ $\C$ or $\R$ is considered not with the discrete topology).
However, from topological point of view these are completely legitime (not ``formal'') notions.

Similarly, if $\Mmod$ is an $\Aalg$--module
and we have a sequence of $\Aalg$--submodules (a \REMemph{decreasing filtration}):
$$
\Mmod 
\,\supseteq\, 
\Mmod_1 
\,\supseteq\, 
\Mmod_2
\,\supseteq\,
\,\cdots\,
\,\supseteq\,
\Mmod_n 
\,\supseteq\,
\,\cdots\,
\,,
$$
then we consider it as a sequence of neighbourhoods of zero in $\Mmod$ and get a \REMemph{linear topology} on $\Mmod$.
Again, the completion of $\Mmod$ with respect to the above linear topology is the inverse limit
$$
\COMPLET{\Mmod}
\,=\,
\mathop{\lim}\limits_{\longleftarrow} \,
\Mmod
\hspace{1pt}\bigl/\hspace{1pt}
\Mmod_n
$$
and the natural map
$\Mmod \to \COMPLET{\Mmod}$
is an inclusion when
$\mathop{\cap}\limits_{n \,=\, 1}^{\infty} \Mmod_n = 0$.

In the case when
\beq\label{examp111}
\Mmod_n \,=\, \Jalgn{n} \cdot \Mmod
\eeq
($n=1,2,\dots$)
the $\Aalg$--module action $\Aalg \times \Mmod$ $\to$ $\Mmod$ is (formally) \REMemph{continuous}.
It follows then that $\COMPLET{\Mmod}$ is an $\COMPLET{\Aalg}$--module.	

\medskip

In out applications $\Jalg$ will be a \REMemph{maximal} ideal in $\Aalg$.
Thus, $\Jalg$ represents a ``point of the space corresponding to $\Aalg$'', or, more precisely, it is a point in the so called \REMemph{maximal spectrum} of $\Aalg$.
The homomorphism 
$\Aalg$ $\to$ $\Aalg/\Jalg$ $=$ $\GRF$
can be called an ``evaluation'' at the point corresponding to $\Jalg$.

\medskip

\begin{EXMP}{ExamFPS}
Let $\Aalg$ $=$ $\GRF[t_1,\dots,t_N]$ as in Example~\ref{ex2.1} and 
$$
\Jalg \,:=\, t_1 \Jalg + \cdots + t_N \Jalg
\,=\, \{f \,\in\, \Aalg \,|\, f(0) \,=\, 0 \}
\,.
$$
Then, 
$$
\COMPLET{\Aalg} \,=\,
\GRF[[t_1,\dots,t_N]]
\,
$$
is the space of formal power series.
A sequence $(c_j)_{j \,=\, 1}^{\infty}$ $\subseteq$ $\COMPLET{\Aalg}$ is \REMemph{convergent} in the $\Jalg$--adic topology iff every sequence of coefficients with respect to a fixed monomial in $t_1,$ $\dots,$ $t_N$ has only finite number of different members.
Also, let $V$ be a vector space and 
$$
\Mmod \,:=\, \Aalg \otimes V \,\equiv\,
\GRF[t_1,\dots,t_N] \otimes V \,=\,
V[t_1,\dots,t_N]
$$
be a \REMemph{free} $\Aalg$--module, which also is the space of polynomials with coefficients in $V$.
Assume that $\Mmod$ is equipped with linear topology provided by the filtration $(\Mmod_n)_{n \,=\, 1}^{\infty}$ in Eq.~(\ref{examp111}).
Then
$$
\COMPLET{\Mmod} \,=\, V[[t_1,\dots,t_N]]
$$
is the space of formal power series in $t_1,$ $\dots,$ $t_N$ with coefficients in $V$.
\end{EXMP}

\medskip

\begin{EXMP}{192710-ex1}
Let $\Aalg$ be a unital commutative algebra with augmentation 
$\AUGM: \Aalg \to \GRF$.
Then $J = \text{ker} \, \AUGM$ is called \DEFemph{augmentation ideal} and can be used for introducing linear topology on 
$\Aalg$.
The ideal of Example~\ref{ExamFPS} is an augmentation ideal with respect to the augmentation
$\AUGM : t_j \mapsto 0$.
\end{EXMP}

\medskip

\begin{EXMP}{192710-ex2}
If $\Aalg$, $\AUGM: \Aalg \to \GRF$ and $\Aalg'$, $\AUGM': \Aalg' \to \GRF$
are two unital commutative associative algebras with augmentations and 
$\vartheta: \Aalg \to\Aalg'$ is an algebra homomorphism that preserves the augmentations, i.e.,
$\AUGM \circ \vartheta$ $=$ $\AUGM'$, then $\vartheta$ is formally continuous
with respect to the linear topologies induced by the corresponding augmentation ideals.
\end{EXMP}

\medskip

\begin{THEOR}{ThFS1}
{\it
$(a)$
Let $\DOpu$ $\in$ $\DfrOp{k}{\Aalg}(\Mmod,\Nmod)$, where $\Mmod$ and $\Nmod$ are $\Aalg$--modules with linear topologies provided by the decreasing filtrations $(\Mmod_n$ $=$ $\Jalgn{n} \cdot \Mmod)_{n \, = \, 1}^{\infty}$ and 
$(\Nmod_n$ $=$ $\Jalgn{n} \cdot \Nmod)_{n \, = \, 1}^{\infty}$ of $\Aalg$--submodules, respectively.
Then $\DOpu$ is \THEemph{continuous} in the sense that for every $n=1,2,\dots$ there exists $n'=1,2,\dots$ such that $\DOpu \bigl(\Mmod_{n'}\bigr)$ $\subseteq$ $\Nmod_n$.
Hence, $\DOpu$ extends to a \THEemph{continuous} $\GRF$--linear map $\COMPLET{\DOpu}$ $:$ $\COMPLET{\Mmod}$ $\to$ $\COMPLET{\Nmod}$.

$(b)$
In addition, it follows that
$\COMPLET{\DOpu}$ $\in$ $\DfrOp{}{\Aalg}(\COMPLET{\Mmod},\COMPLET{\Nmod})$.
The order of the differential operator $\COMPLET{\DOpu}$ is the same as those of $\DOpu$.

$(c)$
The differential operator 
$\COMPLET{\DOpu}$ $\in$ $\DfrOp{}{\Aalg}(\COMPLET{\Mmod},\COMPLET{\Nmod})$
is the unique extension of $\DOpu$ as a differential operator.

$(d)$
One has $\COMPLET{\DOpu}$ $\in$ $\DfrOp{}{\,\COMPLET{\Aalg}\,}(\COMPLET{\Mmod},\COMPLET{\Nmod})$, i.e., $\COMPLET{\DOpu}$ is a differential operator also for $\COMPLET{\Aalg}$--modules.}
\end{THEOR}

\medskip

\begin{LMA}{lmvgvuj-11}
{\it
Under the conditions of Theorem~\ref{ThFS1}, if the differential operator $\DOpu$ has an order not exceeding $k$, then
\beq\label{TemEqu-1}
\DOpu\bigl(j^{k+\ell} \cdot m\bigr) \,\in\, \Jalgn{\ell} 
\cdot \Nmod
\ (\,\equiv\, \Nmod_{\ell})
\eeq
for every $j \in \Jalg$, $m \in \Mmod$ and $\ell=0,1,\dots$.
In other words,
$\DOpu\bigl(\Mmod_{k+\ell}\bigr)$ $\subseteqq$ $\Nmod_{\ell}$.}
\end{LMA}

\medskip

One {\it proves} the lemma by induction in 
$k = 0,1,\dots$
and within this induction making another induction in $\ell = 0,1,\dots$.
For $k=0$ the statement is obvious and for $k\geqslant 1$ one 
uses the identity
$$
\DOpu\bigl(j^{k+(\ell+1)} \cdot m\bigr)
\,=\,
j \cdot \DOpu\bigl(j^{k+\ell} \cdot m\bigr)
-
\DAl{\DOpu}{j}\bigl(j^{(k-1)+(\ell+1)} \cdot m\bigr)
\,.
$$

We continue with the {\it proof} of Theorem~\ref{ThFS1}.
By Lemma~\ref{lmvgvuj-11} we have already established the (formal) continuity of $\DOpu$ and thus, we obtain its continuation 
$\COMPLET{\DOpu}$ $:$ $\COMPLET{\Mmod}$ $\to$ $\COMPLET{\Nmod}$
as a $\GRF$--linear map.
Next, statement $(b)$ follows by Eq.~(\ref{MissEq1}) since the latter extends, by continuity, for the arguments of $\DOpu$.
As we can extend, by continuity, Eq.~(\ref{MissEq1}) also in the elements $a_1,\dots,a_{k+1}$ of $\Aalg$ to $\COMPLET{\Aalg}$ we also obtain part $(d)$.
Finally, to prove $(c)$ let us assume that $\DOpv$ $\in$ $\in$ $\DfrOp{}{\Aalg}(\COMPLET{\Mmod},\COMPLET{\Nmod})$ be such that $\DOpv \bigl|_{\Mmod}$ $=$ $0$.
We need to show then that $\DOpv = 0$.
But applying Lemma~\ref{lmvgvuj-11} to $\DOpv$ we conclude that $\DOpv$ is (formally) continuous.
Since $\Mmod$ is dense in $\COMPLET{\Mmod}$ and $\DOpv \bigl|_{\Mmod}$ $=$ $0$ it follows that $\DOpv = 0$.$\quad\Box$

\medskip

\begin{COROL}{jhewaq}
{\it
Let us consider $\Aalg$ $=$ $\GRF[t_1,\dots,t_N]$ as an $\Aalg$--module with respect to the multiplication as a module action
and let
$\COMPLET{\Aalg}$ be its $($formal$)$ completion considered as an $\Aalg$--module
$($as in Examples~$\ref{ex2.1}$ and $\ref{ExamFPS}$$)$.
Let $\Nmod$ be a linearly topologized $\Aalg$--module with respect to the filtration
$(\Nmod_n$ $=$ $\Jalgn{n} \cdot \Nmod)_{n \, = \, 1}^{\infty}$ and $\COMPLET{\Nmod}$ be its $($formal$)$ completion.
Then if $\DOpu$ $\in$ $\DfrOp{k}{\Aalg}\bigl(\COMPLET{\Aalg},\COMPLET{\Nmod}\bigr)$, it follows that $\DOpu$ has the form of Eq.~$(\ref{repU})$, where now
$U^{(0)}$ and $U^{(n)}_{\mu_1,\dots,\mu_n}$ $\in$ $\COMPLET{\Nmod}$ and $f \in \COMPLET{\Aalg}$.}
\end{COROL}

\medskip

\noindent
{\it Proof.}
If we restrict $\DOpv$ $:=$ $\DOpu\bigl|_{\Aalg}$ we get an element $\DOpv$ $\in$ $\DfrOp{k}{\Aalg}\bigl(\Aalg,\COMPLET{\Nmod}\bigr)$.
We then apply the result of Example~\ref{ex2.1} and obtain the form of Eq.~(\ref{repU}) for $f \in \Aalg$ with $U^{(0)}$ and $U^{(n)}_{\mu_1,\dots,\mu_n}$ $\in$ $\COMPLET{\Nmod}$.
But since $\DOpu$ is the extension of $\DOpv$ by continuity and the right hand side of Eq.~(\ref{repU}) is continuous in $f$ we obtain the representation (\ref{repU}) for all $f \in \COMPLET{\Aalg}$.$\quad\Box$

\section{\TSedi operators}\label{SectADiffOp}
\SETCNTR

In this section we introduce a generalization of the notion of differential operators,
which we call ``\Tsedi operators''.
These are linear operators between modules of a commutative algebra $\Aalg$ such that when we multiply their arguments by elements of the algebra $\Aalg$ they act on these elements as differential operators.

Our construction is facilitated by the presence of a unit in the algebra.
Therefore, in this section we assume everywhere that $\Aalg$ is a commutative associative algebra 
with \REMemph{unit}
over a ground field $\GRF$ of characteristics $0$.
Here is the main observation that motivates our generalization.

\medskip

\begin{PROPO}{pr4.1mm}
{\it Let us set for a $\GRF$--linear map
$\DOpu \in \Hom_{\GRF} (\Mmod,\Nmod)$
between two modules $\Mmod$ and $\Nmod$ of a unital $($commutative$)$ algebra $\Aalg$$:$
\beq\label{DMo-def}
\DMo{\DOpu}{m} \, : \, \Aalg \, \to \, \Nmod \, : a \, \mapsto \, \DMo{\DOpu}{m}(a) \, := \, \DOpu(a \cdot m)
\,,
\eeq
where $a \in \Aalg$, $m \in \Mmod$.
Then,
\beqs
\podr
\DOpu \in \DfrOp{k}{\Aalg}(\Mmod,\Nmod)
\quad
\bnn
\podr
\Longleftrightarrow \quad
\text{for all} \quad m \in \Mmod \quad \text{we have} \quad
\DMo{\DOpu}{m} \in \DfrOp{k}{\Aalg}(\Aalg,\Nmod)
\,.
\eeqs}%
\end{PROPO}%
\PRF
One checks that
$$
\DAl{\bigl(\DMo{\DOpu}{m}\bigr)}{a} \, = \, \DMo{\bigl(\DAl{\DOpu}{a}\bigr)}{m}
$$
for all $a \in \Aalg$ and $m \in \Mmod$
and then the proposition follows by induction in~$k$.
Note that the presence of algebra unit guarantees that
$$
\DOpu \, = \, 0 
\quad \Longleftrightarrow \quad
\DMo{\DOpu}{m} \, = \, 0
\ \ (\forall \, m \,\in\, \Mmod)
$$
since $\DOpu(m)$ $=$ $\DMo{\DOpu}{m}(1)$.\quad$\Box$

\medskip

\begin{COROL}{2.2}
{\it For a unital algebra $\Aalg$ one has
\beqs
\podr
\DfrOp{}{\Aalg}(\Mmod,\Nmod)
\, = \,
\bigl\{
\DOpu \in \Hom_{\GRF} (\Mmod,\Nmod)
\, \bigl| \,
(\exists k \in \N)
\,
(\forall m \in \Mmod)
\,
\nnb
\podr
\phantom{\DfrOp{}{\Aalg}(\Mmod,\Nmod)
\, = \,
\bigl\{
\DOpu \in \Hom_{\GRF} (\Mmod,\Nmod)
\, \bigl| \,}
\DMo{\DOpu}{m} \in \DfrOp{k}{\Aalg}(\Aalg,\Nmod)
\bigr\} \,.
\eeqs}%
\end{COROL}%
\begin{DEFIN}{df4.1}
Exchanging the places of the quantifiers
``$\exists k \in \N$'' and
``$\forall m \in \Mmod$''
in the above equation we set:
\beqa
\ADfrOp{\Aalg} (\Mmod,\Nmod)
\, := \podr
\bigl\{
\DOpu \in \Hom_{\GRF} (\Mmod,\Nmod)
\, \bigl| \,
(\forall m \in \Mmod)
\,
(\exists k \in \N)
\,
\nnbnn
\podr
\phantom{\bigl\{
\DOpu \in \Hom_{\GRF} (\Mmod,\Nmod)
\, \bigl| \,}
\DMo{\DOpu}{m} \in \DfrOp{k}{\Aalg}(\Aalg,\Nmod)
\bigr\}
\nnbnn
\, \equiv \podr
\bigl\{
\DOpu \in \Hom_{\GRF} (\Mmod,\Nmod)
\, \bigl| \,
(\forall m \in \Mmod)
\,
\DMo{\DOpu}{m} \in \DfrOp{}{\Aalg}(\Aalg,\Nmod)
\bigr\}
\nnbnn
\supseteq \podr \DfrOp{}{\Aalg} (\Mmod,\Nmod) \,,
\eeqa
where $\Aalg$ is a unital algebra and $\Mmod$ and $\Nmod$ are $\Aalg$--modules.
We call the elements of $\ADfrOp{\Aalg} (\Mmod,\Nmod)$ \DEFemph{\Tsedi operators}.
Note that in general these operators do not have an assigned order.
\end{DEFIN}

\medskip

Since
\beq\label{ocf3cg4v}
a \cdot (\DMo{\DOpu}{m}) \,=\, \DMo{(a \cdot \DOpu)}{m}
\,,\quad
(\DMo{\DOpu}{m}) \cdot a  \,=\, \DMo{(\DOpu \cdot a)}{m}
\eeq
for all $a \in \Aalg$, $m \in \Mmod$ and $\DOpu \in \Hom_{\GRF} (\Mmod,\Nmod)$ it follows that
$\ADfrOp{\Aalg}(\Mmod,\Nmod)$ is a $(\Aalg$--$\Aalg)$--sub-bimodule of $\Hom_{\GRF} (\Mmod,\Nmod)$.
Another useful property is
\beq\label{F-liner}
\DMo{\DOpu}{a \,\cdot\, m} \,=\,
(\DMo{\DOpu}{m}) \cdot a
\eeq
($a \in \Aalg$, $m \in \Mmod$ and $\DOpu \in \Hom_{\GRF} (\Mmod,\Nmod)$).

\medskip

\begin{REMAR}{rmihcrqeic}
Definition~\ref{df4.1} can be used also for commutative algebras $\Aalg$ without unit but in order to ensure that
$\ADfrOp{\Aalg} (\Mmod,\Nmod)$ 
$\supseteq$
$\DfrOp{}{\Aalg} (\Mmod,\Nmod)$
we need some extra assumptions like the presence of a unit in $\Aalg$.
A weaker condition is that $\Aalg \cdot \Mmod$ $=$ $\Mmod$.
Then, Proposition~\ref{pr4.1mm} and Corollary~\ref{2.2} are still true.
\end{REMAR}

\medskip

\begin{PROPO}{pr4.3}
{\it
Let $\Mmod$ be a \THEemph{finitely generated} module of a unital algebra $\Aalg$.
Then
\\
${}$ \hfill
$\ADfrOp{\Aalg} (\Mmod,\Nmod)$ $=$ $ \DfrOp{}{\Aalg} (\Mmod,\Nmod) $.
\hfill ${}$}
\end{PROPO}

\PRF
Since $\Mmod$ is a finitely generated $\Aalg$--module there are $m_1,$ $\dots,$ $m_r$ $\in$ $\Mmod$ such that every $m \in \Mmod$ can be represented as
$m$ $=$ $a_1 \cdot m_1$ $+$ $\cdots$ $+$ $a_r \cdot m_r$.
Then if $\DOpu \in \ADfrOp{\Aalg} (\Mmod,\Nmod)$ we have (cf. Eq.~(\ref{F-liner})):
$$
\DMo{\DOpu}{m} \, = \,
(\DMo{\DOpu}{m_1}) \cdot a_1
+ \cdots +
(\DMo{\DOpu}{m_r}) \cdot a_r
\in \DfrOp{k}{\Aalg}(\Mmod,\Nmod)
\,,
$$
where $k$ is the maximum of the orders of the differential operators $\DMo{\DOpu}{m_j}$ for $j=1,\dots,r$.
Thus, the differential operators $\DMo{\DOpu}{m}$ has uniformly bounded orders for all $m \in \Mmod$ and hence, $\DOpu$ $\in$ $\DfrOp{k}{\Aalg} (\Mmod,\Nmod)$.$\quad\Box$

\medskip

\medskip

The main property of our \Tsedi operators is that they stay closed with respect to the composition.
We shall prove this only for the main case of our interest: when the modules are over the polynomial algebra $\GRF[t_1,\dots,t_N]$ 
or its formal completion
$\GRF[[t_1,\dots,t_N]]$.
We leave the generalization of this statement for a future work.

\medskip

\begin{THEOR}{thMain1}
{\it
Let $\Aalg$ $:=$ $\GRF[t_1,\dots,t_N]$
or, 
$\Aalg$ $:=$ $\GRF[[t_1,\dots,t_N]]$ 
and $\Mmod$, $\Nmod$, $\Pmod$ be $\Aalg$--modules.
Then the composition
$$
\Hom_{\GRF} (\Mmod,\Nmod) \times \Hom_{\GRF}(\Nmod, \Pmod) \,\to\, \Hom_{\GRF}(\Mmod, \Pmod) \,:\,
(\DOpu,\DOpv) \,\mapsto\, \DOpv \,\circ\, \DOpu
$$
maps
$\ADfrOp{\Aalg} (\Mmod,\Nmod) \times \ADfrOp{\Aalg} (\Nmod,\Pmod)$
to
$\ADfrOp{\Aalg} (\Mmod,\Pmod)$.}
\end{THEOR}

\medskip

\PRF
Let $\DOpu \in \ADfrOp{\Aalg}(\Mmod,\Nmod)$, $\DOpv \in \ADfrOp{\Aalg}(\Nmod,\Pmod)$, $m \in \Mmod$, $a \in \Aalg$. 
We need to prove that $\DMo{(\DOpv \circ \DOpu)}{m}$ $\in$ $\DfrOp{}{\Aalg}(\Aalg,\Pmod)$.
We start with:
\beq\label{tempe1}
\DMo{(\DOpv \circ \DOpu)}{m} (a) 
\,=\,
\DOpv\bigl(\DOpu(a \cdot m)\bigr)
\,=\,
\DOpv\bigl(\DMo{\DOpu}{m}(a)\bigr)
\,\equiv\,
\DOpv \circ \DMo{\DOpu}{m}(a)
\,.
\eeq
As $\DMo{\DOpu}{m}$ $\in$ $\DfrOp{k}{\Aalg}(\Aalg,\Nmod)$ it is of the form (\ref{repU}) 
(cf. also Corollary~\ref{jhewaq})
for some integer $k$ and coefficients
$U^{(0)}$, $U^{(n)}_{\mu_1,\dots,\mu_n}$ all depending on $m$.
Substituting (\ref{repU}) in (\ref{tempe1}) we then get:
\beq\label{repU-1}
\DOpv \circ \DMo{\DOpu}{m} \,=\,
\DMo{\DOpv}{U^{(0)}} +
\mathop{\sum}\limits_{n \,=\, 1}^{k} \
\mathop{\sum}\limits_{\mu_1,\dots,\mu_n \,=\, 1}^N
\hspace{-4pt}
\DMo{\DOpv}{U^{(n)}_{\mu_1,\dots,\mu_n}}
\circ
\frac{\di^n}{\di t_{\mu_1}\cdots \,\di t_{\mu_n}}
\,.
\eeq
As $\DMo{\DOpv}{U^{(0)}}$ and $\DMo{\DOpv}{U^{(n)}_{\mu_1,\dots,\mu_n}}$ are differential operators we conclude by Proposition~\ref{pr2.1} (a) that the right hand side of Eq.~\ref{repU-1} is a differential operator as well.$\quad\Box$

\medskip

In the remaining part of this section we restore the generality for the underlying commutative associative algebras: they will be arbitrary such algebras with unit.

\medskip

\begin{PROPO}{pr4.7}
{\it
Let $\DOpu_1$ $\in$ $\ADfrOp{\Aalg_1}(\Mmod_1,$ $\Nmod_1)$,
$\DOpu_2$ $\in$ $\ADfrOp{\Aalg_2}(\Mmod_2,$ $\Nmod_2)$.
Then, under the natural inclusion~$(\ref{isom1a})$ it follows that
$$
\DOpu_1 \otimes \DOpu_2 \,\in\,
\ADfrOp{\Aalg_1\otimes\Aalg_2}(\Mmod_1\otimes\Mmod_2,\,\Nmod_1\otimes\Nmod_2).
$$%
}%
\end{PROPO}%
\PRF
The proposition follows by the observation that
$$
\DMo{(\DOpu_1 \otimes \DOpu_2)}{m_1 \otimes m_2}
\,=\,
\DMo{(\DOpu_1)}{m_1}
\otimes
\DMo{(\DOpu_1)}{m_1}
$$
and Proposition~\ref{pr2.1} (b).$\quad\Box$

\medskip

\begin{THEOR}{th4.5}
{\it
Under the assumptions of Theorem~$\ref{th3.1}$ 
let us assume additionally that the algebra $\Aalg$ has a unit $($and $\Mmod$, and $\Nmod$ are $\Aalg$--modules with a $\Mmod$-- and $\Nmod$--regular element $\FrE \in \Aalg$$)$. 
It follows that every \Tsedi operator
$\DOpu \in \ADfrOp{\Aalg} (\Mmod,\Nmod)$
possess a unique extension
$\EXTOP{\DOpu} \in \ADfrOp{\Aalg} (\Mmod[1/\FrE],\Nmod[1/\FrE])$ such that
$\EXTOP{\DOpu} (m)$ $=$ $\DOpu (m)$ for every $m \in \Mmod$.}
\end{THEOR}

\medskip

\PRF
a) We first construct $\EXTOP{u}$ as an element of $\Hom_{\GRF} (\Mmod[1/\FrE],\Nmod[1/\FrE])$.
We set
\beq\label{Ext2}
\EXTOP{\DOpu}\Bigl(\frac{m}{\FrE^n}\Bigr)
\,:=\,
\EXTOP{(\DMo{\DOpu}{m})} \Bigr(\frac{\FrE}{\FrE^{n+1}}\Bigr)
\,.
\eeq
We check the correctness: if $\TXDfrac{m_1}{\FrE^{n_1}}$ $=$ $\TXDfrac{m_2}{\FrE^{n_2}}$ for $n_1$ $>$ $n_2$ then $m_1$ $=$ $\FrE^{n_{12}} \cdot m_2$ for $n_{12} := n_1-n_2$ and hence,
\beqs
\podr
\EXTOP{\DOpu}\Bigl(\frac{m_1}{\FrE^{n_1}}\Bigr)
\,=\,
\EXTOP{(\DMo{\DOpu}{m_1})} \Bigr(\frac{\FrE}{\FrE^{n_1+1}}\Bigr)
\,=\,
\EXTOP{(\DMo{\DOpu}{\FrE^{n_{12}} \cdot m_2})} \Bigr(\frac{\FrE}{\FrE^{n_1+1}}\Bigr)
\nnb
\podr=\,
\EXTOP{\bigl((\DMo{\DOpu}{m_2}) \cdot \FrE^{n_{12}}\bigr)} \Bigr(\frac{\FrE}{\FrE^{n_1+1}}\Bigr)
\,=\,
\bigl((\EXTOP{(\DMo{\DOpu}{m_2})}) \cdot \FrE^{n_{12}}\bigr) \Bigr(\frac{\FrE}{\FrE^{n_1+1}}\Bigr)
\nnb
\podr=\,
\EXTOP{(\DMo{\DOpu}{m_2})}
\Bigr(\FrE^{n_{12}}\, \frac{\FrE}{\FrE^{n_1+1}}\Bigr)
\,=\,
\EXTOP{(\DMo{\DOpu}{m_2})} \Bigr(\frac{\FrE}{\FrE^{n_2+1}}\Bigr)
\,=\,
\EXTOP{\DOpu}\Bigl(\frac{m_2}{\FrE^{n_2}}\Bigr)
\,,
\eeqs
where in the third equality we used Eq.~(\ref{ocf3cg4v}) and in the fourth equality we used Eq.~(\ref{AAmod1}).
Thus, $\EXTOP{\DOpu}$ is correctly defined as a map $\Mmod[1/\FrE]$ $\to$ $\Nmod[1/\FrE]$.
It is also a $\GRF$--linear map as
the right hand side of Eq.~(\ref{Ext2}) is linear in $m$.

b) Next, we claim that
\beq\label{Eq-7}
\DMo{(\EXTOP{\DOpu})}{m/\FrE^n}
\,=\,
\DMo{(\EXTOP{(\DMo{\DOpu}{m})})}{\FrE/\FrE^{n+1}}
\,.
\eeq
Indeed,
$\DMo{(\EXTOP{\DOpu})}{m/\FrE^n} (a)$
$=$
$\EXTOP{\DOpu}\Bigl(
\TXDfrac{am}{\FrE^{n}}
\Bigr)$
$=$
$\EXTOP{(\DMo{\DOpu}{am})} \Bigr(\TXDfrac{\FrE}{\FrE^{n+1}}\Bigr)$
\\
$=$
$\EXTOP{((\DMo{\DOpu}{m}) \cdot a)} \Bigr(\TXDfrac{\FrE}{\FrE^{n+1}}\Bigr)$
$=$
$(\EXTOP{(\DMo{\DOpu}{m})} \cdot a) \Bigr(\TXDfrac{\FrE}{\FrE^{n+1}}\Bigr)$
$=$
$\EXTOP{(\DMo{\DOpu}{m})} \Bigr(\TXDfrac{\FrE a}{\FrE^{n+1}}\Bigr)$
\\
$=$
$\DMo{(\EXTOP{(\DMo{\DOpu}{m})})}{\FrE/\FrE^{n+1}} (a)$.
Then, Eq.~(\ref{Eq-7}) implies that $\EXTOP{\DOpu}$ defined by (\ref{Ext2}) is \anTsedi operator, i.e.,
$\EXTOP{\DOpu}$ $\in$ $\ADfrOp{\Aalg}(\Mmod[1/\FrE],\Nmod[1/\FrE])$.

c)
Why $\EXTOP{\DOpu}$ is an extension of $\DOpu$:
\beqs
\podr
\EXTOP{\DOpu}\Bigl(\frac{\FrE \cdot m}{\FrE}\Bigr)
\,=\,
\EXTOP{(\DMo{\DOpu}{\FrE m})} \Bigr(\frac{\FrE}{\FrE^{2}}\Bigr)
\,=\,
\EXTOP{(\DMo{\DOpu}{m} \cdot \FrE)} \Bigr(\frac{\FrE}{\FrE^{2}}\Bigr)
\nnb
\podr=\,
(\EXTOP{(\DMo{\DOpu}{m})} \cdot \FrE) \Bigr(\frac{\FrE}{\FrE^{2}}\Bigr)
\,=\,
\EXTOP{(\DMo{\DOpu}{m})} \Bigr(\frac{\FrE^2 \, 1}{\FrE^{2}}\Bigr)
\,=\,
\DMo{\DOpu}{m}(1) \,=\, \DOpu(m)
\,.
\eeqs

d) Let us prove the \textit{uniqueness} of $\EXTOP{\DOpu}$.
As in Theorem~\ref{th3.1} it is enough to prove that if
$\DOpv \in \ADfrOp{\Aalg}(\Mmod[1/\FrE],\Nmod[1/\FrE])$ is such that it is zero on $\Mmod \subset \Mmod[1/\FrE]$, i.e., $v \Bigl(\TXUfrac{\FrE \cdot m}{\FrE}\Bigr)$ $=$ $0$ ($\forall m \in \Mmod$), then $\DOpv=0$.
Indeed, let $n \in \{1,2,\dots\}$ and $m \in \Mmod$.
We want to show that 
$\DOpv\Bigl(\TXDfrac{m}{\FrE^n}\Bigr)$ $=$ $0$.
Let $\widetilde{\DOpv}$ $:=$ $\DMo{\DOpv}{m/\FrE^{n+1}}$.
Then $\widetilde{\DOpv}$ $\in$ $\DfrOp{}{\Aalg} (\Aalg,\Nmod[1/\FrE])$ is such that $\widetilde{\DOpv} (\FrE^{n+1} a)$ $=$ $0$ for all $a \in \Aalg$.
By Lemma~\ref{Lm4.XX} it follows that $\widetilde{\DOpv} = 0$ and hence
$0$ $=$
$\widetilde{\DOpv} (\FrE)$ $=$ $\DMo{\DOpv}{m/\FrE^{n+1}} (\FrE)$ $=$
$\DOpv\Bigl(\TXDfrac{m}{\FrE^n}\Bigr)$.

This completes the proof of Theorem~\ref{th4.5}.$\quad\Box$

\medskip

\begin{PROPO}{pr4.8}
{\it
$(a)$
Let $\DOpu$ $\in$ $\ADfrOp{\Aalg}(\Mmod,$ $\Nmod)$,
$\DOpv$ $\in$ $\ADfrOp{\Aalg}(\Nmod,$ $\Pmod)$,
$\EXTOP{\DOpu}$ $\in$ $\ADfrOp{\Aalg}(\Mmod[1/\FrE],$ $\Nmod[1/\FrE])$ and $\EXTOP{\DOpv}$ $\in$ $\ADfrOp{\Aalg}(\Nmod[1/\FrE],$ $\Pmod[1/\FrE])$
satisfy the conditions of Theorem~$\ref{th4.5}$.
Then,
\beq\label{ExtCompPr}
\EXTOP{(\DOpv \circ \DOpu)} \,=\,
\EXTOP{\DOpv} \circ \EXTOP{\DOpu}
\,.
\eeq

$(b)$ 
Let $\DOpu_j$ $\in$ $\ADfrOp{\Aalg_j}(\Mmod_j,$ $\Nmod_j)$,
$\EXTOP{\DOpu_j}$ $\in$ $\ADfrOp{\Aalg_j}(\Mmod_j[1/\FrE_j],$ $\Nmod_j[1/\FrE_j])$
satisfy the conditions of Theorem~$\ref{th4.5}$
for $j=1,2$.
Them
\beq\label{ExtTensPr}
\EXTOP{(\DOpu_1 \otimes \DOpu_2)} \,=\,
\EXTOP{\DOpu_1} \otimes \EXTOP{\DOpu_2}
\,,
\eeq
which belongs to \\
${}$ \hfill
$\ADfrOp{\Aalg_1\otimes\Aalg_2}\bigl((\Mmod_1\otimes\Mmod_2)[1/(\FrE_1\otimes\FrE_2)],\,(\Nmod_1\otimes\Nmod_2)[1/(\FrE_1\otimes\FrE_2)]\bigr)$.
\hfill ${}$}
\end{PROPO}

\medskip

\PRF
Both statements follow by the uniqueness of the extensions from Theorem~\ref{th4.5} as the right hand sides of Eqs.~(\ref{ExtCompPr}) and (\ref{ExtTensPr}) provide the required extensions in the left hand sides.$\quad\Box$

\medskip

In our application we will not need a direct analog of 
Theorem~\ref{ThFS1}
and so we skip this and continue with considering the pull--backs and their relation to \Tsedi operators.

\medskip

\begin{DEFIN}{1-df2.2}
For an $\Aalg$--module $\Mmod$, $\Balg$--module $\Nmod$ and an algebra homomorphism $\vartheta : \Aalg \to \Balg$
let us set (as in Definition~\ref{df2.2} $(b)$),
\beq\label{1-theta-diff}
\ADfrOp{\,\vartheta} (\Mmod,\Nmod)
\,:=\,
\ADfrOp{\Aalg} (\Mmod, \vartheta^* \, \Nmod)
\,,
\eeq
as a \REMemph{right} $\Aalg$--module (cf.~(\ref{Bi-act})) and
$\ADfrOp{\,\vartheta} (\Mmod,$ $\Nmod)$
is additionally a left $\Balg$--module
(i.e., it is a $(\Balg$--$\Aalg)$--bimodule), where the left $\Balg$--module of 
$\ADfrOp{\,\vartheta} (\Mmod,$ $\Nmod)$
is pulled back via $\vartheta$
to a left $\Aalg$--module in
$\ADfrOp{\Aalg} (\Mmod,$ $\vartheta^* \, \Nmod)$.
The elements of 
$\ADfrOp{\,\vartheta\,} (\Mmod,\Nmod)$
we call also
\DEFemph{\Tsemi{$\vartheta$} operators}.
\end{DEFIN}

\medskip

For \Tsemi{$\vartheta$} operators we have the following reformulations of Theorem~\ref{thMain1}, Proposition~\ref{pr4.7} and Theorem~\ref{th4.5}.

\medskip

\begin{COROL}{2-pr2.1}
{\it Under the same conditions as in Proposition~$\ref{1-pr2.1}$ $(a)$, $(b)$ and $(c)$, we have respectively$:$

$(a)$
The composition map
$$
\Hom_{\GRF} (\Mmod,\Nmod) \times \Hom_{\GRF}(\Nmod, \Pmod) \,\to\, \Hom_{\GRF}(\Mmod, \Pmod) \,:\,
(\DOpu,\DOpv) \,\mapsto\, \DOpv \,\circ\, \DOpu
$$
maps
$\ADfrOp{\,\vartheta} (\Mmod,\Nmod) \times \ADfrOp{\,\psi} (\Nmod,\Pmod)$
to
$\ADfrOp{\,\psi \,\circ\, \vartheta} (\Mmod,\Pmod)$
if we additionally assume that $\Aalg$, $\Balg$ and $\Calg$ are polynomial algebras.

$(b)$
The natural inclusion
$(\ref{isom1a})$
restricts to an inclusion
\beq\label{isom1b-3-1103}
\hspace{0pt}
\ADfrOp{\,\vartheta_1}(\Mmod_1,\Nmod_1) \,\otimes\, \ADfrOp{\,\vartheta_2}(\Mmod_2,\Nmod_2) \hookrightarrow
\ADfrOp{\,\vartheta_1\otimes\,\vartheta_2}(\Mmod_1\otimes\Mmod_2,\Nmod_1\otimes\Nmod_2)\,.
\eeq%

$(c)$
Every $\DOpu \in \ADfrOp{\,\vartheta}(\Mmod,\Nmod)$ posses a unique extension
\\
$\EXTOP{\DOpu} \in \ADfrOp{\,\vartheta}(\Mmod[1/\FrE],\Nmod[1/\vartheta(\FrE)])$
in the sense that
$\EXTOP{\DOpu}(m)$ $=$ $\DOpu(m)$ for every $m \in \Mmod$.}%
\end{COROL}%

\section[The residue as \anTsedi operator\vspace{4.4444pt}]{The residue as \anTsedi operator}\label{S5}
\SETCNTR

By Proposition~\ref{pr4.3}
we saw examples when
the \Tsedi operators are only the differential operators, i.e.,
$ \DfrOp{}{\Aalg} (\Mmod,\Nmod)$
$=$ 
$\ADfrOp{\Aalg} (\Mmod,\Nmod)$.
In this section we give another extreme example when
\beq\label{extriso}
0 \,=\, \DfrOp{}{\Aalg} (\Mmod,\Nmod) \,\subsetneqq\, \ADfrOp{\Aalg} (\Mmod,\Nmod)
\,\subsetneqq\, \Hom_{\GRF} (\Mmod,\Nmod)
\,.
\eeq

\begin{THEOR}{Thm5.1}
{\it
Let 
$\Aalg$ $=$ $\GRF[t]$,
and let $\Mmod$ $:=$ $\GRF[t,1/t]$ $(\equiv$ $(\GRF[t])[1/t])$ be the algebra of all \REMemph{Laurent polynomials} considered as a module over its subalgebra $\Aalg$.
Let $\Nmod$ $:=$ $\GRF$ with the $\Aalg$--action$:$
\(
f \cdot c \,:=\, f(0)c
\,,\,
\)
for $f \in \Aalg$ $(= \GRF[t])$, $c \in \GRF$.\footnote{%
i.e., the action on $\Nmod$ is trivial in the sense of Definition~\ref{df2.2}~(c)}
Then Eq.~$(\ref{extriso})$ holds.}
\end{THEOR}

\medskip

\PRF
First, note that
\beq\label{e5.1}
\Hom_{\GRF} (\Mmod,\Nmod) 
\,\equiv\,
\Hom_{\GRF}\bigl(\GRF[t,1/t],\GRF\bigr) \,\cong\,
_{\GRF}(\GRF[[t,1/t]])_{\Aalg}
\,,
\eeq
as ($\Aalg$--$\Aalg$)--bimodules, where:
\beq\label{e5.2}
\GRF[[t,1/t]]
\,:=\,
\biggl\{\mathop{\sum}\limits_{n \,\in\, \Z} 
c_n \, t^n \,\biggl|\, c_n \,\in\, \GRF \ (\forall n \,\in\, \Z)\biggr\}
\,,
\eeq
is the vector space of all double infinite formal power series (cf. \cite[Sect. 2.1]{K98});
we turn the vector space $\GRF[[t,1/t]]$ into an ($\Aalg$--$\Aalg$)--bimodule 
$_{\GRF}(\GRF[[t,1/t]])_{\Aalg}$ on which $\Aalg$ acts trivially from the left (i.e., it is a left $\GRF$--module) and the right action of $\Aalg$ is generated by the natural multiplication rule:
$$
\biggl(\mathop{\sum}\limits_{n \,\in\, \Z} c_n t^n \,\biggl) \cdot \, t^k
\,=\,
\mathop{\sum}\limits_{n \,\in\, \Z}c_{n-k}\, t^n \,,
$$
($k \in \Z$, so this is actually an action of $\GRF[t,1/t]$).
In the category of vector spaces, the isomorphism (\ref{e5.1}) follows just by the fact that the vector space $\GRF[t,1/t]$ is a countable direct \REMemph{sum} of copies of $\GRF$ and hence, its (algebraic) dual 
$\Hom_{\GRF}\bigl(\GRF[t,1/t],\GRF\bigr)$ is isomorphic to a countable direct \REMemph{product} that in turn is isomorphic to $\GRF[[t,1/t]]$ (Eq.~(\ref{e5.2})).
As an isomorphism of ($\Aalg$--$\Aalg$)--bimodules (for $\Aalg$ $=$ $\GRF[t]$) we use the so called \REMemph{residue} functional on 
\beq\label{res-dfn}
\RES \,:\, \GRF[[t,1/t]] \to \GRF \,:\, 
\mathop{\sum}\limits_{n \,\in\, \Z} c_n \, t^n
\,\mapsto\,
c_{-1} 
\,.
\eeq
Then, the map in Eq.~(\ref{e5.1}) is:
\beqa\label{Kernel}
&
\Ker 
\,:\, 
\Hom_{\GRF}\bigl(\GRF[t,1/t],\GRF\bigr) 
\,\cong\, 
\GRF[[t,1/t]]
\,:\,
\DOLP \,\mapsto\,\KER{\DOLP}
\,,
& \qquad \bnn &
\text{defined by:}
\quad
\DOLP(\LorPol)
\,=\,
\RES \left(\KER{\DOLP} \cdot \LorPol\right)
\quad
\text{for all} 
\quad
\LorPol \,\in\, \GRF[t,1/t]
\,,
&
\eeqa
and it is obviously an isomorphism of $(\Aalg$--$\Aalg)$--bimodules; note also that we have:
\beq\label{e11110-6.1}
\KER{\DOLP} \,=\,
\mathop{\sum}\limits_{n \,\in\, \Z}
\DOLP(t^{-n-1})\,t^n
\,.
\eeq
In the subsequent part of the proof we shall write $\GRF[[t,1/t]]$ also for the ($\Aalg$--$\Aalg$)--bimodule $_{\GRF}(\GRF[[t,1/t]])_{\Aalg}$ in the isomorphism~(\ref{e5.1}) and~(\ref{Kernel}).

Now, for $\DOLP \in \Hom_{\GRF}(\GRF[t,1/t],\GRF)$, $n \in \Z$ and $k=0,1,\dots$ one has
$$
\DMo{\DOLP}{t^n} (t^k)
\,=\,
\DOLP(t^{k+n})
\,=\,
\text{the $(-k-n-1)$th coefficient in }
\KER{\DOLP}
\,
$$
(cf. Eq.~(\ref{e11110-6.1})).
Hence, 
$\DMo{\DOLP}{t^n}$ will be a differential operator 
belonging to
$\DfrOp{k}{\Aalg}(\Aalg,\Nmod)$
iff all the coefficients of the formal series vanish $\KER{\DOLP}$ below $-n-k-1$.\footnote{%
i.e., $\DMo{\DOLP}{t^n}$ will be a linear combination of $\left.\frac{\di^{\ell}}{(\di t)^{\ell}}\right|_{t \, = \, 0}$ for $\ell=0,1,\dots,k$}
This implies that under the isomorphism~(\ref{Kernel})
we have:
$$
\ADfrOp{\Aalg}(\Mmod,\Nmod) 
\,(\,\equiv
\ADfrOp{\GRF[t]}(\GRF[t,1/t],\GRF) )
\,\cong\,
\GRF[[t,1/t]
\,\subsetneqq\,
\GRF[[t,1/t]]
\,,
$$
where $\GRF[[t,1/t]$ is the space of \REMemph{Laurent power series}:
\beqs
\GRF[[t,1/t]
\,:=\podr
\biggl\{\mathop{\sum}\limits_{n \,=\, -N}^{\infty} 
c_n \, t^n \,\in\, \GRF[[t,1/t]] \,\biggl|\, N \,=\,0,1,\dots\biggr\}
\nnb
=\podr
\bigl(\GRF[[t]]\bigr)[1/t]
\,.
\eeqs
Indeed, if all the coefficients of the series $\KER{\DOLP}$ are zero below $n=-N$ then the order of $\DMo{\DOLP}{t^{n_1}}$ is not exceeding 
$\max\{0,N-n_1-1\}$, where $n_1 \in \Z$.
But this implies also that no element of $\ADfrOp{\Aalg}(\Mmod,\Nmod)$ is a differential operator except~$0$.
The proof of the theorem is completed.$\quad\Box$

\medskip

\def\EVAL{\text{eval}}
\begin{REMAR}{rm5.01}
It follows by Theorem~\ref{Thm5.1} that
the restricted residue functional (\ref{res-dfn}) on Laurent polynomials only, i.e., 
$\RES \hspace{-2pt}\left.\raisebox{9pt}{}\right|_{\GRF[t,1/t]}$,\raisebox{-8pt}{}
 is \anTsedi operator.
There is a formula that makes manifest this fact:
\beq\label{res-dfn11}
\RES \, = \,
\mathop{\text{w-$\lim$}}\limits_{k \,>\hspace{-3pt}>\, 0} \,\frac{1}{k!} \, \EVAL_{(t \, = \, 0)} \circ \Bigl(\frac{\di}{\di t}\Bigr)^k \circ t^{k+1}
\,,
\eeq
where the ``weak limit'' $\mathop{\text{w-$\lim$}}\limits_{k \,>\hspace{-3pt}>\, 0}$ means that the subsequent sequence of linear functionals stabilizes for sufficiently large $k$ when it is evaluated on an arbitrary element $g(t)$ $\in$ $\GRF[t,1/t]$.
In other words,
for every $g(t)$ $\in$ $\GRF[t,1/t]$ there exists $K_g$ $\in$ $\{0,1,2,\dots\}$ such that
\beq\label{res-dfn11-x1110}
\RES (g) \, = \,
\frac{1}{k!} \, \left.\left(\Bigl(\frac{\di}{\di t}\Bigr)^k \bigl(t^{k+1} \, g(t)\bigr)\right)\right|_{t \, = \, 0}
\quad
\text{for all $k > K_g$}
\,.
\eeq
\end{REMAR}

\begin{REMAR}{rm5.1}
There is a straightforward generalization of Theorem~\ref{Thm5.1} using residues in higher dimensions
introduced in \cite[Sect.~III]{BN06}.
\end{REMAR}

\addcontentsline{toc}{section}{\it Part II. \TSediPRODterm operations and OPE algebras\vspace{4.4444pt}}

\section{The operads of differential\PRODterm and \TsediPRODterm operations}\label{s6}
\SETCNTR

From this section on, the operad theory (cf.~\cite{LV12} and \cite[Sect. B]{N14}) will start to play role.
We use the so called ``classical definition'' of an operad presented in Sect. 5.3 of \cite{LV12}.

\medskip

In this section we shall fix a initial commutative algebra that is a polynomial algebra.
We shall denote this algebra by $\Salg$.
In a more general situation this algebra can be the coordinate ring of the space--time, where our quantum fields live.
A \REMemph{motivation} why we consider only polynomials in the coordinate ring but not more general smooth functions is that for the models of vertex algebras in one and higher dimensions one needs only such regular (smooth) functions.
This is because the correlation functions are rational functions (\cite{NT01}).
Furthermore, the denominators in the correlation functions are products of two--point polynomials, which we also use in the construction of the operad for the vertex algebras.
Nevertheless, the latter two specifics are not crucial for our construction although they make significant simplification of the considerations.

\SSECSPA

\subsection{Some notations}\label{s6.0}
\ADDCONT{Some notations}

\noindent
Let us denote the collection of the space--time coordinates  $x^1$, $\dots$, $x^{\DMN}$ with a roman letter,
\beq\label{s6.1-1}
\x \,=\, (x^1,\dots,x^{\DMN})
\,.
\eeq
Thus, we fix in this section
\beq\label{e6.2-1}
\Salg \,:=\, \GRF [\x] \,:=\,
\GRF [x^1,\dots,x^{\DMN}]
\,.
\eeq
Then, for $n=1,2,\dots$:
\beq\label{e6.3-1}
\hspace{3pt}
\Salgn{n} \,:=\,
\Salg^{\otimes n} 
\,\cong\,
\GRF[\x_1,\dots,\x_n]
\,:=\,
\GRF[x_1^1,\dots,x_1^{\DMN};\ldots;x_n^1,\dots,x_n^{\DMN}]
\,.
\hspace{-10pt}
\eeq
Let us introduce the algebra homomorphism
\beq\label{eq6.5}
\mu_n  : \Salgn{n} \,\rightarrow\, \Salg :
F_1 \otimes \cdots \otimes F_n
\,\longmapsto\,
F_1 \cdots F_n
\qquad
(\mu_1 \,:=\, \textrm{id}_{\Salg})
\eeq
that is the \REMemph{$n$-tuple product}.
Note that $\mu_n$ also represents the \REMemph{evaluation at the total diagonal}: $\mu_n$ $:$ $F$ $\mapsto$ $F(\x_1,\dots,\x_n)\bigl|_{\x_1 \,=\cdots=\, \x_n \,=\, \x}$ for $F \in \Salgn{n} = \GRF[\x_1,\dots,\x_n]$.
We have
\beq\label{diagon-11}
\mu_n \, =\, \mu_k \circ (\mu_{j_1} \otimes \cdots \otimes \mu_{j_k})
\,,
\eeq
where 
$n=j_1+\cdots+j_k$.
The algebra homomorphism
$\mu_{j_1} \otimes \cdots \otimes \mu_{j_k}$
acts as an evaluation at the \REMemph{partial diagonal}
$\{\x_1 =\cdots= \x_{j_1} $,
$\x_{j_1+1} =\cdots= \x_{j_1+j_2}$,
$\dots$, 
$\x_{j_1+\cdots+j_{k-1}+1} =\cdots= \x_n\}$.

\SSECSPA

\subsection{The operad $\OpDiff$ of differential\PRODterm operations}\label{s6.1}
\ADDCONT{The operad $\OpDiff$ of differential\PRODterm operations}

\noindent
We denote this symmetric operad by 
$\OpDiff$ $=$ $\left(\OpDiff (n)\right)_{n \,=\, 1}^{\infty}$.
It is defined by:
\beq\label{eq6.4}
\OpDiff(n) \,:=\,
\DfrOp{}{\,\mu_n}
\bigl(
\Salgn{n},\,\Salg
\bigr)
\,.
\eeq
The action of permutation group $\PermGr_n$ is induced by the natural action on $\Salgn{n}$ $\equiv$ $\Salg^{\otimes n}$ provided by the exchange of the copies of $\Salg$.
The operadic compositions are
\beqa\label{eq6.6}
\OperCmp{j_1,\dots,j_k}
\,:\,&
\hspace{-0pt}
\OpDiff(k) \otimes \OpDiff(j_1) \otimes \cdots \otimes \OpDiff(j_k)
\hspace{-0pt}
&
\,\to\,
\OpDiff(n)
\bnn
&
\hspace{16pt}
\PDOa \otimes \PDOa'_1 \otimes \cdots \otimes \PDOa'_k
&
\,\mapsto\,
\OperComp{j_1,\dots,j_k}{\PDOa}{\PDOa'_1,\dots,\PDOa'_k}
\,,
\bnn
& &
\phantom{\,\mapsto\,}
\,:=\,
\PDOa
\circ \bigl(\PDOa'_1 \otimes \cdots \otimes \PDOa'_k\bigr)
\,,
\eeqa
for $n,k,j_1,\dots,j_k$ $\in$ $\{1,2,\dots\}$, $n=j_1+\cdots+j_k$.
The operadic unit is $1_{\Salg}$ $\in$ $\DfrOp{}{\,\Salg}(\Salg,\Salg)$ $\equiv$~$\OpDiff(1)$.
Diagrammatically the operadic composition~(\ref{eq6.6}) is given below:
$$
\begin{array}{cccc}
\Salg_{j_1} \otimes \cdots \otimes \Salg_{j_k} 
& 
&
\hspace{-17pt}
\mathop{-\hspace{-5pt}-\hspace{-5pt}-\hspace{-5pt}-\hspace{-5pt}-\hspace{-5pt}-\hspace{-5pt}\longrightarrow}\limits^{\PDOa'_1 \otimes \cdots \otimes \PDOa'_k}
\
\Salg^{\otimes k} \, = \, \Salg_k 
\
\mathop{\longrightarrow}\limits^{\PDOa}
& 
\Salg
\phantom{\,.}
\\
|\hspace{-1pt}|
&
&
&
|\hspace{-1pt}|
\phantom{\,.}
\\
\Salg_{j_1+\cdots+j_k}
&
\hspace{-12pt}
\, = \, \Salg_n
&
\mathop{-\hspace{-5pt}-\hspace{-5pt}-\hspace{-5pt}-\hspace{-5pt}-\hspace{-5pt}-\hspace{-5pt}-\hspace{-5pt}-\hspace{-5pt}-\hspace{-5pt}-\hspace{-5pt}-\hspace{-5pt}-\hspace{-5pt}-\hspace{-5pt}-\hspace{-5pt}-\hspace{-5pt}\longrightarrow}\limits_{\OperComp{j_1,\dots,j_k}{\PDOa}{\PDOa'_1,\dots,\PDOa'_k}}
&
\Salg
\,.
\end{array}
$$

The proof that Eqs.~(\ref{eq6.4})--(\ref{eq6.6}) define an operad structure on 
$\OpDiff$ $=$ $\bigl(\OpDiff (n)\bigr)_{n \,=\, 1}^{\infty}$
follows by the fact that this operad is a \REMemph{suboperad} of the \REMemph{endomorphism operad} (\cite[Secr.~5.2.11]{LV12}) over $\Salg$, 
$\OpEnd_{\Salg}$ $=$
$(\OpEnd_{\Salg}(n))_{n \, =\, 1}^{\infty}$:
\beq\label{eq6.7}
\OpDiff (n) \,\subset\,
\OpEnd_{\Salg}(n) \,:=\, \Hom_{\GRF}\bigl(\Salgn{n},\,\Salg\bigr)
\,.
\eeq
Indeed, the composition law (\ref{eq6.6}), the actions of permutation groups and the operadic unit are exactly those that come from the operad $\OpEnd_{\Salg}$ if we consider the $\mu_n$--differential operators just as $\GRF$--linear maps.
So, it remains to prove that the actions of permutations and the compositions do not bring out of the corresponding spaces of differential operators.
The latter, in turn,
is implied by Proposition~\ref{1-pr2.1} $(a)$ and $(b)$,
and Eq.~(\ref{diagon-11}).\footnote{%
Equation~(\ref{diagon-11}) guarantees that $\OperComp{j_1,\dots,j_k}{\PDOa}{\PDOa'_1,\dots,\PDOa'_k}$, which by construction is a $(\mu_k \circ (\mu_{j_1} \otimes \cdots \otimes \mu_{j_k}))$--differential operator, is also a $\mu_n$--differential operator.}

Let us give the explicit form of the differential\PRODterm operations, i.e., of the elements of the operad $\OpDiff$.
If $\PDOa$ $\in$ $\OpDiff(n)$ $(=$ $\DfrOp{}{\,\mu_n}(\Salgn{n},$ $\Salg))$ then according to Example~\ref{ex2.1} its action on $F$ $\in$ $\Salg_n$ is of a form:\footnote{%
These operators are also called \REMemph{poly-differential operators}.}
\beqa\label{repU-0913}
\PDOa(F) \,=\podr
\Bigl(
\mathop{\sum}\limits_{\r_1,\dots,\r_n}
f_{\r_1,\dots,\r_n}(\x) .\, \di_{\x_1}^{\,\r_1} \cdots \di_{\x_n}^{\,\r_n} F(\x_1,\dots,\x_n) 
\Bigr)\Bigl|_{\x_1 \,=\, \cdots \,=\, \x_n \,= \, \x}
\,.
\qquad
\eeqa
Here, 
$f_{\r_1,\dots,\r_n}$ $\in$ $\Salg$;
the sum is finite and runs over multiindices
$$
\r_j \,=\, (r_j^1,\dots,r_j^{\DMN}) \, \in \, \{0,1,\dots\}^{\times \DMN}
\quad (j \, = \, 1,\dots n)
\,;
$$
finally, we use the multi-index notations:
$$
\di_{\x_j}^{\,\r_j}
\, := \,
\Bigl(\frac{\di}{\di x_j^1}\Bigr)^{r_j^1}
\cdots \,
\Bigl(\frac{\di}{\di x_j^{\DMN}}\Bigr)^{r_j^{\DMN}}
\,.
$$
Equation~(\ref{repU-0913}) follows from Eq.~(\ref{repU})
in which we replace the coefficients $U^{(n)}_{\mu_1,\dots,\mu_n}$ by $f_{\r_1,\dots,\r_n}$ $\in$ $\Salg$ and since $\Salg_n$ acts on $\Salg$ via the pull-back  over $\mu_n$ (\ref{eq6.5}) we get from (\ref{repU}) first
$$
\PDOa(F) \,=\,
\mathop{\sum}\limits_{\r_1,\dots,\r_n}
\mu_n \bigl(
\di_{\x_1}^{\,\r_1} \cdots \di_{\x_n}^{\,\r_n} F
\bigr)
\cdot
f_{\r_1,\dots,\r_n}
\,,
$$
which is exactly (\ref{repU-0913}).
Note that $\ProdOper_n \in \OpDiff (n)$ is a particular example of Eq.~(\ref{repU-0913}).
Then Eq.~(\ref{diagon-11}) takes the operadic form:
\beq\label{diagon-11-913}
\ProdOper_n \, =\, \OperComp{j_1,\dots,j_k}{\ProdOper_k}{\ProdOper_{j_1},\dots,\ProdOper_{j_k}}
\,.
\eeq
It follows that the sequence of subspaces $\GRF \,.\,\ProdOper_n$ $\subseteq$ $\OpDiff (n)$ form a \REMemph{suboperad} in $\OpDiff$ and the composition law (\ref{diagon-11-913})  indicates that this suboperad is exactly the operad $\OpCom$ $(\OpCom(n)$ $=$ $\GRF \,.\,\ProdOper_n$) that governs the \REMemph{commutative associative algebras} (cf. \cite[Sect. 5.2.10]{LV12}).
In this way we get an operadic inclusion:
\beq\label{eq9.3nn0913nn1}
\OpCom \, \hookrightarrow \, \OpDiff
\,.
\eeq
Further consequence of Eq.~(\ref{repU-0913}) is that if we take in addition the elements
\beqa\label{NewGen-0913-0}
x^{\alpha} \,\in\, \OpDiff (1) \podr\quad
(\text{the operator of multiplication by $x^{\alpha}$}),
\\ \label{NewGen-0913}
\Drv_{\alpha} \,\in\, \OpDiff (1) \podr\quad
(\text{the partial derivative } \frac{\di}{\di x^{\alpha}})
\eeqa
($\alpha = 1,\dots,\DMN$) we get the relations:
\beqa\label{NewRel-0913-0}
\OperComp{1}{x^{\alpha}}{\ProdOper_n}
\,=\podr
\OperComp{1,\dots,1}{\ProdOper_n}{1,\dots,1,x^{\alpha},1,\dots,1}
\quad \text{for all positions of $x^{\alpha}$}
\,,\qquad\quad
\\ \label{NewRel-0913}
\OperComp{1}{\Drv_{\alpha}}{\ProdOper_n}
\,=\podr
\mathop{\sum}\limits_{\text{all positions of $\Drv_{\alpha}$}}
\OperComp{1,\dots,1}{\ProdOper_n}{1,\dots,1,\Drv_{\alpha},1,\dots,1}
\,,
\\ \label{NewRel-0913-2}
\OperComp{1}{\Drv_{\alpha}}{x^{\beta}} 
\,=\podr
\OperComp{1}{x^{\beta}}{\Drv_{\alpha}} 
+ \delta_{\alpha}^{\beta}
\,
\eeqa
($\alpha,\beta = 1,\dots,\DMN$, $\delta_{\alpha}^{\beta}$ being the Kr\"onecker delta symbol)).
One recognizes in the last two lines operadic forms of the \REMemph{Leibniz rule}, which we shall further comment in Sect.~\ref{Se8nn}.

\SSECSPA

\subsection{The operad $\OpADiff$ of \TsediPRODterm operations}\label{s6.2}
\ADDCONT{The operad $\OpADiff$ of \TsediPRODterm operations}

\noindent
Let us fix an element
\beq\label{e6.8}
\BQu_{1,2} \,\in\, \Salgn{2} \,(\equiv\, \Salg^{\otimes 2})
\,.
\eeq
and denote:
\beqa\label{e6.9}
\BQu_{j,k} \,\in\, \Salgn{n}
\ \
\podr 
-\ \parbox[t]{180pt}{the image of $q_{1,2}$ under the embedding $\Salg^{\otimes 2} \,\hookrightarrow\, \Salg^{\otimes n}$ on the $(j,k)$ position,\raisebox{-7pt}{}}
\\ \label{eq6.10-0}
\TQu_n 
\ \
\podr
:= \
\mathop{\text{\Large$\textstyle \prod$}}\limits_{1 \,\leqslant\, j \,\leqslant\, k \,\leqslant\, n} \BQu_{j,k}
\ \in\ \Salgn{n}
\ \ (n \,\geqslant\, 2,\, \TQu_1 \,:=\, 1)
\,,\qquad\quad
\\ \label{eq6.10}
\TQu_{j_1|\dots|j_k}
\in \Salgn{n}
\ \,
\podr 
-\ \parbox[t]{200pt}{defined by\raisebox{10pt}{} \\
\(
\TQu_n \,=\, \TQu_{j_1|\dots|j_k}
\,.\,
\bigl(\TQu_{j_1} \otimes \cdots \otimes \TQu_{j_k}\bigr)
\)}
\eeqa
for $n,k.j_1,\dots,j_k$ $\in$ $\{1,2,\dots\}$, $n=j_1+\cdots+j_k$.
Let us set
\beqa\label{eq6.12}
\Oalg_n \,:=\podr \Salgn{n}[1/\TQu_n]
\bnn
\,(\, \equiv\podr
\GRF[\x_1\,\dots,\x_n]
\biggl[\biggl(
\mathop{\text{\Large$\textstyle \prod$}}\limits_{1 \,\leqslant\, j \,\leqslant\, k \,\leqslant\, n} \BQu_{j,k}
\biggr)^{-1}\biggr]
\,)\,,
\eeqa
which is an algebra containing $\Salgn{n}$ as a subalgebra and hence, is a $\Salgn{n}$--module.
We assume the symmetry
\beq\label{q-symm}
\BQu_{2,1} \,=\, \BQu_{1,2}
\,,
\eeq
which allows us to extend the natural action of the permutation group $\PermGr_n$ on~$\Oalg_n$.

In our applications it will happen that
\beq\label{eq6.14nn1}
\mu_2 (\BQu_{1,2}) \,=\, 0
\eeq
and hence,
\beq\label{eq6.14nn1-1}
\mu_n(\TQu_n) \,=\, 0
\,.
\eeq
The latter means that $\BQu_{1,2}$ and $\TQu_n$ vanish on the diagonals (i.e., vanish if some pair of their arguments coincide).
However,\footnote{%
Similarly to $\mu_n$ in Eq.~(\ref{eq6.5}), $\mu_{j_1} \otimes \cdots \otimes \mu_{j_k}$ acts as the evaluation at the partial diagonal $\x_1$ $=$ $\cdots$ $=$ $\x_{j_1}$, $\x_{j_1+1}$ $=$ $\cdots$ $=$ $\x_{j_1+j_2}$, \dots.}
\beqa\label{eq6.15nn1}
\TQu_{k;\,j_1,\dots,j_k} 
&:=&
(\mu_{j_1} \otimes \cdots \otimes \mu_{j_k})
\bigl(\TQu_{j_1|\cdots|j_k}\bigr) \,\neq\, 0
\,,\quad
\bnn
\TQu_{k;\,j_1,\dots,j_k} 
&\in &
\underbrace{\Salg \otimes \cdots \otimes \Salg}_{\text{
\normalsize 
$\Salgn{k}$}}
\,,\
\eeqa
since
\(
\TQu_{j_1|\dots|j_k} 
\,=\,
\prod
\raisebox{-2pt}{\raisebox{9pt}{\hspace{-1pt}}}'_{i' \,<\, i''} 
\
\BQu_{i',i''}
\,,
\)
where the product 
$\prod\raisebox{-2pt}{\raisebox{9pt}{\hspace{-1pt}}}'_{i' \,<\, i''}$\raisebox{-6pt}{}
stands for the product over all ordered pairs $i'<i''$ such that $i'$ and $i''$ do not belong to one and the same set in the partition:
\beq\label{standpart}
\{1,\dots,j_1\},\,
\{j_1+1,\dots,j_1+j_2\},\,
\ldots,\,
\{j_1+\cdots+j_{k-1}+1,\dots,n\}.\,
\eeq
It will be important that
\beq\label{ImpEqxx}
\Salgn{k}[1/\TQu_{k;\,j_1,\dots,j_k}] \,=\,
\Salgn{k}[1/\TQu_k]
\,
\eeq
as 
$\TQu_{k;\,j_1,\dots,j_k}$
(\ref{eq6.15nn1}) is equal to a similar product to (\ref{eq6.10}) for $\TQu_k$ but with larger powers of the $q$'s.

Now we introduce the symmetric operad of \TsediPRODterm operations, which we denote by
$\OpADiff$ $=$ $\left(\OpADiff (n)\right)_{n \,=\, 1}^{\infty}$.
We set:
\beq\label{eq6.13}
\OpADiff (n) \,:=\,
\ADfrOp{\,\mu_n}
\bigl(
\Oalg_n,\, \Salg
\bigr)
\,,
\eeq
where the permutation group $\PermGr_n$ acts again via its action on 
$\Hom_{\GRF}\bigl(\Oalg_n,$ $\Salg \bigr)$ that is induced 
by the natural action of $\PermGr_n$ on $\Oalg_n$ (i.e., permutation of the arguments of the functions $\in \Oalg_n$).
The operadic compositions~are
\beqa\label{eq6.6v1-1}
\OperCmp{j_1,\dots,j_k}
:&
\hspace{-0pt}
\OpADiff(k) \otimes \OpADiff(j_1) \otimes \cdots \otimes \OpADiff(j_k)
\hspace{-0pt}
&
\to
\OpADiff(n)
\bnn
&
\hspace{16pt}
\APDOa \otimes \APDOa'_1 \otimes \cdots \otimes \APDOa'_k
&
\mapsto
\OperComp{j_1,\dots,j_k}{\APDOa}{\APDOa'_1,\dots,\APDOa'_k}
\,,
\bnn 
& &
\phantom{\,\mapsto\,}
:=\,
\APDOa
\circ \EXTOP{(\APDOa'_1 \otimes \cdots \otimes \APDOa'_k)}
\,,
\eeqa
for $n,k,j_1,\dots,j_k$ $\in$ $\{1,2,\dots\}$, $n=j_1+\cdots+j_k$;
here:
(1)
$\APDOa'_1$ $\otimes$ $\cdots$ $\otimes$ $\APDOa'_k$ is \anTsedi operator due to Corollary~\ref{2-pr2.1} $(b)$;
(2)
the extension 
$(\APDOa'_1$ $\otimes$ $\cdots$ $\otimes$ $\EXTOP{\APDOa'_k)}$
is according to Corollary~\ref{2-pr2.1} $(c)$:
\beqa\label{eq6.21}
\podr
\APDOa'_1 \otimes \cdots \otimes \APDOa'_k
\,:\,
\Oalg_{j_1} \otimes \cdots \otimes \Oalg_{j_k}
\,\to\
\,
\underbrace{
\Salg 
\otimes \cdots \otimes
\Salg
}_{
\text{
\small 
$\Salgn{k}$}
}
\,,\,
\\
\podr
\EXTOP{(\APDOa'_1 \otimes \cdots \otimes \APDOa'_k)}
\,:\,
\left(\Oalg_{j_1} \otimes \cdots \otimes \Oalg_{j_k}\right)[1/\TQu_{j_1|\dots|j_k}]
\bnn
\podr
\phantom{\APDOa'_1 \otimes \cdots \otimes \APDOa'_k
\,:\,
\Oalg_{j_1} \otimes \cdots \otimes \Oalg_{j_k}}
\,\to\
\,
\underbrace{\Salgn{k}[1/\TQu_{k;\,j_1,\dots,j_k}]}_{\text{
\small 
$\Salgn{k}[1/\TQu_k]\,=\,\Oalg_k$}}
\,,\,
\eeqa

\vspace{-5pt}

\noindent
(cf. Eqs.~(\ref{eq6.15nn1}) and (\ref{ImpEqxx}));
finally,
recall that
the composition of \Tsedi operators is again \anTsedi operator as we established in Corollary \ref{2-pr2.1}~$(a)$.

As $\OpADiff(1)$ $=$ $\OpDiff(1)$ $\equiv$
$\DfrOp{}{\,\Salg}(\Salg,\Salg)$ we have the same operadic unit $1_{\Salg}$ $\in$ $\DfrOp{}{\,\Salg}(\Salg,\Salg)$. 
This completes the construction of the operad $\OpADiff$.

The proof that this is  indeed a (symmetric) operad is straightforward and we present it only schematically.
The permutation equivariance is manifest in the construction.

Before continuing with the operadic associativity laws we make
an important observation. 

\medskip

\begin{THEOR}{SupOpera}
{\it
When we restrict 
the \Tsedi operators 
$\Oalg_n$ $\to$ $\Salg$ from 
$\Oalg_n$ to its $\Salgn{n}$--submodule
$\Salgn{n}$ we get differential operators
$\Salgn{n}$ $\to$ $\Salg$
$($cf.~Proposition~\ref{pr4.3}$)$.
In this way we obtain a linear surjection
\beq\label{SupOper}
\begin{array}{ccl}
\OpADiff (n) & \repimorf & \OpDiff (n)
\\
\UPin & & \UPin
\\ 
\APDOa & \ \longmapsto\ & 
\APDOa\hspace{1pt}\bigl|_{\,\Salgn{n}}
\ \,,
\end{array}
\eeq
which \ makes \ $\OpDiff$ a \ \THEemph{sup}operad \ of \ $\OpADiff$.}
\end{THEOR}

\medskip

{\it Indeed}, when we restrict the composition (\ref{eq6.6v1-1}) on $\Salgn{n}$ the extensions $\EXTOP{(\cdots)}$ will be removed and we end up with the composition law (\ref{eq6.6}).

\medskip

Now, the operadic associativity laws state:
\beqa\label{eq6.23nn}
\podr
\Bigl(
\APDOa \, \circ \,
\EXTOP{\Bigl(\mathop{\bigotimes}\limits_{i'} \APDOa_{i'}'\Bigr)}
\Bigr)
\circ\,
\EXTOP{\Bigl(\mathop{\bigotimes}\limits_{i',\,i''} \APDOa_{i',\,i''}''\Bigr)}
\nnb
\podr=\,
\APDOa 
\, \circ \,
\EXTOP{\Bigl(\mathop{\bigotimes}\limits_{i'}
\
\Bigl(
\APDOa_{i'}' 
\circ\,
\EXTOP{\Bigl(\mathop{\bigotimes}\limits_{i''}
\APDOa_{i',\,i''}''\Bigr)}\Bigr)
\Bigr)}
\,,
\quad
\eeqa
where $i'$ runs over $\{1,\dots,k\}$, $i''$ runs over $\{1,\dots,j_{i'}\}$ (the range depending on $i'$).
We use the following integral partitions:
\beq\label{subpatitions}
n\,=\,j_1\,+\,\cdots\,+\,j_k\,,\quad 
N\,=\,J_1\,+\,\cdots\,+\,J_k\,,\quad 
J_{i'}\,=\,\ell_{i',1}\,+\,\cdots\,+\,\ell_{i',j_{i'}}\,,\quad 
\eeq
so that both sides of Eq.~(\ref{eq6.23nn}) belong to $\OpADiff(N)$,
$\APDOa \in \OpADiff (k)$,
$\APDOa_{i'}'  \in  \OpADiff (j_{i'})$ and
$\APDOa_{i',\,i''}'' \in \OpADiff(\ell_{i',i''})$.
One proves Eq.~(\ref{eq6.23nn}) by showing that both sides are equal~to:
\beq\label{zzziii}
\hspace{-0pt}
\APDOa 
\,\circ\,
\EXTOP{\Bigl(\mathop{\bigotimes}\limits_{i'} \APDOa_{i'}'\Bigr)}
\,\circ \,
\EXTOP{\Bigl(\mathop{\bigotimes}\limits_{i',\,i''} \APDOa_{i',\,i''}''\Bigr)}
\,
\eeq
(that is obvious for the left hand side of (\ref{eq6.23nn})).
To this end one notes first that if we remove all the extensions then we obtain the identity:
\beqa\label{zzziii1}
&
\hspace{-0pt}
\Bigl(
\APDOa 
\,\circ\,
\Bigl(\mathop{\bigotimes}\limits_{i'} \APDOa_{i'}'\Bigr)
\Bigr)
\,\circ \,
\Bigl(\mathop{\bigotimes}\limits_{i',\,i''} \APDOa_{i',\,i''}''\Bigr)
\, = \,
\APDOa 
\,\circ\,
\Bigl(
\Bigl(\mathop{\bigotimes}\limits_{i'} \APDOa_{i'}'\Bigr)
\,\circ \,
\Bigl(\mathop{\bigotimes}\limits_{i',\,i''} \APDOa_{i',\,i''}''\Bigr)
\Bigr)
& \nnb &
= \,
\APDOa 
\,\circ\,
\Bigl(\mathop{\bigotimes}\limits_{i'} \APDOa_{i'}'\Bigr)
\,\circ \,
\Bigl(\mathop{\bigotimes}\limits_{i',\,i''} \APDOa_{i',\,i''}''\Bigr)
\,,
&
\eeqa
and this is what happens when we restrict Eqs.~(\ref{eq6.23nn}) and (\ref{zzziii}) on
\beq\label{vkbfw}
\mathop{\bigotimes}\limits_{i',\,i''} 
\,
\Oalg_{\,\ell_{i',\,i''}}
\, \subseteq \,
\Oalg_N
\,
\eeq
(i.e., we consider them just as identities of $\GRF$--linear maps).
It remains to pass to the full domain for Eq.~(\ref{eq6.23nn}), which is the right hand side $\Oalg_N$ in (\ref{vkbfw}).
But $\Oalg_N$ is obtained by localization with
$\TQu_{\ell_{1,1}|\dots|\ell_{1,\ell_1}|\dots|\ell_{k,\ell_k}}$.
Uniqueness of the extensions established in Corollary~\ref{2-pr2.1} $(c)$ guarantees that the equality (\ref{eq6.23nn}) is extended on $\Oalg_N$.
In addition, one should take into account the identity
$$
\TQu_{\ell_{1,1}|\dots|\ell_{1,\ell_1}|\ldots|\ell_{k,1}|\dots|\ell_{k,\ell_k}}
\,=\,
\TQu_{J_1|\dots|J_k}
\cdot\
(
\TQu_{\ell_{1,1}|\dots|\ell_{1,j_1}}
\otimes \cdots \otimes
\TQu_{\ell_{k,1}|\dots|\ell_{k,j_k}}
)
\,,
$$
as in the right hand side of (\ref{eq6.23nn}) we have successive localizations.

\section{Associative algebras with derivations, \DMoD--modules and $\OpDiff$--algebras}\label{Se8nn}
\SETCNTR

According to the operad theory (cf.~\cite[Sect.~5.2.12]{LV12}) an algebra $V$ over a symmetric operad $\OpGen$ $=$ $(\OpGen (n))_{n \, = \, 1}^{\infty}$ 
is a vector space $V$ together with an operadic morphism 
\beq\label{eq7.1nn913-1}
\OpGen \, \to \, \OpEnd_V \,:\,
\OpGen (n) \,\to\, \OpEnd_V (n)
\quad (n\,=\,1,2,\dots)
\,.
\eeq
Recall that the $n$-th operadic space of the endomorphism operad $\OpEnd_V$ is $\OpEnd_V\hspace{-1.7pt} (n)$ $=$ $\Hom_{\GRF}(V^{\otimes \, n},$ $V)$, i.e., the space of all possible $n$-linear operations on $V$ and so, the map (\ref{eq7.1nn913-1}) assigns to every abstract $n$-ary operation $\in \OpGen (n)$ an actual $n$-linear operation on $V$.
The class of algebras over the operad $\OpGen$ form a category and in this way, one can think of an operad as presenting a type (i.e., a category) of algebras (as we mentioned in Sect.~\ref{ssec1.2nn0913n1}).

Our main interest in this paper is on what kind of algebras are the algebras over the operad $\OpADiff$ constructed in Sect.~\ref{s6.2}.
However, the operadic morphism constructed in Theorem~\ref{SupOpera} makes every algebra over the operad $\OpDiff$ (of Sect.~\ref{s6.1}) also an algebra over $\OpADiff$.
Indeed, composing the sequence of operadic morphisms $\OpDiff$ $\to$ $\OpADiff$ $\to$ $\OpEnd_V$ we obtain a $\OpDiff$--algebra $\OpDiff$ $\to$ $\OpEnd_V$.
So, we first look at the algebras over $\OpDiff$ and then, the algebras over $\OpADiff$ can be thought of as a sort of ``extension'' of the class of $\OpDiff$--algebras  (e.g., by a``deformation'' of the latter algebras).

As we shall explain now the algebras over the operad $\OpDiff$ are the so called \REMemph{\DMoD--module commutative associative algebras}.
To begin with, notice that for every operad $\OpGen$ the first operadic space $\OpGen(1)$ is an associative algebra (in general, noncommutative) and for every $\OpGen$--algebra $V$, the linear space $V$ becomes a module of this algebra.
For the operad $\OpDiff$, the associative algebra 
$$
\OpDiff(1) \,=\, \DfrOp{}{\GRF[\x]} (\GRF[\x]\,,\,\GRF[\x])
$$
is the so called also the \REMemph{Weyl algebra}.
Modules of Weyl algebra are called in mathematics also \DEFemph{\DMoD--modules}.
In particular, on a \DMoD--module we have an action of the commutative (associative) algebra of polynomials 
$\GRF[\x]$ ($=$
$\GRF[x^1,\dots,x^{\DMN}]$)
and also, we have mutually commuting endomorphisms $\Drv_1$, $\dots,$ $\Drv_{\DMN}$ of $V$ satisfying the Leibniz rule 
\beq\label{eq9.2nn0913nn1}
\Drv_{\alpha}(x^{\beta} \cdot \FldEla)
\,=\, 
\Drv_{\alpha}(x^{\beta}) \cdot \FldEla + 
x^{\beta} \cdot \Drv_{\alpha}(\FldEla)
\quad \text{and} \quad
\Drv_{\alpha}(x^{\beta}) \, = \, \delta_{\alpha}^{\beta} \, 1
\eeq
($\delta_{\alpha}^{\beta}$ being the Kr\"onecker delta symbol)
for every $\FldEla \in V$.
The relations (\ref{eq9.2nn0913nn1}) completely determine a \DMoD--module.
However, an algebra $V$ over the operad $\OpDiff$ is more than a \DMoD--module.
As we mentioned at the end of Sect.~\ref{s6.1} the operad $\OpDiff$ contains as a \REMemph{suboperad} the  operad $\OpCom$ of commutative associative algebras.
It follows that under the inclusion (\ref{eq9.3nn0913nn1}) every $\OpDiff$--algebra $V$ becomes a commutative associative algebra.
According to operadic relations~(\ref{NewRel-0913-0})--(\ref{NewRel-0913-2}) between the operadic elements (\ref{NewGen-0913-0}), (\ref{NewGen-0913}) we get that
\beqa\label{Leib-0913-2}
&
\Drv_{\alpha} (\FldEla\FldElb) \,=\,
\Drv_{\alpha}(\FldEla)\,\FldElb+\FldEla\,\Drv_{\alpha}(\FldElb)
\,, &
\\ \label{Leib-0913-2a}
&
x^{\alpha} \cdot (\FldEla \, \FldElb)
\,=\,
(x^{\alpha} \cdot \FldEla) \, \FldElb
\,=\,
\FldEla \, (x^{\alpha} \cdot \FldElb)
&
\eeqa
$(\alpha=1,\dots,\DMN,$ $\FldEla,\FldElb \in V)$.
By definition, a \DMoD--module $V$, which is also a commutative associative algebra that satisfy the relations (\ref{Leib-0913-2}) is called \DEFemph{\DMoD--module commutative associative algebra}
(over the Weyl algebra
$\DfrOp{}{\GRF[\x]} (\GRF[\x],$ $\GRF[\x])$). 

Thus, let us summarize

\medskip

\begin{COROL}{Cr8.1hjgfhj}
{\it 
The category of all algebras over the operad $\OpDiff$ can be naturally identified with the category of the \DMoD--module commutative associative algebras over the Weyl algebra
$\DfrOp{}{\GRF[\x]} (\GRF[\x],$ $\GRF[\x])$.}
\end{COROL}

\medskip

Dropping the 
generators $x^{\alpha}$ $\in$ $\OpDiff(1)$ (Eq.~(\ref{NewGen-0913-0})) and keeping only $\Drv_{\alpha}$ $\in$ $\OpDiff(1)$ and $\ProdOper_n$ $\in$ $\OpDiff(n)$, subject to the relations of Eqs.~(\ref{diagon-11-913}) and~(\ref{NewRel-0913})
we get another suboperad of $\OpDiff$ that contains $\OpCom$.
This is the operad $\OpComDer$ of \REMemph{commutative associative algebras with derivations}.
The latter type of algebras contain only actions of mutually commuting endomorphisms $\Drv_1,$ $\dots,$ $\Drv_{\DMN}$ that satisfy the Leibniz rule (\ref{Leib-0913-2}).
The obtained operad was described by Loday in \cite{L10} as a \REMemph{nonsymmetric} operad under the name $\textit{AsDer}$.
The latter operad is later generalized in \cite{L10} by extending it to more general operads that include also ``integration'' operation.
However, in the present paper we shall be interested mainly in another extension of the operad of associative algebras with derivations: the operad $\OpADiff$ of \TsediPRODterm operations.
In the next section we shall identify the suboperad $\OpComDer$ of $\OpDiff$ with the suboperad of \REMemph{translation invariant} differential\PRODterm operations.

\section{Separation of translations. Translation invariance}\label{Se9nn}
\SETCNTR

We shall introduce now two subspaces
\beq\label{eq9.1}
\hspace{5pt}
\TISalgn{n}
\,\subset\,
\Salgn{n}
\,(\,=\, \GRF[\x_1,\dots,\x_n])
\,,\quad
\TIOalgn{n}
\,\subset\,
\Oalg_n
\,(\,=\, \Salgn{n}[1/\TQu_n])
\eeq
(recall Eqs.~(\ref{e6.3-1}) and (\ref{eq6.12}))
such that
$\TISalgn{n}$ is a unital subalgebra of $\Salgn{n}$ and 
$\TIOalgn{n}$ is an $(\TISalgn{n})$--submodule of $\Oalg_n$.
This will be the spaces of \REMemph{translation invariant} (``t.i.'') elements.

To this end let us set for a given $n = 2,3,\dots$:
\beq\label{eq9.2}
\DXn{j} \,:=\, \x_j -\x_n
\quad
(j \,=\, 1,\dots,n-1)
\,,\quad
\TISalgn{n}
\,:=\,
\GRF[\DXn{1},\dots,\DXn{n-1}]
\,\cong\,
\Salgn{n-1}
\ \
\eeq
(i.e., $\DXn{j}$ $:=$ $(\Dxn{j}^{\alpha})_{\alpha \,=\, 1}^{\DMN}\,$,
$\Dxn{j}^{\alpha} := x_j^{\alpha}-x_n^{\alpha}$).
Assuming further that
\beq\label{eq9.3}
\BQu_{1,2} \,\in\, \TISalgn{2}
\,,\quad
\text{i.e.,}
\quad
\BQu_{1,2} (\x_1,\x_2)
\,=\,
\HQu (\x_1-\x_2)
\,,
\eeq
where $\HQu \in \Salg$, we 
have that
$$
\TQu_n 
\,=\,
\Bigl(\mathop{\prod}\limits_{j \,=\, 1}^{n-1}
\HQu (\DXn{j})\Bigr)
\Bigl(\mathop{\prod}\limits_{1 \,\leqslant \, k \,<\, \ell \,\leqslant n-1}
\HQu (\DXn{k}-\DXn{\ell})\Bigr)
\,\in\, \TISalgn{n}
$$
and so we can set also
\beq\label{eq9.4}
\TIOalgn{n}
\,:=\,
\TISalgn{n}[1/\TQu_n]
\,.
\eeq

Note that $\TISalgn{n}$ and $\TIOalgn{n}$ consists of all translation invariant elements of $\Salgn{n}$ and $\Oalg_n$, respectively, i.e.,
\beqa\label{eq9.5}
\TISalgn{n}
=\,
\bigl\{&\hspace{-2pt}
F \,\in\,
\Salgn{n}
&\bigl|\,
\bigl(\di_{x_1^{\alpha}} + \cdots + \di_{x_n^{\alpha}}\bigr) \, F = \, 0
\ \
(\alpha = 1,\dots,\DMN)
\bigr\}
\,,
\nnb
\TIOalgn{n}
=
\bigl\{&\hspace{-2pt}
G \,\in\,
\Oalg_n
&\bigl|\,
\bigl(\di_{x_1^{\alpha}} + \cdots + \di_{x_n^{\alpha}}\bigr) \, G = \, 0
\ \
(\alpha = 1,\dots,\DMN)
\bigr\}
\,
\qquad
\eeqa
$(\di_{x_j^{\alpha}} = \TXDfrac{\di}{\di x_j^{\alpha}})$.
It follows that $\TISalgn{n}$ and $\TIOalgn{n}$ stay invariant under the actions of the permutation group $\PermGr_n$ on $\Salgn{n}$ and $\Oalg_n$, respectively.
Furthermore, as the operadic compositions (\ref{eq6.6}) and (\ref{eq6.6v1-1}) are translation equivariant we get

\medskip

\begin{THEOR}{Th9.1}
{\it
The vector spaces
\beqa\label{eqn9.6khdlea}
\TIOpDiff (n)
\,=\,
\bigl\{&\hspace{-2pt}
\PDOa \,\in\,
\OpDiff (n)
&\bigl|\,
\PDOa\bigl((\di_{x_1^{\alpha}} + \cdots + \di_{x_n^{\alpha}}) \, F\bigr) = \, 
\di_{x^{\alpha}} \, \PDOa(F)
\ \
(\forall \alpha)
\bigr\}
\,,
\nnb
\TIOpADiff (n)
\,=\,
\bigl\{&\hspace{-2pt}
\APDOa \,\in\,
\OpADiff (n)
&\bigl|\,
\APDOa\bigl((\di_{x_1^{\alpha}} + \cdots + \di_{x_n^{\alpha}}) \, G\bigr) = \, 
\di_{x^{\alpha}} \, \APDOa(G)
\ \
\nnb
&&
\phantom{
\APDOa\bigl((\di_{x_1^{\alpha}} + \cdots + \di_{x_n^{\alpha}}) \, G\bigr)}
(\alpha = 1,\dots,\DMN)
\bigr\}
\eeqa
form suboperads $\TIOpDiff$ and $\TIOpADiff$ in
$\OpDiff$ and $\OpADiff$, respectively.
The operadic epimorphism~$(\ref{SupOper})$ restricts to: 
\beq\label{SupOper-1}
\TIOpADiff \, \repimorf \, \TIOpDiff
\eeq
$($again an operadic epimorphism$)$.}
\end{THEOR}

\medskip

\begin{PROPO}{pr9.2-0916n}
{\it
The suboperad $\TIOpDiff$ of $\OpDiff$ is generated, under the operadic compositions, by the elements $\Drv_1,$ $\dots,$ $\Drv_{\DMN}$ $\in$ $\OpDiff(1)$ $($Eq.~$\ref{NewGen-0913})$ and $\ProdOper_n$ $\in$ $\OpDiff(n)$ for $n = 2,3,\dots$.}
\end{PROPO}

\medskip

\noindent
{\it Proof.}
According to Eq.~(\ref{Th9.1}) and Eq.~(\ref{repU-0913}), one has that $\PDOa$ $\in$ $\TIOpDiff(n)$ iff the coefficients $f_{\r_1,\dots,\r_n}$ in Eq.~~(\ref{repU-0913}) are constant functions for all multi-indices $\r_1,$ $\dots,$ $\r_n$.
Then the proposition follows by the operadic composition law~(\ref{eq6.6}) since
$$
\di_{\x_1}^{\,\r_1} \cdots \di_{\x_n}^{\,\r_n}
\Bigl|_{\x_1 \,=\, \cdots \,=\, \x_n \,= \, \x}
\, = \,
\OperComp{1,\dots,1}{\ProdOper_n}{\Drv^{\r_1},\dots,\Drv^{\r_n}}
\,,
$$
and
$\Drv^{\r_j}$ $=$ $\OperComp{1}{\Drv_{\beta_1}}{\OperComp{1}{\Drv_{\beta_2}}{\dots}}$.$\quad\Box$

\medskip

In order the describe explicitly the structure of the spaces $\TIOpDiff(n)$ and $\TIOpADiff (n)$
let us introduce for every $n=2,3,\dots$: 
\beq\label{eq9.8}
\Salgn{n} \,=\,
\GRF[\x_1,\dots,\x_n]
\,=\,
\GRF[\DXn{1},\dots,\DXn{n-1},\x_n]
\,\cong\,
\TISalgn{n} \otimes \Salg
\,.
\eeq
Then, the multiplication homomorphism $\mu_n$,
\beqa\label{eq9.9}
\TISalgn{n} \otimes \Salg
&\ \cong\,
\Salgn{n}
\ \mathop{\longrightarrow}\limits^{\mu_n}\
\Salg
\,\cong\ &
\GRF \otimes \Salg
\quad
\text{factorizes}
\quad
\nnb
\TISalgn{n} \otimes \Salg
&\ 
\mathop{-\hspace{-5pt}-\hspace{-5pt}-\hspace{-5pt}-\hspace{-5pt}-\hspace{-5pt}\longrightarrow}\limits_{\AUGM \,\otimes\, \textrm{id}}\ &
\GRF \otimes \Salg
\,,
\eeqa
where $\AUGM$ is the
the character
\beq\label{eq9.10}
\AUGM
\,:\,
\TISalgn{n} \,\to\, \GRF
\,:\,
F \,\mapsto\, F \left.\raisebox{-4pt}{\hspace{-1pt}}\right|_{\DXn{1} \,=\,\cdots\,=\, \DXn{n-1} \,=\, 0}
\,
\eeq
($\AUGM$ is also an augmentation of $\TISalgn{n}$ in the sense of Definition~\ref{df2.2}(c)).
We shall consider the isomorphism
$\Salgn{n}$ $\cong$ $\TISalgn{n} \otimes \Salg$ of Eq.~(\ref{eq9.8}) together with the induced isomorphism
\beq\label{eq9.11}
\Oalg_n \,\cong\, \TIOalgn{n} \otimes \Salg
\,
\eeq
as identifications (with a slight abuse in the notations).

\medskip

\begin{THEOR}{th9.xx-0915}
{\it
Under the above identification we have that the natural linear isomorphisms of the type:
\beq\label{eq9.12}
\begin{array}{rcl}
\Hom_{\GRF} (U \otimes V , W)
& \ \cong \ &
\Hom_{\GRF} \bigl(U,\Hom_{\GRF}(V , W)\bigr)
\\
\UPin \raisebox{10pt}{}\hspace{51pt} & & \hspace{19pt}\UPin
\\
\bigl(u \otimes v \,\mathop{\longmapsto}\limits^{A}\, A(u \otimes v) \bigr)
&
\mapsto
&
\bigl(
u \,\mathop{\longmapsto}\limits^{\TRANS{A}}\, \bigl(v \,
\raisebox{1.5pt}{\text{\tiny $|$}}
\hspace{-4pt}-\hspace{-5pt}-\hspace{-5pt}-\hspace{-5pt}-\hspace{-5pt}\rightarrow
{\hspace{-27pt}}^{\TRANS{A}(u)\raisebox{-4.8pt}{}}
\hspace{9pt} A(u \otimes v)\bigr)
\bigr)
\,,
\end{array}
\eeq
for arbitrary vector spaces $U,V,W$, restrict to isomorphisms:
\beqa\label{eq9.13}
&
\OpDiff (n)
\ (\,=\,\DfrOp{}{\mu_n} (\Salgn{n},\,\Salg)\,)
& \cong\,
\DfrOp{}{\AUGm} \bigl(\TISalgn{n},\DfrOp{}{\Salg}(\Salg,\Salg)\bigr)
\\ \label{eq9.14}
&
\OpADiff (n)
\ (\,=\,\DfrOp{}{\mu_n} (
\Oalg_n
,\,\Salg)\,)
& \cong\,
\ADfrOp{\AUGm} \bigl(\TIOalgn{n},\DfrOp{}{\Salg}(\Salg,\Salg)\bigr)
\,.
\eeqa}
\end{THEOR}

In order to 
{\it prove} the above isomorphisms we start first with establishing Eq. (\ref{eq9.13}).
If we take the representation (\ref{repU-0913}) of an element $\PDOa$ $\in$ $\OpDiff (n)$ and make the change of variables
$(\x_1,$ $\dots,$ $\x_n)$ $\to$ $(\DXn{1},$ $\dots,$ $\DXn{n-1},$ $\x_n)$
then $\PDOa$ can be viewed as a $\AUGM$--differential operator with coefficients in $\DfrOp{}{\Salg}(\Salg,\Salg)$.
This is exactly what is stated in 
(\ref{eq9.13}).
Under the above arguments we get also 
an isomorphism
\beq\label{eq9.15}
\OpDiff (n)
\ (\,=\,\DfrOp{}{\mu_n} (\Salgn{n},\,\Salg)\,)
\,\cong\,
\DfrOp{}{\AUGm} \bigl(\TISalgn{n},\GRF\bigr)
\otimes
\DfrOp{}{\Salg}(\Salg,\Salg)
\,.
\eeq
Next, let us write the map of Eq.~(\ref{eq9.14}) in an explicit form
(according to the general correspondence in (\ref{eq9.12})):
\beq\label{eq9.16}
\begin{array}{ccl}
\OpADiff (n) & 
\longrightarrow
& 
\ADfrOp{\AUGm} \bigl(
\TIOalgn{n}
,\,
\DfrOp{}{\Salg} (\Salg,\Salg)
\bigr)
\\
\UPin & & \hspace{5pt}\UPin \raisebox{10pt}{}
\\ 
\APDOa & \ \longmapsto\ & 
\Bigl(
\TIOalgn{n}\,\ni\, G
\,\mathop{\longmapsto}\limits^{\TRANS{\APDOa}}\,
\bigl(\DMo{\APDOa}{G \,\otimes\, 1}
\left.\raisebox{-4pt}{\hspace{-1pt}}\right|_{\,1\,\otimes\, \Salg\,}
\bigr)
\,\in\,
\DfrOp{}{\Salg} (\Salg,\Salg)
\Bigr)
\ \,
\end{array}
\eeq
(note that 
$\TRANS{\APDOa}(G)(F')$ $=$
$\APDOa\bigl(G \otimes F'\bigr)$ $=$
$\DMo{\APDOa}{G \,\otimes\, 1}(1 \otimes F')$
since by Eq.~(\ref{DMo-def}),\\
$\DMo{\APDOa}{G \,\otimes\, 1}(F \otimes F')$ $=$
$\APDOa\bigl((F\cdot G) \otimes F'\bigr)$, for
$F \in \TISalgn{n}$, $G \in \TIOalgn{n}$ and $F' \in \Salg$).
In~order to prove that the above assignment
$\TRANS{\APDOa}$ $:$ $G$ $\mapsto$ 
\(\bigl(\DMo{\APDOa}{G \,\otimes\, 1}
\left.\raisebox{-4pt}{\hspace{-1pt}}\right|_{\,1\,\otimes\, \Salg\,}
\bigr)\)
is indeed an element of
$\ADfrOp{\AUGm} \bigl(\TIOalgn{n},$
$\DfrOp{}{\Salg} (\Salg,\Salg)\bigr)$
we need to prove
that for every $G \in \TIOalgn{n}$ we have
\beqs
\podr
\DMo{\TRANS{\APDOa}}{G}  
\,\in\,
\DfrOp{}{\AUGm} \bigl(\TIOalgn{n},
\DfrOp{}{\Salg} (\Salg,\Salg)\bigr)
\,,\quad\text{where}\quad
\\
\podr
\DMo{\TRANS{\APDOa}}{G} 
\,:\,
F 
\,\mapsto\,
\Bigl(
\TRANS{\APDOa}(FG)
\,:\,
F' 
\,\mapsto\,
\DMo{\APDOa}{G\,\otimes\,1}
(F\otimes F')
\Bigr)
\,
\eeqs
($F \in \TISalgn{n}$, $F' \in \Salg$).
But $\DMo{\APDOa}{G\,\otimes\,1}$ $\in$
$\DfrOp{}{\,\AUGM \,\otimes\, \textrm{id}} (\TISalgn{n}\otimes\Salg,$ $\Salg)$
$\cong$
$\DfrOp{}{\,\mu_n} (\Salgn{n},$ $\Salg)$
and under the isomorphism (\ref{eq9.13}) we have
$\DMo{\TRANS{\APDOa}}{G}$ $=$
$\TRANs{\DMo{\APDOa}{G\,\otimes\,1}}$.
The latter equality guarantees also that if conversely one has an element
$\TRANS{\APDOa}$ $\in$ $\DfrOp{}{\AUGm} \bigl(\TIOalgn{n},$
$\DfrOp{}{\Salg} (\Salg,$ $\Salg)\bigr)$ then it corresponds to
$\APDOa$ $\in$ $\OpADiff (n)$
$(=\,\DfrOp{}{\mu_n} (\Oalg_n,$ $\Salg))$.

\medskip

This completes the proof of Theorem~\ref{th9.xx-0915}.$\quad\Box$

\medskip

\begin{THEOR}{TITH}
{\it
The isomorphism $(\ref{eq9.14})$ restricts to a natural isomorphism:
\beqa\label{eq9.17}
\TIOpADiff (n)
\,\cong\podr
\ADfrOp{\AUGm}
\bigl(
\TIOalgn{n}
,\,
\GRF[\TRa]
\bigr)
\bnn
\ (\ \equiv\podr
\ADfrOp{\AUGm}
\bigl(
\GRF[\DXn{1},\dots,\DXn{n-1}]\bigl[1/\TQu_n\bigr]
,\,
\GRF[\TRa]
\bigr)
\,)
\,,
\eeqa
where $\GRF[\TRa] = \GRF[\Tra_1,\dots,\Tra_{\DMN}]$ is a polynomial algebra over the set of new formal variables $\TRa$ $=$ $(\Tra_1,\dots,\Tra_n)$.}
\end{THEOR}

\medskip

\noindent
{\it Proof.}
We note first that if $\APDOa \in \TIOpADiff (n)$ it is mapped via (\ref{eq9.16}) to a map
$\TRANS{\APDOa}$ 
that assigns to every $G \in \TIOalgn{n}$ a differential operator
$\TRANS{\APDOa}(G)$ $\in$
$\DfrOp{}{\Salg}(\Salg,\Salg)$, which commutes with every partial derivative $\di_{x^{\alpha}}$ on $\Salg$.
Hence, $\TRANS{\APDOa}(G)$ is a polynomial in $\di_{x_1}$ $=:$ $\Tra_1$, $\dots,$ $\di_{x^{\DMN}}$ $=:$ $\Tra_{\DMN}$ with constant coefficients (i.e., $\in \GRF$).$\quad\Box$

\medskip

\begin{COROL}{Cr9.3}
{\it
$(a)$
The isomorphisms $(\ref{eq9.17})$ and $(\ref{eq9.14})$ can be transformed to the following form$:$\footnote{%
``loc. fin.''  stands for ``locally finite'', i.e., the infinite sum
$\sum_{\r \,\geqslant\, 0}
\TRa^{\r} \otimes \APDOa_{\r}$ becomes finite when is applyed to some $G$ $\in$ $\TIOalgn{n}$}
\beqa\label{ewwejnxd}
\podr
\TIOpADiff (n)
\,\cong\,
\bigl(
\ADfrOp{\AUGm}
(
\TIOalgn{n}
,\,
\GRF)
\bigr)
[[\TRa]]^{\text{\rm loc. fin.}}
\bnn
\podr
:=\,
\biggl\{
\mathop{\sum}\limits_{\r \,\geqslant\, 0}
\TRa^{\r} \otimes \APDOa_{\r}
\,
\biggl|
\,
\APDOa_{\r} \,\in\,
\ADfrOp{\AUGm}
(
\TIOalgn{n}
,\,
\GRF)
\,,\
\forall G \in\TIOalgn{n}:\,
\APDOa_{\r} (G) = 0 
\ \text{if} \ \r >\!\!>0 
\biggr\}
\,,\hspace{7pt}
\eeqa

\vspace{-20pt}

\beqa\label{ewwejnxd-1}
\podr
\OpADiff (n)
\,\cong\,
\bigl(
\ADfrOp{\AUGm}
(
\TIOalgn{n}
,\,
\GRF [\x])
\bigr)
[[\TRa]]^{\text{\rm loc. fin.}}
\bnn
\podr
:=\,
\biggl\{
\mathop{\sum}\limits_{\r \,\geqslant\, 0}
\TRa^{\r} \otimes \APDOa_{\r}
\,
\biggl|
\,
\APDOa_{\r} \,\in\,
\ADfrOp{\AUGm}
(
\TIOalgn{n}
,\,
\GRF[\x])
\,,\
\forall G \in\TIOalgn{n}:\,
\APDOa_{\r} (G) = 0 
\ \text{if} \ \r >\!\!>0 
\biggr\}
\,
\eeqa
$($with the multiindex notations$:$ $\r := (r_1,\dots,r_{\DMN})$ $\in$ $\{0,1,\dots\}^{\DMN}$, $\r!$ $:=$ $r_1!$ $\cdots$ $r_{\DMN}!$, $\TRa^{\r}$ $:=$ $\Tra_1^{r_1}$ $\cdots$ $\Tra_{\DMN}^{r_{\DMN}}$$)$.

$(b)$
Under the isomorphism $(\ref{eq9.17})$ the permutation group $\PermGr_n$ acts on $\TIOpADiff(n)$ in the following way.
Let $G \in \TIOalgn{n}$ and $G$ $=$ $G (\DXn{1},\dots,\DXn{n-1})$ $($cf. Eq.~$(\ref{eq9.2})$$)$ and 
let $\APDOa \in \TIOpADiff (n)$.
Then the action of a permutation $\sigma \in \PermGr_n$ that permutes only the indices $\{1,$ $\dots,$ $n-1\}$
is induced by its action on $G$ via permuting $\DXn{1},$ $\dots,$ $\DXn{n-1}$.
For the transposition $\tau_j \in \PermGr_n$ that exchanges $j$ and $n$ one has
\beqa\label{cffctrbvhy}
\APDOa^{\tau_j} (G) (\DXn{1},\dots,\DXn{n-1})
\,=\podr
\mathop{\sum}\limits_{\r \, \geqslant \, 0}
\, \frac{1}{\r!} \
\TRa^{\r} \cdot
\APDOa\bigl(\DXn{j}^{\,\r} \cdot G'\bigr) 
\,,
\quad
\nnb
G'
\,:=\podr
G\,\Biggl|\raisebox{-8pt}{$
{\,}^{\DXn{j} \,\mapsto\,-\DXn{j}\,,}_{\DXn{\ell} \,\mapsto\, \DXn{\ell}-\DXn{j} \ (\ell \neq j)}$}
\,.
\eeqa}
\end{COROL}

\section{Vertex algebras and $\OpADiff$--algebras}\label{Se10nn}
\SETCNTR

We shall follow the definition of a vertex algebra as given in \cite{N05} and \cite{BN06}.
In dimension $\DMN=1$ this definition is equivalent to the usual definition (see, e.g., \cite{K98}).
A brief comparison between the definitions of vertex algebras in one and higher dimensions is given in \cite[Sect.~1]{BN08}.
For the sake of completeness we shall include in Sect.~\ref{sse10.2} some basic facts about vertex algebras.
But before that we shall fix some notations.

\SSECSPA

\subsection{Preliminaries}
\ADDCONT{Preliminaries}

\noindent
We follow the notations and conventions of Sect.~\ref{s6} and \ref{Se9nn}.
In particular,
as in Sect.~\ref{s6.0}
let $\Salg$ $=$ $\GRF[\x]$.
In the theory of vertex algebras we choose the polynomial
$\BQu_{1,2} \in \Salgn{2}$ (\ref{e6.8}) to be translation invariant in the form (\ref{eq9.3}) for a \REMemph{nondegenerate quadric} $\HQu \in \Salg$, say,
\beq\label{HQu-def}
\HQu (\z) \,=\,
(z^1)^2 +\,\cdots\,+(z^{\DMN})^2
\,=:\,
\z^{\hspace{1pt}2}
\,,
\quad
\BQu_{1,2} \, = \, \HQu (\x_1-\x_2)
\,.
\eeq
Thus,
\beqa\label{eq10.2-1103}
\Oalg_n
\,=\,
\GRF [\x_1\,\dots,\x_n]
\biggl[\biggl(
\mathop{\text{\Large$\textstyle \prod$}}\limits_{1 \,\leqslant\, j \,\leqslant\, k \,\leqslant\, n} 
(\x_j-\x_k)^{\hspace{0pt}2}
\biggr)^{-1}\biggr]
\,\qquad\quad
\bnn
((\x_j-\x_k)^{\hspace{0pt}2} \,:=\,
(x_j^1-x_k^1)^2 +\,\cdots\,+(x_j^{\DMN}-x_k^{\DMN})^2
\,(=\, \BQu_{j,k}))
\,.\qquad\quad\raisebox{12pt}{}
\eeqa
For a vector space $V$ over $\GRF$ one has the identifications,\footnote{%
As in Example~\ref{ExamFPS} we write $V$ on the right hand side of the tensor product as it is the vector space that generates a free module over a commutative associative algebra acting on the left.}
\beqa\label{ident1}
\Salgn{n} \otimes V
= \hspace{-2pt}\podr
\GRF[\x_1\,\dots,\x_n] \otimes V
=
V[\x_1\,\dots,\x_n]
\,,
\\ \label{ident2}
\Oalg_n \otimes V
=
\Salgn{n}[1/\TQu_n] \otimes V
= \hspace{-2pt}\podr
\GRF[\x_1\,\dots,\x_n]
\biggl[\biggl(
\mathop{\text{\Large$\textstyle \prod$}}\limits_{1 \,\leqslant\, j \,\leqslant\, k \,\leqslant\, n} \BQu_{j,k}
\biggr)^{-1}\biggr] \otimes V
\nnb
= \hspace{-2pt}\podr
V[\x_1\,\dots,\x_n]
\biggl[\biggl(
\mathop{\text{\Large$\textstyle \prod$}}\limits_{1 \,\leqslant\, j \,\leqslant\, k \,\leqslant\, n} \BQu_{j,k}
\biggr)^{-1}\biggr]
\,.\qquad\quad
\eeqa
Here, $V[\x_1\,\dots,\x_n]$ has also meaning of the vector space of all polynomials in the formal variables of the list $\x_1,\dots,\x_n$ (cf. Eq.~(\ref{e6.3-1})) with coefficients in $V$ 
(as in Example~\ref{ExamFPS}).
Similarly,
\(
V[[\x_1\,\dots,\x_n]]
\)
stands for the space of all formal power series in $\x_1,\dots,\x_n$ with coefficients in $V$.
As in Example~\ref{ExamFPS} 
\(
V[[\x_1\,\dots,\x_n]]
\)
is a formal completion of
\(
V[\x_1\,\dots,\x_n]
\).

\medskip

\begin{THEOR}{ThmX1}
{\it
Every \Tsedi operator 
$$
\APDOa \,\in\,
\OpADiff(n)
\ (\,=\, \ADfrOp{\,\mu_n}
\bigl(\Oalg_n,\,\Salg\bigr)
)
$$
has a unique extension
$$
\APDOa_V
\,\in\,
\ADfrOp{\,\mu_n}\bigl(
V [[\x_1,\dots,\x_n]]\bigl[1/\TQu_n\bigr]
,\,V[[\x]]
\bigr)
$$
such that $\APDOa_V \left.\raisebox{-4pt}{\hspace{-1pt}}\right|\raisebox{-3pt}{\hspace{-1pt}}_{\Oalg_n \,\otimes\, V}$
$=$ $\APDOa \otimes \mathrm{id}_V$ $:$
$\Oalg_n\otimes V$ $\to$ $\Salg \otimes V$ $(\,$$=$ $V[\x]$$\,)$.}
\end{THEOR}

\medskip

\noindent
{\it Proof.}
Let us consider for arbitrary $G \in \Oalg_n$
the differential operator
$\DMo{\APDOa}{G} \otimes \mathrm{id}_V$ $:$
$V[\x_1,$ $\dots,$ $\x_n]$ $\to$ $V[\x]$
belonging to $\DfrOp{}{\,\mu_n} (V[\x_1,$ $\dots,\x_n],$ $V[\x])$.
According to Theorem~\ref{ThFS1} there is a unique extension
$\COMPLET{(\DMo{\APDOa}{G} \otimes \mathrm{id}_V)}$
$:$ $V[[\x_1,$ $\dots,$ $\x_n]]$ $\to$ $V[[\x]]$.
Now, for arbitrary $\TQu^{-N} \cdot u$ $\in$ 
$V [[\x_1,$ $\dots,$ $\x_n]]\bigl[1/\TQu_n\bigr]$
(where $u$ $\in$ $V [[\x_1,$ $\dots,$ $\x_n]]$) we set
\beq\label{eq10.1}
\APDOa_V \bigl(\TQu^{-N} \cdot u\bigr)
\,:=\,
\COMPLEt{(\DMo{\APDOa}{\TQu^{-N}} \otimes \mathrm{id}_V)}
(u)
\,.
\eeq
The \REMemph{correctness}: if
$\TQu^{-N_1} \cdot u_1$ $=$ $\TQu^{-N_2} \cdot u_2$ for $N_1 > N_2$ then we have $u_1$ $=$ $\TQu^{N_{1,2}} \cdot u_2$ with $N_{1,2}$ $=$ $N_1-N_2$.
Hence,
\beqs
\podr
\APDOa_V \bigl(\TQu^{-N_1} \cdot u_1\bigr)
\,:=\,
\COMPLEt{(\DMo{\APDOa}{\TQu^{-N_1}} \otimes \mathrm{id}_V)}
(u_1)
\nnb
\podr
=\,
\COMPLEt{(\DMo{\APDOa}{\TQu^{-N_1}} \otimes \mathrm{id}_V)}
(\TQu^{N_{1,2}} \cdot u_2)
\,=\,
\Bigl(\COMPLEt{(\DMo{\APDOa}{\TQu^{-N_1}} \otimes \mathrm{id}_V)}
\hspace{-5pt}\cdot \TQu^{N_{1,2}}
\Bigr)
(u_2)
\nnb
\podr
=\,
\COMPLEt{\bigl((\DMo{\APDOa}{\TQu^{-N_1}} \otimes \mathrm{id}_V) \cdot \TQu^{N_{1,2}}\bigr)}
(u_2)
\,=\,
\COMPLEt{(\DMo{\APDOa}{\TQu^{-N_2}} \otimes \mathrm{id}_V)}
(u_2)
\,.
\eeqs
Then, since
\beqs
\DMo{(\APDOa_V) }{\TQu^{-N} \cdot u} (F)
\,=\podr
\COMPLEt{(\DMo{\APDOa}{\TQu^{-N}} \otimes \mathrm{id}_V)}
(F \cdot u)
\nnb
\,=\podr
\DMo{\bigl(\COMPLEt{(\DMo{\APDOa}{\TQu^{-N}} \otimes \mathrm{id}_V)}
\bigr)}{u}
(F)
\eeqs
it follows that 
$\APDOa_V$
$\in$
$\ADfrOp{\,\mu_n}\bigl(V [[\x_1,$ $\dots,$ $\x_n]]\bigl[1/\TQu_n\bigr],$ $V[[\x]]\bigr)$.
Finally, for the uniqueness we note that the condition
$\APDOa_V \left.\raisebox{-4pt}{\hspace{-1pt}}\right|\raisebox{-3pt}{\hspace{-1pt}}_{\Oalg_n \,\otimes\, V}$
$=$ $\APDOa \otimes \mathrm{id}_V$
implies that Eq.~(\ref{eq10.1}) is satisfied without an extension for $u \in V[\x_1,$ $\dots,$ $\x_n]$.
Then the uniqueness of the extension in the right hand side of Eq.~(\ref{eq10.1}) implies the uniqueness of $\APDOa_V$.$\quad\Box$

\SSECSPA

\subsection{Brief reminder on vertex algebras}\label{sse10.2}
\ADDCONT{Brief reminder on vertex algebras}

\noindent
For the sake of simplicity we shall restrict our considerations only to purely bosonic (i.e., even) vertex algebras but this is not an important restriction.
According to \cite[Definition~2.1]{N05} and \cite[Definition~4.1]{BN06} it is a vector space $\VA$ equipped with a map
\beq\label{101-n1}
\VA \otimes \VA \to \VA[[\z]][1/\HQu] \,(\, = \, \VA[[\z]][1/\z^{\hspace{1pt}2}])
\,,\quad
a \otimes b \mapsto \YB(a,\z)b
\eeq
($\HQu = \z^{\hspace{1pt}2}$).
Additionally, there is a set of mutually commuting linear endomorphisms $\Tra_{\alpha} : \VA \to \VA$ ($\alpha = 1,\dots,\DMN$) and a given vector $\vac \in \VA$.

According to the theory of vertex algebras (cf.~\cite[Theorem~5.1]{BN06}), 
for every $a_1,$ $\dots,$ $a_n$ $\in$ $\VA$ ($n=2,3,\dots$)
there is an element
\beq\label{101-n2}
\YM_{a_1,\dots,a_n} (\DXn{1},\dots,\DXn{n-1})
\,\in\,
\VA [[\DXn{1},\dots,\DXn{n-1}]][1/\TQu_n]
\,,
\eeq
where we have embedded
\beq\label{101-n3}
\VA [[\DXn{1},\dots,\DXn{n-1}]][1/\TQu_n]
\,\hookrightarrow\,
\VA [[\x_1,\dots,\x_n]][1/\TQu_n]
\eeq
according to Eq.~(\ref{eq9.2})
(remember, $\TQu_n$ depends on $\DXn{1},$ $\dots,$ $\DXn{n-1}$ trough the variables $\x_1,$ $\dots,$ $\x_n$).
For $n=2$,
\beq\label{101-n8}
\YM_{a,b} (\z) \,\equiv\, \YB(a,\z)b
\eeq
and for $n>2$ ,
$\YM_{a_1,\dots,a_n} (\DXn{1},$ $\dots,$ $\DXn{n-1})$ is related to
$\YB(a_j,\DXn{j})a_{j+1}$
via \REMemph{successive expansions} in $\DXn{1},$ $\dots,$ $\DXn{n-1}$.
Further properties are:
(1) the \REMemph{permutation equivariance}:
\beq\label{101-n10}
\YM_{a_1,\dots,a_n} (\DXn{1},\dots,\DXn{n-1})
\,=\,
\YM_{a_{\sigma(1)},\dots,a_{\sigma(n)}} (\z'_1,\dots,\z'_{n-1})
\,,\quad
\z'_j \,:=\, \x_{\sigma(j)} - \x_{\sigma(j+1)}
\eeq
for every permutation $\sigma \in \PermGr_n$;
(2) the \REMemph{translation invariance}:
\beqa\label{101-n101}
\podr
\hspace{-15pt}
\YM_{a_1,\dots,\Tra_{\alpha}(a_j),\dots,a_n} (\DXn{1},\dots,\DXn{n-1})
=
\di_{\Dxn{j}^{\alpha}}\,
\YM_{a_1,\dots,a_n} (\DXn{1},\dots,\DXn{n-1})
\,,
\\ \label{101-n11}
\podr
\hspace{-15pt}
\Tra_{\alpha}\,
\YM_{a_1,\dots,a_n} (\DXn{1},\dots,\DXn{n-1})
=
\mathop{\sum}\limits_{k \,=\, 1}^n
\YM_{a_1,\dots,\Tra_{\alpha}(a_k),\dots,a_n} (\DXn{1},\dots,\DXn{n-1})
\,,
\qquad
\eeqa
where $j=1,\dots,n-1$, $\alpha=1,\dots,\DMN$ and
$\Tra_{\alpha}$ in the left hand side
acts on the coefficients $(\in \VA)$ of the series belonging to the spaces
$\VA[[\DXn{1},$ $\dots,$ $\DXn{n-1}]]$ and their localizations;
(3)~the \REMemph{vacuum property}:
\beqa\label{101-n141}
\podr
\YM_{a_1,\dots,a_{n-1},\vac\,}
(\DXn{1},\dots,\DXn{n-1})
\bnn
\podr=\,
\mathop{\sum}\limits_{k \, = \, 0}^{\infty}
\,
\frac{1}{k!} \,
\Bigl(\mathop{\sum}\limits_{\alpha \,=\, 1}^{\DMN}\, \Dxn{n-1}^{\alpha} \, \Tra_{\alpha}\Bigr)^k
\YM_{a_1,\dots,a_{n-1}}
(\DXn{1}-\DXn{n-1},\dots,\DXn{n-2}-\DXn{n-1})
\bnn
\podr=:\,
e^{\DXn{n-1} \,\cdot\, \TRa} \
\YM_{a_1,\dots,a_{n-1}}
(\DXn{1}-\DXn{n-1},\dots,\DXn{n-2}-\DXn{n-1})
\qquad
\eeqa
and in particular,
\beqa\label{101-n121}
\YM_{a_1,\dots,a_{n-1},\vac\,} 
\podr
\,\in\,
\VA[[\DXn{1},\dots,\DXn{n-1}]][1/\TQu_{n-1}]
\nnb 
\podr
\,\subsetneqq\,
\VA[[\DXn{1},\dots,\DXn{n-1}]][1/\TQu_n]
\,,
\qquad\qquad
\\ 
\label{101-n131}
\YM_{a_1,\dots,a_{n-1}}
(\DXn{1},\dots,\DXn{n-2})
\podr
\,=\,
\YM_{a_1,\dots,a_{n-1},\vac\,}
(\DXn{1},\dots,\DXn{n-2},\DXn{n-1})
\,\Bigl|\raisebox{-6pt}{}_{\DXn{n-1} \,=\, 0}
\,
\nnb
\eeqa
(note that Eq.~(\ref{101-n121}) allows us to specialize $\DXn{n-1}$ $=$ $0$ in (\ref{101-n131})).

Since for the $n$--point vector functions $\YM_{a_1,\dots,a_n}$ we needed $n-1$ differences obtained by the change (\ref{eq9.2}) of formal variables
$(\x_1,$ $\dots,$ $\x_n)$ $\mapsto$ $(\DXn{1},$ $\dots,$ $\DXn{n-1},$ $\x_n)$,
then for $n+1$ one points, when we consider
$\YM_{a_1,\dots,a_{n+1}}$,
one needs $n$ differences,
\beq\label{eqn10.15}
\DXsn{j} \,=\, \ \x_j - \x_{n+1}
\quad \text{for} \quad
j \,=\, 1,\dots,n
\,.
\eeq
Setting $a_{n+1} = \vac$, since the 
$\YM_{a_1,\dots,a_n,\vac}$ depends only on $n$ formal vector variables we shall denote with a slight abuse in the notations these variables with $\x_1,$ $\dots,$ $\x_n$
instead of $\DXsn{1},$ $\dots,$ $\DXsn{n}$.
In this way,
Eq.~(\ref{101-n141}) for $n+1$ reads
\beqa\label{eqn10.16}
\YM_{a_1,\dots,a_n,\vac\,}
(\x_1,\dots,\x_n)
\,=\podr
e^{\x_n \cdot\, \TRa} \
\YM_{a_1,\dots,a_n}
(\DXn{1},\dots,\DXn{n-1})
\,,
\\
\label{eqn10.16-1-1028}
\YM_{a_1,\dots,a_n,\vac\,}
(\x_1,\dots,\x_n)
\,\in\podr
\VA[[\x_1,\dots,\x_n]][1/\TQu_n]
\eeqa
with the usual relation $\DXn{j} = \x_j-\x_n$ ($j=1,\dots,n-1$).
Now, the permutation equivariance (\ref{101-n10}) take more simple form
\beq\label{eq10.17nn1809m}
\YM_{a_1,\dots,a_n,\vac\,}
(\x_1,\dots,\x_n)
\,=\,
\YM_{a_{\sigma(1)},\dots,a_{\sigma(n)},\vac\,}
\bigl(\x_{\sigma(1)},\dots,\x_{\sigma(n)}\bigr)
\,.
\eeq
As a consequence of Eqs.~(\ref{101-n101}) and (\ref{101-n11}) one has
\beq\label{eqn10.18nnnn}
\Tra_{\alpha}\,
\YM_{a_1,\dots,a_n,\vac\,} (\x_1,\dots,\x_n)
\,=\,
\mathop{\sum}\limits_{j \,=\, 1}^n
\di_{x_j^{\alpha}}\,
\YM_{a_1,\dots,a_n,\vac\,} (\x_1,\dots,\x_n)
\,.
\eeq
A particular case of (\ref{eqn10.18nnnn}) is
\beq\label{eqn1-.10ixde}
\Tra_{\alpha} \, 
\YM_{a,\vac\,} (\x)
\,=\,
\di_{x^{\alpha}} \,
\YM_{a,\vac\,} (\x)
\qquad
(\YM_{a,\vac\,} (\x) \,=\, \YB(a,\x)\vac)
\,.
\eeq
The assignment
\beq\label{eqn10.20xid2}
\VA \,\ni\, a \,\mapsto\,
\YM_{a,\vac\,} (\x) \,\in\, \VA[[\x]]
\eeq
is an injection and the image of (\ref{eqn10.20xid2}) consists of all series $u (\x) \in \VA[[\x]]$ that are \REMemph{translation invariant}:
\beq\label{eqn10.2123c3ixde}
\Tra_{\alpha} \, 
u (\x)
\,=\,
\di_{x^{\alpha}} \,
u(\x)
\,.
\eeq

\SSECSPA

\subsection{Vertex algebras as $\OpADiff$--algebras}
\ADDCONT{Vertex algebras as $\OpADiff$--algebras}

\noindent
The vertex algebras describe the OPE of translation invariant local fields in GCI QFT.
Therefore, they are directly related with the operad $\TIOpADiff$ (introduced in Theorem~\ref{Th9.1}), where we separate the translations.

\medskip

\begin{THEOR}{the101-n2aa}
{\it
With every vertex algebra $\VA$ we can associate 
an algebra 
over
the operad $\TIOpADiff$ in the following way. 
For every $n=1,2,\dots$ and 
$a_1,\dots,a_n \in \VA$ we set$:$
\beqa\label{Eq10.14nnaa}
&
\VaRep{\VA} \,:\, 
\TIOpADiff (n) \,\to\, 
\OpEnd_{\vA}(n)
\,(\,=\, \Hom_{\GRF}(\vA^{\otimes n},\vA)
\,:\,
\APDOa \,\mapsto\,
\VaRep{\VA}(\APDOa)
\,,
&\qquad\qquad
\\ \label{Eq10.15nnaa}
&
\VaRep{\VA} (\APDOa) 
\bigl(a_1,\dots,a_n\bigr)
\,:=\,
\APDOa_{\VA} \bigl(\YM_{a_1,\dots,a_n,\vac\,}\bigr)
\,\Bigl|\raisebox{-6pt}{\,}_{\x \,=\, 0}
\,\in\,
\VA
\,,
&
\eeqa
where $\APDOa_{\VA}$ is provided by Theorem~$\ref{ThmX1}$.}
\end{THEOR}

\medskip

As the {\it proof} is more technical we leave it to \ref{ApSe-1}.

\medskip

There is a natural generalization of Theorem~\ref{the101-n2aa} to general local fields that are not translation invariant.
It is based on the following observation.
According to translation invariant properties (\ref{eqn10.18nnnn}) and (\ref{eqn9.6khdlea}) it follows that 
\beq\label{12WF43Q66}
\APDOa_{\VA} \bigl(\YM_{a_1,\dots,a_n,\vac\,}\bigr)
\,\in\,
\VA [[\x]]
\eeq
is a translation invariant vector formal power series for $\APDOa$ $\in$ $\TIOpADiff (n)$.
As we have mentioned above (cf. Eq.~(\ref{eqn10.2123c3ixde})) the latter are in one-to-one correspondence with the elements of $\VA$.
Then,
let us relax the assumptions of
Theorem~\ref{the101-n2aa} by removing the condition of translation invariance and passing to the full operad~$\OpADiff$.
To this end let us make some conventions.
For a vertex algebra $\VA$ we set
\beq\label{eqn10.17}
\vA
\,:=\,
\VA[[\x]]
\,
\eeq
and we consider the unique extension of the map
\beq\label{eqn10.18}
\begin{array}{ccccc}
\VA & \otimes\cdots\otimes & \VA & \to & \VA [[\x_1,\dots,\x_n]][1/\TQu_n]
\,,
\raisebox{-6pt}{}
\\
a_1 & \otimes\cdots\otimes & a_n
& \mapsto &
\YM_{a_1,\dots,a_n,\vac} (\x_1,\dots,\x_n)
\phantom{\,,}
\end{array}
\eeq
to a $\GRF[\x_1,\dots,\x_n]$--linear map
\beq\label{eqn10.19}
\VA [\x_1]  \otimes\cdots\otimes \VA[\x_n] \, \to \, \VA [[\x_1,\dots,\x_n]][1/\TQu_n]
\,,
\eeq
which further possess a unique extension 
to a $\GRF[[\x_1]]$ $\otimes$ $\dots$ $\otimes$ $\GRF[[\x_n]]$--linear map
\beq\label{eqn10.20}
\begin{array}{ccccc}
\vA & \otimes\cdots\otimes & \vA & \to & \VA [[\x_1,\dots,\x_n]][1/\TQu_n]
\,,
\raisebox{-6pt}{}
\\
u_1 & \otimes\cdots\otimes & u_n
& \mapsto &
\TYM_{u_1,\dots,u_n,\vac} (\x_1,\dots,\x_n)
\,.
\end{array}
\eeq
Let us also introduce the space of translation invariant vector formal power series
\beq\label{eq10.30r23}
\TAvA
\,:=\,
\bigl\{
u(\x) \,\in\, \vA 
\,\bigl|\,
\di_{x^{\alpha}} u(\x) \,=\, \Tra_{\alpha} u(\x)
\ (\forall \alpha)
\bigr\}
\,,
\eeq
which, as we have pointed out above, is naturally isomorphic to $\VA$:
\beq\label{eqc3v10.29xfdv3}
\TAvA \,\cong\, \VA
\,:\quad
\TAvA \,\ni\, u(\x) 
\,\mapsto\, u(\x)
\hspace{1pt}\bigl|\raisebox{-5pt}{\hspace{1pt}}_{\x \,=\, 0} \,\in\, \VA
\,,\quad
\VA \,\ni\, a
\,\mapsto\, e^{\x \cdot \TRa} a
\,\in\, \TAvA
\,.
\eeq

With this conventions we state the following:

\nopagebreak\medskip\nopagebreak

\begin{COROL}{the101-n2}
{\it
With every vertex algebra $\VA$ we can associate 
on $\vA$ $(= \VA[[\x]])$
a structure of
an algebra 
over 
the operad $\OpADiff$ in the following way. 
For every $n=1,2,\dots$ and 
$u_1,\dots,u_n \in \vA$ we set$:$%
\beqa\label{Eq10.14nn}
&
\VaRep{\VA} \,:\, 
\OpADiff (n) \,\to\, 
\OpEnd_{\vA}(n)
\,(\,=\, \Hom_{\GRF}(\vA^{\otimes n},\vA)
\,:\,
\APDOa \,\mapsto\,
\VaRep{\VA}(\APDOa)
\,,
&\qquad\qquad
\\ \label{Eq10.15nn}
&
\VaRep{\VA} (\APDOa) 
\bigl(u_1,\dots,u_n\bigr)
\,:=\,
\APDOa_{\VA} \bigl(\TYM_{u_1,\dots,u_n,\vac\,}\bigr)
\,\in\,
\vA
\,.
&
\eeqa
Then, the subspace of $\vA$ consisting of translation invariant vector formal power series $\TAvA$ $(\ref{eq10.30r23})$ is an algebra for the suboperad $\TIOpADiff$, which is isomorphic to the algebra constructed in Theorem~\ref{the101-n2aa}.}
\end{COROL}

\section{Outlook}\label{Se13nn}
\SETCNTR

The main focus of this work is on the new generalization of the notion of a differential operator and how this concept serves to describe the OPE operations and their structure. We will list now some topics that have fallen outside the scope of our research here, as well as plans for future developments.

\medskip

a) We left open the question whether there are more general algebras over the operad $\TIOpADiff$, which do not correspond to vertex algebras under Theorem~\ref{the101-n2aa}. Here are some guidelines to work towards.
One can try to use a truncated operad structure up to the third operadic space, i.e., one can consider only binary operations and the relations among them that appear in the third operadic space. This is motivated by the fact that in the vertex algebras one has a generalized associativity that allows us to consider these algebras as algebras defined by binary operations and quadratic relations among them. According to this strategy one can focus on the structures on commutative associative algebras and modules of two and three point functions, which is facilitated by some techniques of Conformal Field Theory.

\medskip

b) One can axiomatize the construction of the operad $\OpADiff$ in Sect.~\ref{s6.2}.
It is based on considering
more general sequences $\COalgn{n}$ ($n=1,2,\dots$) of modules, each $\COalgn{n}$ being a $(\CSalg^{\,\otimes\hspace{1pt}n})$--module for some unital commutative associative algebra $\CSalg$.
The $(\CSalg^{\,\otimes\hspace{1pt}n})$--module $\COalgn{n}$ should contain the algebra $(\CSalg^{\,\otimes\hspace{1pt}n})$ as a submodule and
the key property that it should obey is the uniqueness of the extension of the differential operators from $(\CSalg^{\,\otimes\hspace{1pt}n})$ to 
$\COalgn{n}$ (analogously to Theorem~\ref{th3.1}).
We call the spaces $\COalgn{n}$ ($n=1,2,\dots$) \DEFemph{OPE function spaces} and they can be thought as specifying a class of QFT models.
A first application of such more general theory can be a local theory of OPE algebras over curved space--time (as a fixed background).
By ``local theory'' we mean that one can work with germs at a point
(i.e., in a \REMemph{formal} neighborhood of the point). 
Then the starting commutative associative algebra $\CSalg$ will be the algebra of all formal power series of the coordinates around the fixed point.

\medskip

c) Another application of an axiomatized OPE algebras  (according to the above project in ``b)'') will be a construction of OPE function spaces that allows us to solve fields' equation for vertex algebras.
We propose the following point of view.
We consider the OPE function spaces $\Oalg_n$ of Eq.~(\ref{eq10.2-1103}) as a ``minimal physical choice''.
It defines the class of QFT models with GCI (\cite{N05,NT01}).
These are quantum fields for which all the signals and correlations can propagate exactly with the speed of light.
Extensions for this class of QFT models can be obtained via some extensions of the OPE function spaces, $\COalgn{n}$ $\supseteq$ $\Oalg_n$.
These can be algebraic extensions or differential algebraic extensions.
The latter case namely will appear if we want to solve fields' equations using as a source for the free field equations composite fields that are defined via OPE operations from the initial fields.

\medskip

d) One can continue project ``a)'' by developing a structure theory for the algebras over the operad $\TIOpADiff$ (truncated up to the third operadic space).
One aspect of such a theory is to develop a \REMemph{cohomological investigation} of the \TsediPRODterm operations.
According to \cite[Sect.~3.2]{N09} the de Rham cohomologies of \Tsedi operators over the spaces $\Oalg_n$ (\ref{eq10.2-1103}) are finite dimensional and they are dual to the algebraic de Rham cohomology spaces of the commutative associative algebras $\Oalg_n$ (cf. also \cite{N10}).
This will lead to a finite dimensional operad defined as a certain cohomology operad for the operad $\OpADiff$.
One can thought of the latter cohomology operad as a ``master operad'' that can generate models for quantum fields in higher dimensions similarly to the situation in two dimensional Conformal Field Theory where there are plenty of models induced by certain Lie algebras and their representations.

\begin{acknowledgements}
This work began as a joint project with Jean--Louis Loday. 
His demise ended the fruitful cooperation that had begun. 
The author owes Jean--Louis Loday the idea of ​​using operad theory in this study. 
This played a crucial role in all subsequent development of ideas.

The author is grateful to Maxim Kontsevich for the useful discussions on various topics related to this work during his visits at Institut des Hautes \'Etudes Scientifiques (IH\'ES, Bures-sur-Yvette, France).
The author is grateful also to Bojko Bakalov and Milen Yakimov for many useful discussions on this work.
The author thanks
Lilia Angelova, Ludmil Hadjiivanov, Petko Nikolov, Todor Todorov and Svetoslav Zahariev for their reading of the manuscript and for suggesting several improvements and corrections.
This work was supported in part by the Bulgarian National Science Fund under research grant DN-18/3.
\end{acknowledgements}

\setcounter{section}{0}

\renewcommand{\thesection}{Appendix~\Alph{section}}
\renewcommand{\theequation}{\Alph{section}.\arabic{equation}}
\renewcommand{\theSTATEM}{\Alph{section}.\arabic{STATEM}}

\section[${}$\hspace{45pt}Proof of Theorem~\ref{the101-n2aa}]{Proof of Theorem~\ref{the101-n2aa}}\label{ApSe-1}
\SETCNTR
\renewcommand{\thesection}{\Alph{section}}
\renewcommand{\theSTATEM}{\Alph{section}.\arabic{STATEM}}
\renewcommand{\theRemk}{\Alph{section}.\arabic{Remk}}

We need to prove that the map $\VaRep{\VA}$ (\ref{Eq10.14nnaa}) is a morphism of (symmetric) operads.
This includes the compatibility (preservation) of the operadic composition, preservation of units and permutation equivariance.
The latter is manifest according to (\ref{eq10.17nn1809m}).
The preservation of the operadic units is also obvious.

We continue with the compatibility of the operadic compositions.
The proof is straightforward, but there are several  hidden natural maps in our constructions, which makes the proof complicated.
The fact that that the map $\VaRep{\VA}$ (\ref{Eq10.14nnaa}) preserves the operadic compositions means that
\beqa\label{eqA.1-nn20190923}
\podr
\VaRep{\VA} (\APDOa') 
\Bigl(
\VaRep{\VA} (\APDOa''_1)\bigl(\DNu{a}{1,1},\dots,\DNu{a}{1,j_1}\bigr),\,
\dots,\,
\VaRep{\VA} (\APDOa''_k)\bigl(\DNu{a}{k,1},\dots,\DNu{a}{k,j_k}\bigr)
\Bigr)
\nnb
\podr = \,
\VaRep{\VA} (\OperComp{j_1,\dots,j_k}{\APDOa'}{\APDOa''_1,\dots,\APDOa''_k})
\bigl(\DNu{a}{1,1},\dots,\DNu{a}{1,j_1},\,
\ldots,\,
\DNu{a}{k,1},\dots,\DNu{a}{k,j_k}\bigr)
\,.
\nnb
\eeqa
In other words, if
\beqa\label{eqA.2-20190923}
b_{\ell} \, :=\podr
\VaRep{\VA} (\APDOa''_{\ell})\bigl(\DNu{a}{\ell,1},\dots,\DNu{a}{\ell,j_{\ell}}\bigr)
\quad (\ell \, = \, 1,\dots,k)
\,,
\nnb
c \, :=\podr
\VaRep{\VA} (\APDOa')\bigl(b_1,\dots,b_k\bigr)
\,,
\nnb
\APDOa
\, := \podr
\OperComp{j_1,\dots,j_k}{\APDOa'}{\APDOa''_1,\dots,\APDOa''_k}
\,,
\quad \text{then:}
\nnb
c \, = \podr
\VaRep{\VA} (\APDOa)\bigl(\DNu{a}{1,1},\dots,\DNu{a}{1,j_1},\,
\ldots,\,
\DNu{a}{k,1},\dots,\DNu{a}{k,j_k}\bigr)
\,.
\eeqa
Now, combining the definition (\ref{Eq10.15nnaa}) of $\VaRep{\VA}$ and the construction of 
$\APDOa_{\VA}$
(\ref{eq10.1}) one has\footnote{%
recall, $\COMPLEt{\bigl(\cdots\bigr)}$ stands for the formal continuation according to Theorem~\ref{ThFS1}}
\beqa\label{eqA.4nn2-20190923}
\podr
b_{\ell} 
\, = \,
\COMPLEt{\bigl((\AUGM_{(\x_{\ell} = 0)} \circ \DMo{(\APDOa''_{\ell})}{{\TQu''_{\ell}}^{-1}}) \otimes \textrm{id}_{\VA}\bigr)}
\bigl(
\FM_{\DNu{a}{\ell,1},\dots,\DNu{a}{\ell,j_{\ell}}}
\bigr)
\,,
\nnb
\podr
\FM_{\DNu{a}{\ell,1},\dots,\DNu{a}{\ell,j_{\ell}}} (\DNu{\x}{\ell,1},\dots,\DNu{\x}{\ell,j_{\ell}})
\, := \,
\TQu''_{\ell} \cdot
\YM_{\DNu{a}{\ell,1},\dots,\DNu{a}{\ell,j_{\ell}},\vac\,} (\DNu{\x}{\ell,1},\dots,\DNu{\x}{\ell,j_{\ell}})
\nnb
\podr
\phantom{\FM_{\DNu{a}{\ell,1},\dots,\DNu{a}{\ell,j_{\ell}}} (\DNu{\x}{\ell,1},\dots,\DNu{\x}{\ell,j_{\ell}}) \,:}
\in \,
\VA[[\DNu{\x}{\ell,1},\dots,\DNu{\x}{\ell,j_{\ell}}]]
\,,
\nnb
\podr
\hspace{54pt}
\TQu''_{\ell} 
\,
:=\,
\bigl(
\TQu_{j_\ell} (\DNu{\x}{\ell,1},\dots,\DNu{\x}{\ell,j_{\ell}})
\bigr)^{N_{\DNu{a}{\ell,1},\dots,\DNu{a}{\ell,j_{\ell}}}}
\,,
\eeqa
for $\ell = 1,\dots,k$.
Here: 
\begin{itemize}
\item[(i)]
roughly speaking, $b_{\ell}$ is obtained by applying the differential operator $\AUGM_{(\x_{\ell}=0)} \circ \DMo{(\APDOa''_{\ell})}{{\TQu''_{\ell}}^{-1}}$ on the formal power series 
$\FM_{\DNu{a}{\ell,1},\dots,\DNu{a}{\ell,j_{\ell}}} (\DNu{\x}{\ell,1},$ $\dots,$ $\DNu{\x}{\ell,j_{\ell}})$
without acting on the coefficients $\in \VA$ of that series.
\item[(ii)]
$\AUGM_{(\x_{\ell}=0)}$ $:$ $F(\x_{\ell})$ $\mapsto$ $F(0)$ is the augmentation $\AUGM$ on $\Salg$ with indicated set of generating formal variables $\x_{\ell}$ on which it acts.
\item[(iii)]
In (i) we assumed that after applying the (continued) differential operator $\DMo{(\APDOa''_{\ell})}{{\TQu''_{\ell}}^{-1}}$ to 
$\FM_{\DNu{a}{\ell,1},\dots,\DNu{a}{\ell,j_{\ell}}} (\DNu{\x}{\ell,1},$ $\dots,$ $\DNu{\x}{\ell,j_{\ell}})$
then the resulting variable is $\x_{\ell}$, which afterwards is set to $0$ by $\AUGM_{(\x_{\ell}=0)}$.
\item[(iv)]
$N_{\DNu{a}{\ell,1},\dots,\DNu{a}{\ell,j_{\ell}}}$ are sufficiently large positive integer numbers, which according to the theory of vertex algebras will ensure that 
$\FM_{\DNu{a}{\ell,1},\dots,\DNu{a}{\ell,j_{\ell}}} (\DNu{\x}{\ell,1},$ $\dots,$ $\DNu{\x}{\ell,j_{\ell}})$ $\in$ $\VA[[\DNu{\x}{\ell,1},$ $\dots,$ $\DNu{\x}{\ell,j_{\ell}}]]$ (cf. Eq.~(\ref{eqn10.16-1-1028})).
\end{itemize}
It follows that
\beqa\label{ax1909nn1}
\podr\hspace{-3pt}
\YM_{b_1,\dots,b_k,\vac\,} (\y_1,\dots,\y_k)
\,
\nnb
\podr\hspace{-3pt}
=\Bigl(
\bigl(
(\AUGM_{(\x_1 = 0)} \circ \DMo{(\APDOa''_1)}{{\TQu''_1}^{-1}})
\otimes \cdots \otimes
(\AUGM_{(\x_k = 0)} \circ \DMo{(\APDOa''_k)}{{\TQu''_k}^{-1}})
\bigr)
\nnb
\podr\hspace{-3pt}
\COMPLEtt{
\otimes\,
\text{id}_{\VA[[\y_1,\dots,\y_k]]
[1/\TQu']
} 
\Bigr)}
\hspace{-2pt}
\Bigl(
\YM_{
\FM_{\DNu{a}{1,1},\dots}(\DNu{\x}{1,1},\dots),\,
\ldots
,\,\FM_{\DNu{a}{k,1},\dots,\DNu{a}{k,j_k}}(\DNu{\x}{k,1},\dots,\DNu{\x}{k,j_k})
,\,\vac\,
}
\bigl(\y_1,
\nnb
\podr\hspace{-3pt}
\phantom{
\COMPLEtt{
\otimes\, \text{id}_{\VA[[\y_1,\dots,\y_k]]
[1/\TQu']
} 
\Bigr)}}
\
\dots,\y_k\bigr)
\Bigr)
\,,
\hspace{-20pt}
\qquad
\eeqa
where similarly to 
(\ref{eqA.4nn2-20190923}),
\beqa\label{eqA.3-20190923}
&
\TQu' \,:=\, 
\bigl(
\TQu_k (\y_1,\dots,\y_k)\bigr)^{N_{b_1,\dots,b_k}}
\quad \text{is such that}
\,,
&
\nnb
&
\FM_{b_1,\dots,b_k} (\y_1,\dots,\y_k)
\, := \,
\TQu' \cdot
\YM_{b_1,\dots,b_k,\vac\,} (\y_1,\dots,\y_k)
\, \in \,
\VA[[\y_1,\dots,\y_k]]
\,.
&\qquad\qquad
\eeqa
In Eq.~(\ref{ax1909nn1}) we encounter the first subtlety related to hidden natural maps.
Before substantiating Eq.~(\ref{ax1909nn1}) we will explain the meaning of the substitution process
\[
\YM_{b_1,\dots,b_k,\vac}
\, \mapsto \,
\YM_{
\FM_{\DNu{a}{1,1},\dots,\DNu{a}{1,j_1}},\,
\ldots,\,
\FM_{\DNu{a}{k,1},\dots,\DNu{a}{k,j_k}}
,\,\vac\,
}
\,
\]
and in what space of series the result belongs to.
To this end, note first that the assignment
$b_1 \otimes \cdots \otimes b_k$ $\mapsto$ 
$\YM_{b_1,\dots,b_k,\vac\,} (\y_1,$ $\dots$ $\y_k)$ 
is a linear map
$\VA^{\otimes k}$ $\to$ 
$V [[\y_1,$ $\dots,$ $\y_k]][1/\TQu_k(\y_1,$ $\dots,$ $\y_k)]$
$=$ $V [[\y_1,$ $\dots,$ $\y_k]][1/\TQu']$.
Applying to this map the functorial assignment
$W$ $\mapsto$$W [[(\DNu{\x}{i',i''})_{i',i''}]]$
we lift the map 
$b_1$ $\otimes$ $\cdots$ $\otimes$ $b_k$
$\mapsto$
$\FM_{b_1,\dots,b_k}$ 
to a linear map
\beqs
\bigl(\VA^{\otimes k}\bigr)
[[(\DNu{\x}{i',i''})_{i',i''}]]
\,
\to\,
\bigl(V [[\y_1, \dots, \y_k]][1/\TQu']
\bigr)
[[(\DNu{\x}{i',i''})_{i',i''}]]
\,.
\eeqs
Composing further the latter map on the right by the maps
\beqs
\VA^{\otimes n} 
\hspace{-1pt}
=
\hspace{-1pt}
\mathop{\text{\large $\otimes$}}\limits_{\ell \,=\, 1}^k
\hspace{-3pt}
\VA^{\otimes j_{\ell}}
\mathop{-\hspace{-5pt}-\hspace{-5pt}-\hspace{-5pt}-\hspace{-5pt}-\hspace{-5pt}-\hspace{-5pt}-\hspace{-5pt}-\hspace{-5pt}-\hspace{-5pt}-\hspace{-5pt}-\hspace{-5pt}-\hspace{-5pt}-\hspace{-5pt}-\hspace{-5pt}-\hspace{-5pt}-\hspace{-5pt}-\hspace{-5pt}-\hspace{-5pt}-\hspace{-5pt}-\hspace{-5pt}-\hspace{-5pt}-\hspace{-5pt}-\hspace{-5pt}-\hspace{-5pt}-\hspace{-5pt}-\hspace{-5pt}-\hspace{-5pt}-\hspace{-5pt}\to}
\limits^{
\mathop{\text{\normalsize $\otimes$}}\limits_{\ell \,=\, 1}^k
(\DNu{a}{\ell,1} \otimes \cdots \otimes \DNu{a}{\ell,j_{\ell}} \,\mapsto\,
\FM_{\DNu{a}{\ell,1},\cdots,\DNu{a}{\ell,j_{\ell}}})
}
\hspace{-3pt}
\podr
\mathop{\text{\large $\otimes$}}\limits_{\ell \,=\, 1}^k
\VA [[\DNu{\x}{\ell,1},\dots,\DNu{\x}{\ell,j_{\ell}}]]
\,\hookrightarrow
\\
\raisebox{2.9pt}{\tiny $\subset$} \hspace{-5pt}-\hspace{-5pt}-\hspace{-5pt}-\hspace{-5pt}-\hspace{-5pt}-\hspace{-5pt}-\hspace{-5pt}-\hspace{-5pt}-\hspace{-5pt}-\hspace{-5pt}-\hspace{-5pt}-\hspace{-5pt}-\hspace{-5pt}-\hspace{-5pt}\to
\hspace{-3pt}
\podr
\bigl(\VA^{\otimes \, k}\bigr) [[(\DNu{\x}{i',i''})_{i',i''}]]
\,
\eeqs
we obtain the linear map
\beqs
\DNu{a}{1,1} \otimes \cdots \otimes \DNu{a}{1,j_1} 
\otimes \text{\large $\cdots$} \otimes 
\DNu{a}{k,1} \otimes \cdots \otimes \DNu{a}{k,j_k}
\,\mapsto
\\
\YM_{
\FM_{\DNu{a}{1,1},\dots,\DNu{a}{1,j_1}},\,
\ldots,\,
\FM_{\DNu{a}{k,1},\dots,\DNu{a}{k,j_k}}
,\,\vac\,}
\,.
\eeqs
In particular,
\beq\label{tem20190923-1}
\YM_{
\FM_{\DNu{a}{1,1},\dots,\DNu{a}{1,j_1}},\,
\ldots,\,
\FM_{\DNu{a}{k,1},\dots,\DNu{a}{k,j_k}}
,\,\vac\,}
\,\in\,
\bigl(V [[\y_1, \dots, \y_k]][1/\TQu']
\bigr)
[[(\DNu{\x}{i',i''})_{i',i''}]]
\,.
\eeq
Then the meaning of Eq.~(\ref{ax1909nn1}) is that
the differential operators 
\\
$\AUGM_{(\x_{\ell} = 0)}$ $\circ$ $\DMo{(\APDOa''_{\ell})}{{\TQu''_k}^{-1}}$
act now on variables $(\DNu{\x}{i',i''})$ in the series (\ref{tem20190923-1}) without touching the coefficient series $\in$
$V [[\y_1,$ $\dots,$ $\y_k]][1/\TQu']$.

Having argued (\ref{ax1909nn1}) we now use another important fact from the theory of vertex algebras called ``\REMemph{the associativity theorem}''.
Here we state its general form without a proof, which is a straightforward generalization of \cite[Theorem~5.1]{BN06}.

\medskip

\begin{THEOR}{THa.1-20190923}
{\it
Let $\VA$ be a vertex algebra and $\TQu''_{\ell}$, 
$\FM_{\DNu{a}{\ell,1},\cdots,\DNu{a}{\ell,j_{\ell}}}$
be set according to Eq.~$(\ref{eqA.4nn2-20190923})$ for  $\DNu{a}{i',i''} \in \VA$
$($$i'$ $=$ $1,$ $\dots,$ $k$, $i''$ $=$ $1,$ $\dots,$ $j_{i'}$$)$.
Then, if we set for short
$\DNu{\Ttx}{i',i''}$ $:=$ $\DNu{\x}{i',i''}+\y_{i'}$ we have
\beqa\label{eqA.7-20190923-nn1}
&
\TQu''_1 \cdots \TQu''_k 
&
\nnb
&
\times\,
\YM_{\DNu{a}{1,1},\dots,\DNu{a}{1,j_1},\ldots,\DNu{a}{k,1},\dots,\DNu{a}{k,j_k},\vac\,}
\bigl(\DNu{\Ttx}{1,1},\dots,\DNu{\Ttx}{1,j_1},\ldots,
\DNu{\Ttx}{k,1},\dots,\DNu{\Ttx}{k,j_k}\bigr)
& \nnb &
\in\,
V [[(\DNu{\Ttx}{i',i''})_{i',i''}]]\bigl[1/\TQu_{j_1|\cdots|j_k}\bigl((\DNu{\Ttx}{i',i''})_{i',i''}\bigr)\bigr]
\,,
&
\eeqa
where $\TQu_{j_1|\cdots|j_k}$ is introduced in Eq.~$(\ref{eq6.10})$.
Furthermore, we have
\beqa\label{eqA.8-20190923-nn1}
\podr
\YM_{
\FM_{\DNu{a}{1,1},\dots,\DNu{a}{1,j_1}}(\DNu{\x}{1,1},\dots,\DNu{\x}{1,j_1}),\,
\ldots,\,
\FM_{\DNu{a}{k,1},\dots,\DNu{a}{k,j_k}}(\DNu{\x}{k,1},\dots,\DNu{\x}{k,j_k})
\,,\,\vac\,}
\bigl(\y_1,\dots,\y_k\bigr)
\nnb
\podr 
=\,
\iota_{\DNu{\x}{1,1},\DNu{\x}{1,2}\dots,\DNu{\x}{k,j_k}} \,
\TQu''_1 \cdots \TQu''_k \,
\bnn
\podr
\times \,
\YM_{\DNu{a}{1,1},\dots
,\DNu{a}{1,j_1},
\ldots,\DNu{a}{k,1},\dots,\DNu{a}{k,j_k},\vac\,}
\bigl(\DNu{\Ttx}{1,1},\dots,\DNu{\Ttx}{1,j_1},\ldots,
\DNu{\Ttx}{k,1},\dots,\DNu{\Ttx}{k,j_k}\bigr)
\,,
\eeqa
where $\iota_{\DNu{\x}{1,1},\DNu{\x}{1,2}\dots,\DNu{\x}{k,j_k}}$
stands for the $($formal$)$ Taylor expansion in $\DNu{\x}{i',i''}:$
\beqa\label{eqA.9-20190923-nn1}
\iota_{\DNu{\x}{1,1},\DNu{\x}{1,2}\dots,\DNu{\x}{k,j_k}}
\,:\podr
V [[(\DNu{\Ttx}{i',i''})_{i',i''}]]\bigl[1/\TQu_{j_1|\cdots|j_k}\bigl((\DNu{\Ttx}{i',i''})_{i',i''}\bigr)\bigr]
\nnb
\podr 
\rightarrow\,
\bigl(V [[\y_1, \dots, \y_k]][1/\TQu']
\bigr)
[[(\DNu{\x}{i',i''})_{i',i''}]]
\,.
\eeqa
}
\end{THEOR}

\begin{REMAR}{rmA-1}
Roughly, Eq.~(\ref{eqA.8-20190923-nn1}) can be written as
\beqs
\podr
\YM_{
\YM_{\DNu{a}{1,1},\dots,\DNu{a}{1,j_1},\,\vac\,}(\DNu{\x}{1,1},\dots,\DNu{\x}{1,j_1}),\,
\ldots,\,
\YM_{\DNu{a}{k,1},\dots,\DNu{a}{k,j_k},\,\vac\,}(\DNu{\x}{k,1},\dots,\DNu{\x}{k,j_k})
\,,\,\vac\,}
\bigl(\y_1,\dots,\y_k\bigr)
\nnb
\podr 
\approx\,
\YM_{\DNu{a}{1,1},\dots
,\DNu{a}{1,j_1},
\ldots,\DNu{a}{k,1},\dots,\DNu{a}{k,j_k},\vac\,}
\bigl(\DNu{\Ttx}{1,1},\dots,\DNu{\Ttx}{1,j_1},\ldots,
\DNu{\Ttx}{k,1},\dots,\DNu{\Ttx}{k,j_k}\bigr)
\,,\quad
\eeqs
($\DNu{\Ttx}{i',i''}$ $:=$ $\DNu{\x}{i',i''}+\y_{i'}$),
where ``$\approx$'' stands for the identification after the expansion (\ref{eqA.9-20190923-nn1}).
\end{REMAR}

\medskip

Continuing we the proof of Theorem~\ref{the101-n2aa}, we apply Eq.~(\ref{eqA.8-20190923-nn1}) to the right hand side of Eq.~(\ref{ax1909nn1}).
We note that any differential operator commutes with the Taylor expansion since any operator of multiplication by a formal variable or any partial derivative commute with the Taylor expansion.
As a result, we can move the differential operators $\AUGM_{(\x_{\ell}=0)} \circ \DMo{(\APDOa''_{\ell})}{{\TQu''_{\ell}}^{-1}}$ in (\ref{ax1909nn1}) inside the Taylor expansion on the right hand side of Eq.~(\ref{eqA.8-20190923-nn1}).
We claim that then Eq.~(\ref{ax1909nn1}) takes the following form:
\beqa\label{eqA.10-20190923nn4-1}
\podr
\YM_{b_1,\dots,b_k,\vac\,} (\y_1,\dots,\y_k)
\,=\,
(\AUGM_{(\x_1=0)} \otimes \cdots \otimes \AUGM_{(\x_{\ell}=0)})
\bnn
\podr
\circ\,
\COMPLEt{
\bigl(
\EXTOP{(\DMo{(\APDOa''_1)}{{\TQu''_1}^{-1}} \otimes \cdots \otimes \DMo{(\APDOa''_k)}{{\TQu''_k}^{-1}})} \otimes \textrm{id}_{\VA}\bigr)}
\Bigl(
\TQu''_1 \cdots \TQu''_k \,
\bnn
\podr
\times \,
\YM_{\DNu{a}{1,1},\dots,\DNu{a}{1,j_1},\ldots,\DNu{a}{k,1},\dots,\DNu{a}{k,j_k}\,,\,\vac\,}
\bigl(\DNu{\x}{1,1}+\y_1,
\DNu{\x}{1,2}+\y_1,\ldots,\DNu{\x}{k,j_k}+\y_k\bigr)
\Bigr)
\,.
\eeqa
Indeed: 
\begin{itemize}
\item[$\bullet$]
when we move the differential operator
$\DMo{(\APDOa''_1)}{{\TQu''_1}^{-1}}$ $\otimes$ $\cdots$ $\otimes$ 
$\DMo{(\APDOa''_k)}{{\TQu''_k}^{-1}}$ 
$=$
$(\APDOa''_1$
$\otimes$ $\cdots$ $\otimes$
$\DMo{\APDOa''_1)}{{(\TQu''_1 \cdots \TQu''_k)}^{-1}}$
before the Taylor expansion in all $\DNu{\x}{i',i''}$ it now acts according to its extension $\EXTOP{(\cdots)}$ due to Theorem~\ref{th3.1}.\footnote{%
the extension is because of the additional denominator 
$1/\TQu_{j_1|\cdots|j_k}\bigl((\DNu{\Ttx}{i',i''})_{i',i''}\bigr)$}
\item[$\bullet$]
The differential operators $\DMo{(\APDOa''_{\ell})}{{\TQu''_{\ell}}^{-1}}$ set all the variables $\DNu{\x}{\ell,s}$ equal to $\x_{\ell}$, which then is set to zero by $\AUGM_{(\x_{\ell}=0)}$.
Thus, the Taylor expansion disappears in the right hand side of Eq.~(\ref{eqA.10-20190923nn4-1}).
\end{itemize}
We can further remove all the $\TQu''_{\ell}$ from $\DMo{(\APDOa''_{\ell})}{{\TQu''_{\ell}}^{-1}}$
(following the construction of Theorem~\ref{th4.5}).
Thus, we obtain 
\beqa\label{eqA.10-20190923nn4}
\podr
\YM_{b_1,\dots,b_k,\vac\,} (\y_1,\dots,\y_k)
\,=\,
(\AUGM_{(\x_1=0)} \otimes \cdots \otimes \AUGM_{(\x_{\ell}=0)})
\nnb
\podr
\circ\,
\bigl(
\EXTOP{(\APDOa''_1 \otimes \cdots \otimes \APDOa''_k)}
\bigr)_{\VA}
\Bigl(
\YM_{\DNu{a}{1,1},\dots,\DNu{a}{1,j_1},\ldots,\DNu{a}{k,1},\dots,\DNu{a}{k,j_k}\,,\,\vac\,}
\bigl(\DNu{\x}{1,1}+\y_1,
\nnb
\podr
\phantom{\circ\,
\bigl(
\EXTOP{(\APDOa''_1 \otimes \cdots \otimes \APDOa''_k)}
\bigr)_{\VA}
\Bigl(}
\DNu{\x}{1,2}+\y_1,\ldots,\DNu{\x}{k,j_k}+\y_k\bigr)
\Bigr)
\,,
\eeqa
where
$$
\bigl(
\EXTOP{(\APDOa''_1 \otimes \cdots \otimes \APDOa''_k)}
\bigr)_{\VA}
\,:=\,
\COMPLEt{\bigl(
\EXTOP{(\APDOa''_1 \otimes \cdots \otimes \APDOa''_k)}
\otimes \textrm{id}_{\VA}
\bigr)}
$$
is the same type of extension as those constructed in Theorem~\ref{ThmX1}.

Now, in the right hand side of Eq.~(\ref{eqA.10-20190923nn4}) we can omit all the $\AUGM_{(\x_{\ell}=0)}$ just by replacing $\y_1,$ $\dots,$ $\y_k$ by $\x_1,$ $\dots,$ $\x_k$, respectively.
We obtain:
\beqa\label{eqA.10-20190923nn45}
\podr
\YM_{b_1,\dots,b_k,\vac\,} (\x_1,\dots,\x_k)
\bnn
\podr
=\,
\bigl(
\EXTOP{(\APDOa''_1 \otimes \cdots \otimes \APDOa''_k)}
\bigr)_{\VA}
\Bigl(
\YM_{\DNu{a}{1,1},\dots,\DNu{a}{1,j_1},\ldots,\DNu{a}{k,1},\dots,\DNu{a}{k,j_k}\,,\,\vac\,}
\bigl((\DNu{\x}{i',i''})_{i',i''}\bigr)
\Bigr)
\,.
\eeqa
Finally,
\beqs
c \,=\podr
\bigl(\AUGM_{(\x = 0)} \circ (\APDOa')_{\VA}\bigr)
\Bigl(\YM_{b_1,\dots,b_k,\vac\,} (\x_1,\dots,\x_k)\Bigr)
\nnb
=\podr
\Bigl(\AUGM_{(\x = 0)} \circ
\bigl(
\APDOa' \circ
\EXTOP{(\APDOa''_1 \otimes \cdots \otimes \APDOa''_k)}
\bigr)_{\VA}
\Bigr)
\Bigl(
\YM_{\DNu{a}{1,1},\dots,\DNu{a}{1,j_1},\ldots,\DNu{a}{k,1},\dots,\DNu{a}{k,j_k}\,,\,\vac\,}
\Bigr)
\eeqs
and this is exactly (\ref{eqA.2-20190923}).


\begin{thebibliography}{WW00}

\bibitem[AM69]{AM69}
Atiyah, M.F., and MacDonald, I.G.: {\it Introduction to commutative Algebra}, Addison-Wesley Publishing Complany, Reading, Massachusetts (1969)

\bibitem[BDHK18]{BDHK18}
Bakalov, B., De Sole, A., Heluani, R., and Kac, V.G.:
An operadic approach to vertex algebra and Poisson vertex algebra cohomology,
arXiv:1806.08754

\bibitem[BN06]{BN06}
Bakalov, B., and Nikolov, N.M.: Jacobi Identity for Vertex Algebras in Higher Dimensions, Journ. Math. Phys. {\bf 47}:5,  053505 (2006)

\bibitem[BN08]{BN08}
Bakalov, B., and Nikolov, N.M.: Constructing models of vertex algebras in higher dimensions. In: Dobrev, V.K.  et al, (eds.): \textit{Lie Theory and Its Applications in Physics VII}, Heron Press, Sofia, 2008

\bibitem[BD04]{BD04}
Beilinson, A., and Drinfeld, V.: {\it Chiral algebras}, 
American Mathematical Society Colloquium Publications {\bf 51}, 
American Mathematical Society, Providence, RI, 2004

\bibitem[B86]{B86}
Borcherds, R E.: Vertex algebras, Kac--Moody algebras, and the Monster,” Proc. Natl. Acad. Sci. U.S.A. \textbf{83}, 3068--3071 (1986)

\bibitem[B98]{B98}
Borcherds, R.E.: Vertex algebras.
In: 
Kashiwara, M. (ed.):
\textit{Topological Field Theory, Primitive Forms and Related Topics},
Progr. Math. \textbf{160}, 35--77, Birkh\"auser, Boston, MA, 1998

\bibitem[FLM88]{FLM88}
Frenkel, I.B., Lepowsky, J., and Meurman, A.: 
\textit{Vertex Operator Algebras and the Monster}, 
Pure and Applied Math. \textbf{134}, Academic, Boston, MA, 1988

\bibitem[HS69]{HS69}
Heyneman, R.G., and Sweedler, M.E.:
{Afine Hopf Algebras, I,}
Journ. of Algebra \textbf{13}, 192--241  (1969)

\bibitem[HL93]{HL93}
Huang, Y.-Z., Lepowsky, J.: Vertex operator algebras and operads, arXiv:hep-th/9301009

\bibitem[K98]{K98}
Kac, V.: \textit{Vertex algebras for beginners}, University Lecture Series {\bf 10} (2 ed.), AMS, Providence, Rhode Island (1998)

\bibitem[L10]{L10}
Loday, J.-L.: 
On the operad of associative algebras with derivation, Georgian Mathematical Journal. {\bf 17}, (2010)

\bibitem[LN12]{LN12}
Loday, J.-L., and Nikolov, N.M.:
Operadic construction of the renormalization group,
Springer Proceedings in Mathematics \& Statistics {\bf 36}, 191--211, 
(2013)

\bibitem[LV12]{LV12}
Loday J.-L., and Vallette B.:
\textit{Algebraic Operads}, Springer 2012

\bibitem[M80]{M80}
Matsumura, H.: \textit{Commutative Algebra},
The Benjamin/Cummings publishing company, Inc.,
London -- Amsterdam -- Don Mills, Ontario -- Sydney -- Tokyo 
(1980)

\bibitem[N05]{N05}
Nikolov, N.M.: Vertex algebras in higher dimensions and globally conformal invariant quantum field theory,
Commun. Math. Phys. \textbf{253}, 283--322 (2005) 

\bibitem[N09]{N09}
Nikolov, N.M.:
Anomalies in quantum field theory and cohomologies in configuration spaces, arXiv:0903.0187 [math-ph];
Talk on anomaly in quantum field theory and cohomologies of configuration spaces, arXiv:0907.3735 [hep-th]

\bibitem[N10]{N10}
Nikolov, N.M.: 
Cohomologies of Configuration Spaces and Higher Dimensional Polylogarithms in Renormalization Group Problems, AIP Conf. Proc. {\bf 1243}, 165-178 (2010)

\bibitem[N14]{N14}
Nikolov, N.M.: 
Operadic bridge between renormalization theory and vertex algebras, 
Springer Proceedings in Mathematics \& Statistics \textbf{111}, 457-463 (2014)

\bibitem[N16]{N16}
Nikolov, N.M.:
Vertex Algebras in Higher Dimensions Are Homotopy Equivalent to Vertex Algebras in Two Dimensions, 
Springer Proceedings in Mathematics \& Statistics  \textbf{191}, 523-530 (2016)

\bibitem[NST14]{NST14}
Nikolov, M.N., Stora R., and Todorov, I.: 
Renormalization of massless Feynman amplitudes in configuration space, 
Rev. Math. Phys. \textbf{26}, 1430002 (2014); arXiv:1307.6854 [hep-th]

\bibitem[NT01]{NT01}
Nikolov, N.M., and Todorov I.T.:
Rationality of conformally invariant local correlation functions on compactified Minkowski space, Commun. Math. Phys. \textbf{218}, 417--436 (2001) 

\bibitem[Sm86]{Sm86}
Smith, S.P.: Differential  Operators  on  Commutative  Algebras,
Ring Theory Conf., (Antwerp, 1985), pp. 165-177, Lecture Notes in Math., 1197, Springer, Berlin, 1986

\bibitem[SZ60]{SZ60}
Samuel, O., and Zariski, P.: \textit{Commutative Algebra}, vol. 2, D. Van Nostrand Company, Inc., 
Princeton, New Jersey, New York, London, Toronto
(1960)

\bibitem[W69]{W69}
Wilson, K.G.:
Non-Lagrangian Models of Current Algebra,
Phys. Rev. \textbf{179}, 1499--1512 (1969)

\end{thebibliography}
\end{document}